\documentclass[11pt]{prop} % MM proposal style

\usepackage[super,comma]{natbib} % bibliography
\usepackage{apjfonts} %use times font for math
\usepackage{graphicx} % Graphics
\usepackage{amssymb,amsmath} % Math/Greek symbols
\usepackage{color} % colorize text
\usepackage{enumitem} % better lists
\usepackage{wrapfig} % Wrap text around figures
\usepackage{tikz} % allows inset image in title page

\usepackage{multirow} % combine multiple rows in a table
\usepackage{tabularx} % better tables
\usepackage{array} % for tables

\usepackage{hyperref} % links within the document
\hypersetup{
    colorlinks,
    linkcolor={blue!80!black},
    citecolor={blue!80!black},
    urlcolor={blue!80!black}
}

\usepackage{sidecap}
\sidecaptionvpos{figure}{c} 

\newcolumntype{L}[1]{>{\raggedright\let\newline\\\arraybackslash\hspace{0pt}}p{#1}}
\newcolumntype{C}[1]{>{\centering\let\newline\\\arraybackslash\hspace{0pt}}p{#1}}
\newcolumntype{R}[1]{>{\raggedleft\let\newline\\\arraybackslash\hspace{0pt}}p{#1}}
\renewcommand\arraystretch{1.3}

\definecolor{headcolor}{rgb}{0.65,0.65,0.65}
\usepackage{fancyhdr}
\renewcommand{\topmargin}{-9mm}

\renewcommand{\textfloatsep}{4mm}

\newcommand{\lem}{\textcolor{blue}{{LEM}}}
\newcommand{\LEM}{\textcolor{blue}{{LEM}}}

\newcommand{\bsf}{\sffamily\bfseries}

\definecolor{callout}{rgb}{0.25,0.40,0.85}
\definecolor{synergies}{rgb}{0.20,0.45,0.99}
\definecolor{methods}{rgb}{0.20,0.70,0.45}

\definecolor{calllem}{rgb}{0.20,0.45,0.99}
\definecolor{tabledef}{rgb}{0.95,0.95,0.95}
\definecolor{tablealt}{rgb}{0.77,0.80,1.0}
\definecolor{tablelem}{rgb}{0.80,0.85,1.0}
\definecolor{whitelem}{rgb}{1.0,1.0,1.0}
\definecolor{greenlem}{rgb}{0.7,1.0,0.7}

\usepackage{comment}
\usepackage[export]{adjustbox}

\usepackage{movie15}
\usepackage{color}

\begin{document}

% this is single-space for 11pt font: 11*1.2=13.2
\baselineskip=13.2pt
% (and baselinestretch is 1.0)
\sloppy
\pagenumbering{roman}
\thispagestyle{empty}
%\pagecolor{black}

%%%%%%%%%%%%%%%%%%%%%%%%%%%%%%%%%%%%%%%%%%%%%%%%%%%%%%%%%%%%%%%
%%% cover page
%\begin{figure}[t]
%\vspace*{15mm}
%\centering
%\vspace*{-25mm}
%\hspace*{-26mm}
%\includegraphics[width=8.5in]{figures/cad_overlay3a.pdf}

%\label{fig:cad_ovlay}
%\end{figure}
%%%%%%%%%%%%%%%%%%%%%%%%%%%%%%%%%%%%%%%%%%%%%%%%%%%%%%%%%%%%%%%

%%%%%%%%%%%%%%%%%%%%%%%%%%%%%%%%%%%%%%%%%%%%%%%%%%%%%%%%%%%%
\title{\textcolor{black}{\sf\huge \textcolor{blue}{\sl LEM ALL-SKY SURVEY}\\
\vspace{5mm} Soft X-ray Sky at Microcalorimeter Resolution}\footnote{Corresponding authors: Ildar Khabibullin (ildar@mpa-garching.mpg.de) and Massimiliano Galeazzi (galeazzi@miami.edu)}}
\maketitle

\begin{tikzpicture}[remember picture,overlay]
\node[anchor=north west,yshift=2pt,xshift=2pt]%
    at (current page.north west)
    {\includegraphics[height=20mm]{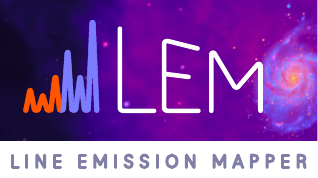}};
\end{tikzpicture}

% \vspace*{2mm}
% \centerline{\textcolor{headcolor}{\sffamily\bfseries\large
% A MISSION CONCEPT FOR THE NASA 2023 ASTROPHYSICS PROBE AO}}

%START HERE TO SWICTH TO ALTERNATE AFFILIATION
%\begin{comment}

% authors: (for some reason, \author{} doesn't work)
\vspace*{-35mm}
\begin{center}
\begin{minipage}{17.5cm}
\hspace{-5mm}
\centering

% Do not worry about the order of authors, just add your name sequentially and we will order it later

Ildar~Khabibullin\footnote{Universitäts-Sternwarte, Fakultät für Physik, Ludwig-Maximilians-Universität München, Scheinerstr.1, 81679 München, Germany}$^,$\footnote{Max Planck Institute for Astrophysics, Karl-Schwarzschild-Str. 1, D-85741 Garching, Germany}$^{,*}$,
Massimiliano~Galeazzi\footnote{~University of Miami, Department of Physics, P.O. Box 248046, Coral Gables, FL 33124, USA}$^{,*}$,
Akos~Bogdan\footnote{Center for Astrophysics, Harvard \& Smithsonian, 60 Garden St., MS-3, Cambridge, MA 02138, USA},
Jenna~M.~Cann\footnote{Oak Ridge Associated Universities, NASA NPP Program, Oak Ridge, TN 37831, USA},
Eugene~Churazov\footnote{Max Planck Institute for Astrophysics, Karl-Schwarzschild-Str. 1, D-85741 Garching, Germany},
Klaus~Dolag\footnote{Universitäts-Sternwarte, Fakultät für Physik, Ludwig-Maximilians-Universität München, Scheinerstr.1, 81679 München, Germany}$^,$\footnote{Max Planck Institute for Astrophysics, Karl-Schwarzschild-Str. 1, D-85741 Garching, Germany}
Jeremy~J.~Drake\footnote{Lockheed Martin Advanced Technology Center, 3251 Hanover St, Palo Alto, CA 94304},
William~Forman\footnote{Center for Astrophysics, Harvard \& Smithsonian, 60 Garden St., MS-3, Cambridge, MA 02138, USA},
Lars~Hernquist\footnote{Center for Astrophysics, Harvard \& Smithsonian, 60 Garden St., MS-3, Cambridge, MA 02138, USA},
Dimitra~Koutroumpa\footnote{LATMOS-OVSQ, CNRS, UVSQ Paris Saclay, Sorbonne Université, 11 Boulevard d'Alembert, 78280, Guyancourt, France},
Ralph~Kraft\footnote{Center for Astrophysics, Harvard \& Smithsonian, 60 Garden St., MS-3, Cambridge, MA 02138, USA},
K.~D.~Kuntz\footnote{Department of Physics and Astronomy, Johns Hopkins University, 3701 San Martin Drive, Baltimore, MD 21218},
Maxim~Markevitch\footnote{NASA Goddard Space Flight Center, Code 662, Greenbelt, MD 20771, USA},
Dan~McCammon\footnote{University of Wisconsin, Madison, WI 53706, USA},
Anna~Ogorzalek\footnote{NASA Goddard Space Flight Center, Code 662, Greenbelt, MD 20771, USA}$^,$\footnote{Department of Astronomy, University of Maryland, College Park, MD 20742-2421, USA},
Ryan~Pfeifle\footnote{Oak Ridge Associated Universities, NASA NPP Program, Oak Ridge, TN 37831, USA},
Annalisa~Pillepich\footnote{Max-Planck-Institut für Astronomie, Königstuhl 17, 69117 Heidelberg, Germany},
Paul~P.~Plucinsky\footnote{Center for Astrophysics, Harvard \& Smithsonian, 60 Garden St., MS-3, Cambridge, MA 02138, USA},
Gabriele~Ponti\footnote{INAF-Osservatorio Astronomico di Brera, Via E. Bianchi 46, 23807 Merate (LC), Italy}$^,$\footnote{Max-Planck-Institut für extraterrestrische Physik, Giessenbachstrasse, 85748, Garching, Germany},
Gerrit~Schellenberger \footnote{Center for Astrophysics, Harvard \& Smithsonian, 60 Garden St., MS-3, Cambridge, MA 02138, USA},
Nhut~Truong\footnote{NASA Goddard Space Flight Center, Greenbelt, MD 20771, USA}$^,$\footnote{Center for Space Sciences and Technology, University of Maryland,Baltimore County,1000 Hilltop Circle,Baltimore,MD 21250, USA},
Milena~Valentini\footnote{Astronomy Unit, Department of Physics, University of Trieste, via Tiepolo 11, I-34131 Trieste, Italy}$^,$\footnote{INAF – Osservatorio Astronomico di Trieste, via Tiepolo 11, I-34131 Trieste, Italy}, 
Sylvain~Veilleux\footnote{University of Maryland, Department of Astronomy, Physical Sciences Complex, 4296 Stadium Drive College Park, MD 20742-2421},
Stephan~Vladutescu-Zopp\footnote{Universitäts-Sternwarte, Fakultät für Physik, Ludwig-Maximilians-Universität München, Scheinerstr.1, 81679 München, Germany},
Q.~Daniel~Wang\footnote{Department of Astronomy, University of Massachusetts, Amherst, MA 01003, USA},
Kimberly~Weaver\footnote{NASA Goddard Space Flight Center, X-ray Astrophysics Laboratory, Code 662, Greenbelt, MD 20771, USA}

%\end{comment}

%\vfill
%\centerline{\em Draft mission concept, March 2023}
%\clearpage
%\pagecolor{white}

%\twocolumn
\begin{comment}
    
{\footnotesize
\noindent
$^1$~~University of Miami, Department of Physics, P.O. Box 248046, Coral Gables, FL 33124, U.S.A. \\
$^{2}$~~LATMOS-OVSQ, CNRS, UVSQ Paris Saclay, Sorbonne Université, 11 Boulevard d'Alembert, 78280, Guyancourt, France\\
$^3$~~NASA Goddard Space Flight Center, Greenbelt, MD 20771, USA \\
$^4$~~Center for Space Sciences and Technology, University of Maryland, Baltimore County, 1000 Hilltop Circle, Baltimore, MD 21250, USA \\
$^5$~Max Planck Institute for Astrophysics, Karl-Schwarzschild-Str. 1, D-85741 Garching, Germany  \\
$^6$~Space Research Institute (IKI), Profsoyuznaya 84/32, Moscow 117997, Russia \\
$^7$~Universitäts-Sternwarte, Fakultät für Physik, Ludwig-Maximilians-Universität München, Scheinerstr.1, 81679 München, Germany \\
$^8$~University of Maryland, Department of Astronomy, Physical Sciences Complex, 4296 Stadium Drive
College Park, MD 20742-2421 \\
$^9$~NASA Goddard Space Flight Center, X-ray Astrophysics Laboratory (Code 662)\\
$^{10}$University of Wisconsin, Madison, WI 53706\\
$^{11}$ Astronomy Unit, Department of Physics, University of Trieste, via Tiepolo 11, I-34131 Trieste, Italy
 $^{12}$INAF – Osservatorio Astronomico di Trieste, via Tiepolo 11, I-34131 Trieste, Italy
  $^{13}$Center for Astrophysics, Harvard \& Smithsonian, 60 Garden St., MS-3, Cambridge, MA 02138, USA 
 
\phantom{${^52}$}~\textcolor{blue}{\bsf \url{lem-observatory.org}}\\
}
\end{comment}

\end{minipage}

%\begin{center}

%text
\vfill
%\vspace*{5mm}
{\small
\phantom{${^52}$}~\textcolor{blue}{\bsf \href{https://lem-observatory.org}{lem-observatory.org}}\\
\phantom{${^52}$}~\textcolor{blue}{\bsf X / twitter: \href{https://www.twitter.com/LEMXray}{LEMXray}}\\
\phantom{${^52}$}~\textcolor{blue}{\bsf facebook: \href{https://www.facebook.com/LEMXrayProbe}{LEMXrayProbe}}}
\vspace*{5mm}

\end{center}

\begin{comment}

\begin{figure*}
    \centering
    \includegraphics[width=1.0\textwidth]{fig/DMC_depleted.pdf}
    %\caption{Caption}
    %\label{fig:enter-label}
\end{figure*}

\end{comment}

%\begin{figure}
%    \includemovie{13cm}{13cm}{fig/vfield_full_5e5_1e6.gif}
%\end{figure}

%add the figure showing a spectrum of a 10 deg$^2$ to be obtained after the complete survey.

% %%%%%%%%%%%%%%%%%%%%%%%%%%%%%%%%%%%%%%%%%%%%%%%%%%%%%%%%%%%%%%%
% \begin{figure}[b]
% %\vspace*{15mm}
% \includegraphics[width=8cm]{figures/3logos_white.pdf}
% 
% %\vspace*{-10mm}
% \label{fig:logos}
% \end{figure}
% %%%%%%%%%%%%%%%%%%%%%%%%%%%%%%%%%%%%%%%%%%%%%%%%%%%%%%%%%%%%%%%

\clearpage
\twocolumn

% \renewcommand{\contentsname}{ }
% {\sf\small\vspace*{-17mm}
% \tableofcontents
% }
% \clearpage

\setcounter{page}{1}
\pagenumbering{arabic}

%%%
\section{SUMMARY}
\label{sec:summary}
The \textcolor{blue}{Line Emission Mapper} (\lem) is an X-ray Probe for the 2030s equipped with a soft X-ray microcalorimeter with spectral resolution $\sim$ 2~eV FWHM from 0.2 to 2.5 keV (Current Best Estimate - CBE) and a focusing X-ray telescope with effective area $>1,500$ cm$^2$ at 0.5 keV and $>2,500$ cm$^2$ at 1~keV (CBE), covering a $33'$~arcmin diameter Field of View with 15$''$ angular resolution, which will be capable of performing efficient scanning observations of very large sky areas\cite{2022arXiv221109827K}. This unique capability will enable the first high spectral resolution survey of the full sky, using 10\% of the observing time budget over the first five years of the mission. 

The \textcolor{blue}{\lem~All-Sky Survey (LASS)} is expected to follow the success of previous all sky surveys such as \textit{ROSAT}\cite{1982AdSpR...2d.241T} and \textit{SRG}/eROSITA\cite{2012arXiv1209.3114M,2021A&A...647A...1P,2021A&A...656A.132S}, adding a third dimension provided by the high resolution microcalorimeter spectrometer, with each 15 arcsec pixel of the survey including a full 1-2 eV resolution energy spectrum that can be integrated over any area of the sky to provide statistical accuracy.
Like its predecessors, \textcolor{blue}{LASS} will provide both long-lasting legacy and open the door to the unknown, enabling new discoveries and providing the baseline for unique GO studies. 
No other current or planned mission has the combination of microcalorimeter energy resolution and large grasp to cover the whole sky 
while maintaining good angular resolution and imaging capabilities.

%%%
\textcolor{blue}{LASS} will be able to probe the physical conditions of the hot phases of the Milky Way at multiple scales, from emission in the solar system due to Solar Wind Charge eXchange (SWCX), to the interstellar and circumgalactic media, including such prominent features as the North Polar Spur and the Fermi/eROSITA bubbles. It will also measure velocities of gas in the inner part of the Galaxy and extract the emissivity of the Local Hot Bubble. 
By maintaining the original angular resolution, \textcolor{blue}{LASS} will also be able to study classes of point sources (e.g., stars, AGNs, distant clusters and group) through stacking. For example for classes with $\sim10^4$ objects, it will provide the equivalent of 1~Ms of high spectral resolution data.
In this White Paper, we describe the technical specifications of \textcolor{blue}{LASS} and highlight the main scientific objectives that will be addressed with the accumulated data. 

\section{INTRODUCTION}
\label{sec:introduction}

\begin{figure}
    \centering
         \includegraphics[angle=0, width=1.0\columnwidth]{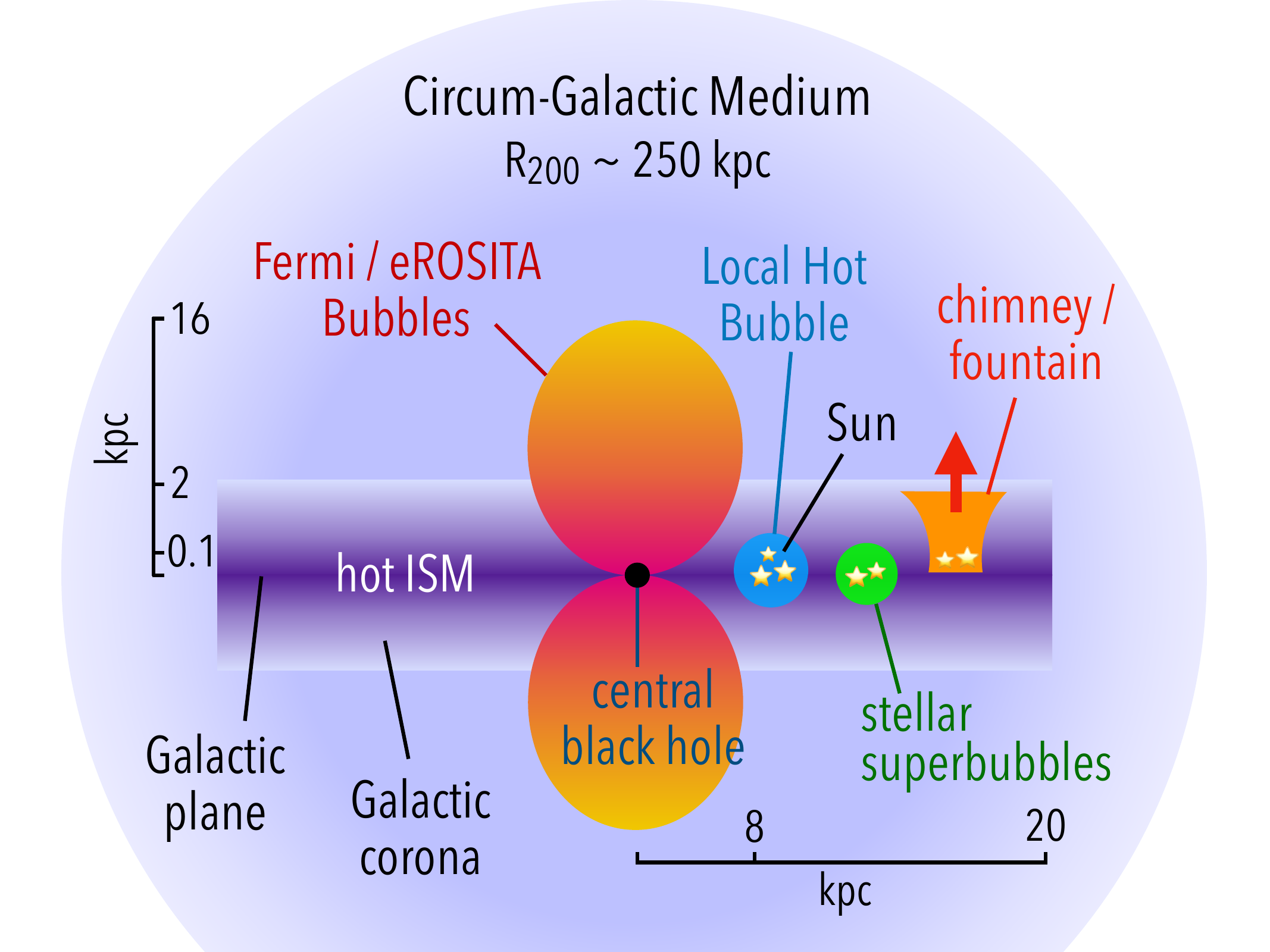}
        \caption{A sketch of the major X-ray emitting components determining the appearance of the diffuse soft X-ray sky: Milky Way's circum-galactic medium (with substructures like the Fermi/eROSITA bubbles), hot interstellar medium in the Galactic disk (with substructures like superbubbles and chimneys), hot gas within Local Cavity/Local Hot Bubble encompassing the Sun, and finally the Solar heliosphere shaped by the Solar wind interaction with the Local Interstellar Cloud.  }
    \label{fig:lass-sketch}
\end{figure}

\begin{figure*}
    \includegraphics[width=\textwidth,trim=0cm 1.7cm 0cm 1.7cm, frame]{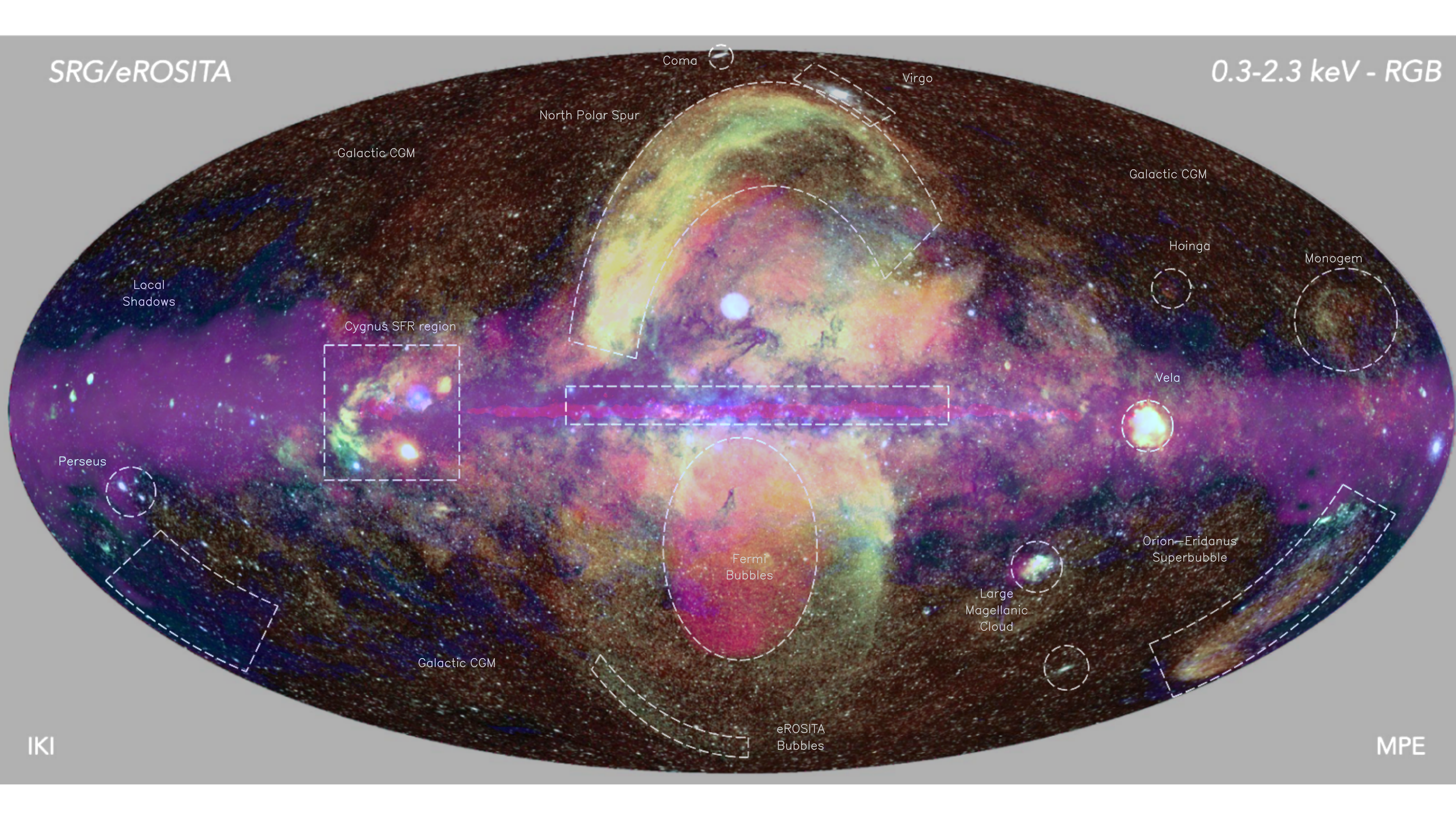}
    \includegraphics[angle=0, width=\textwidth, trim=0cm -0.5cm 0cm 0cm,frame]{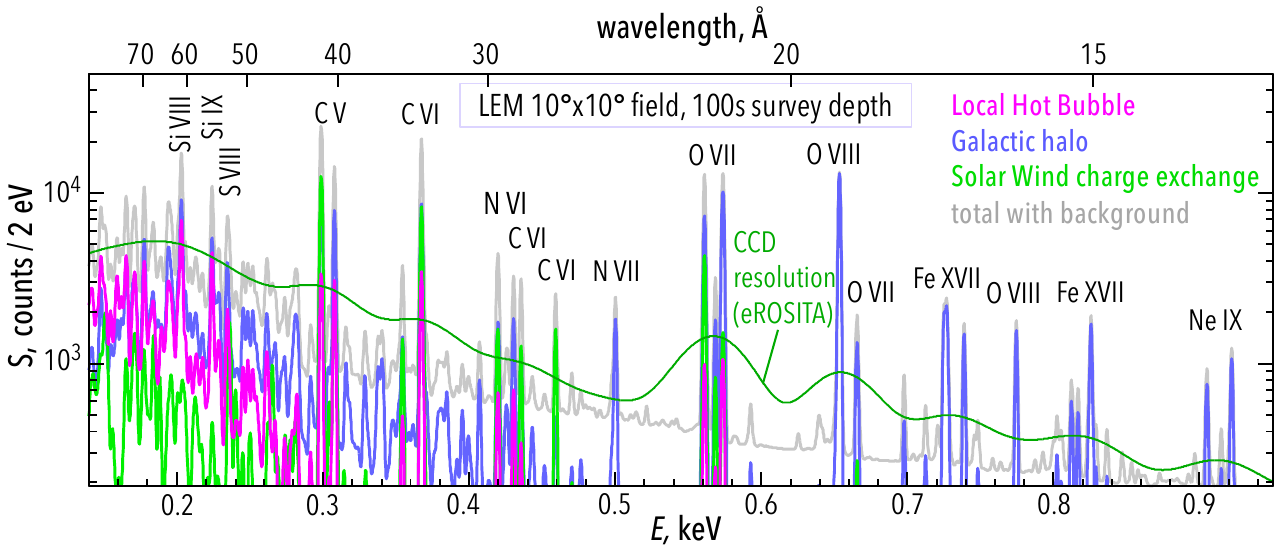}
    \caption{\textbf{Top.} All-sky soft X-ray map obtained after a one half-a-year scan (typical exposure time $\sim240$ s per point) by \textit{SRG}/eROSITA (red: [0.3-0.6] keV, green: [0.6-1.0] keV, blue: [1.0-2.3] keV, after Predehl et al. \cite{2021A&A...647A...1P} and Sunyaev et al. \cite{2021A&A...656A.132S}, complemented with maps of the diffuse gamma-ray emission from \textit{Fermi}\cite{2015A&A...581A.126S} (magenta) and dust emission at 857 GHz by \textit{Planck}\cite{2016A&A...594A...8P} (dark blue). Various remarkable regions visible on the all-sky map are labelled, while the multitude of mildly-extended and point-like sources (mostly stars, active galactic nuclei and galaxy clusters) are also seen.
    \textbf{Bottom.} An example of the spectrum to be obtained by \lem\, for a 10x10 deg$^2$ patch of "blank" sky beyond the Galactic plane by the end of the all-sky survey (grey line). Curves of different colour show contributions of the main diffuse components: Local Hot Bubble (magenta),  Galactic Halo and the hot disk (blue), Solar Wind Charge Exchange (light green) motivated by previous high spectral resolution observations of large sky areas at high Galactic latitudes \cite{2002ApJ...576..188M}. For comparison, the dark green curve shows the same total spectrum convolved with the typical resolution of modern CCD detectors (e.g. \textit{SRG}/eROSITA). \textcolor{blue}{LASS} will be able to resolve with high statistics a large number of emission lines from elements such as Si, S, Mg, C, N, Ar, O, Fe, and Ne. }
    \label{fig:srge_fermi_planck}
\end{figure*}
The X-ray sky below 1 keV differs drastically from the sky at higher energies.   While the latter is composed of a multitude of point and mildly-extended sources, the former is dominated by purely diffuse emission, with some regions being an order of magnitude brighter than others (cf Figure \ref{fig:srge_fermi_planck}). This diffuse emission is produced mostly by the hot gas within our extended Galaxy, from the gas in the Local Cavity within 200 pc of the Sun (the Local Hot Bubble or LHB), to the gas in the hot disk, to the Galactic halo/circumgalactic medium (CGM) at distances from 10 kpc all the way to the virial radius. The bulk of this emission is due to the lines of the most abundant metals (carbon, nitrogen, oxygen, neon, magnesium, silicon, sulphur and iron).  A large grasp soft X-ray spectroscopic mission is ideally suited to explore this emission. % in full detail.

\lem's unprecedented grasp for a spectroscopic mission makes it possible to build sensitive maps for very large sky areas over short periods of time, allowing the construction a full-sky soft X-ray map delineating sharp edges and fine structures with exquisite spectral detail.

Such a map will allow us to probe the physical conditions in the hot phase of the Milky Way's interstellar and circumgalactic media, including such prominent features as the North Polar Spur\cite{2022arXiv220301312L} and the Fermi/eROSITA bubbles\cite{2020Natur.588..227P}, measure velocities of gas in the inner part of the Galaxy, extract the emissivity of the LHB, and explore emission due to solar wind charge exchange (SWCX). The birth and death of stars can be traced by 3D tomography of bright and extended supernova remnants and star formation regions in our own and nearby galaxies, including the Magellanic Clouds, M31, and M33 (Patnaude et al.\cite{2023patnaude}). 

For a multitude of distant extra-galactic sources, mostly active galactic nuclei, galaxy clusters, and galaxy groups, high resolution X-ray spectra will be obtained for the first time, while stacked spectra of fainter sources will allow characterisation of their populations and their intergalactic environments. 
Compared to pointed observations, \textcolor{blue}{LASS} provides access to multiple objects of the same class distributed across the sky, e.g. stars, galaxies, clusters, filaments, whose observations can be combined to alleviate the problem of variance, associated with peculiarities of individual objects and selection biases.  For example, for moderately large samples ($N_{obj}$), the effective  \textcolor{blue}{LASS} exposure will be $N_{obj}\times 100 \,{\rm s}$. In other words, for any sample with $\sim10^4$ objects, \textcolor{blue}{LASS} will automatically provide 1~Ms of high spectral resolution data, characteristic of the class. 

A substantial number of transient sources, either Galactic (e.g. X-ray binaries) or extragalactic (e.g. tidal disruption events and supernova shock break-outs), %or Solar System (e.g. comets)
 will be discovered and studied in great detail. Finally, akin to the data of the \textit{ROSAT} All-Sky Survey (RASS), the all-sky map delivered by \lem\, will provide a high resolution X-ray background/foreground estimation for any position of the sky, invaluable for the in-depth analysis of faint extended objects, such as filaments of the warm-hot intergalactic medium or distant outskirts of massive galaxies, groups and clusters.

\textcolor{blue}{LASS} will provide both broad science return and legacy value, complementing the high scientific value provided by the \textit{SRG}/eROSITA survey.   \lem\ angular resolution will allow point source removal, using the \textit{SRG}/eROSITA catalogue, to flux levels undetectable in the survey itself, will allow the stacking of all classes of objects based on other surveys such as Gaia and eROSITA, and will allow the odd-shaped extraction regions necessary for the analysis of extended sources such as superbubbles, and supernova remnants, etc.

We present an outline of the survey design and parameters and highlight some scientific cases within the grasp of this mission.
        
%
\begin{comment}
    
\begin{figure}
    \centering
    \includegraphics[angle=0, width=1.0\columnwidth,frame]{fig/LEM-apec_lines_tene_all.pdf}
    \label{fig:lass-lines}
    \caption{Lines sensitivity to temperature}
\end{figure}
%
\end{comment}

%
\section{LEM ALL-SKY SURVEY}
\label{sec:lass}
\begin{figure}
    \centering
    \includegraphics[angle=0, trim = 0cm 0cm 0cm 0cm,width=0.8\columnwidth]{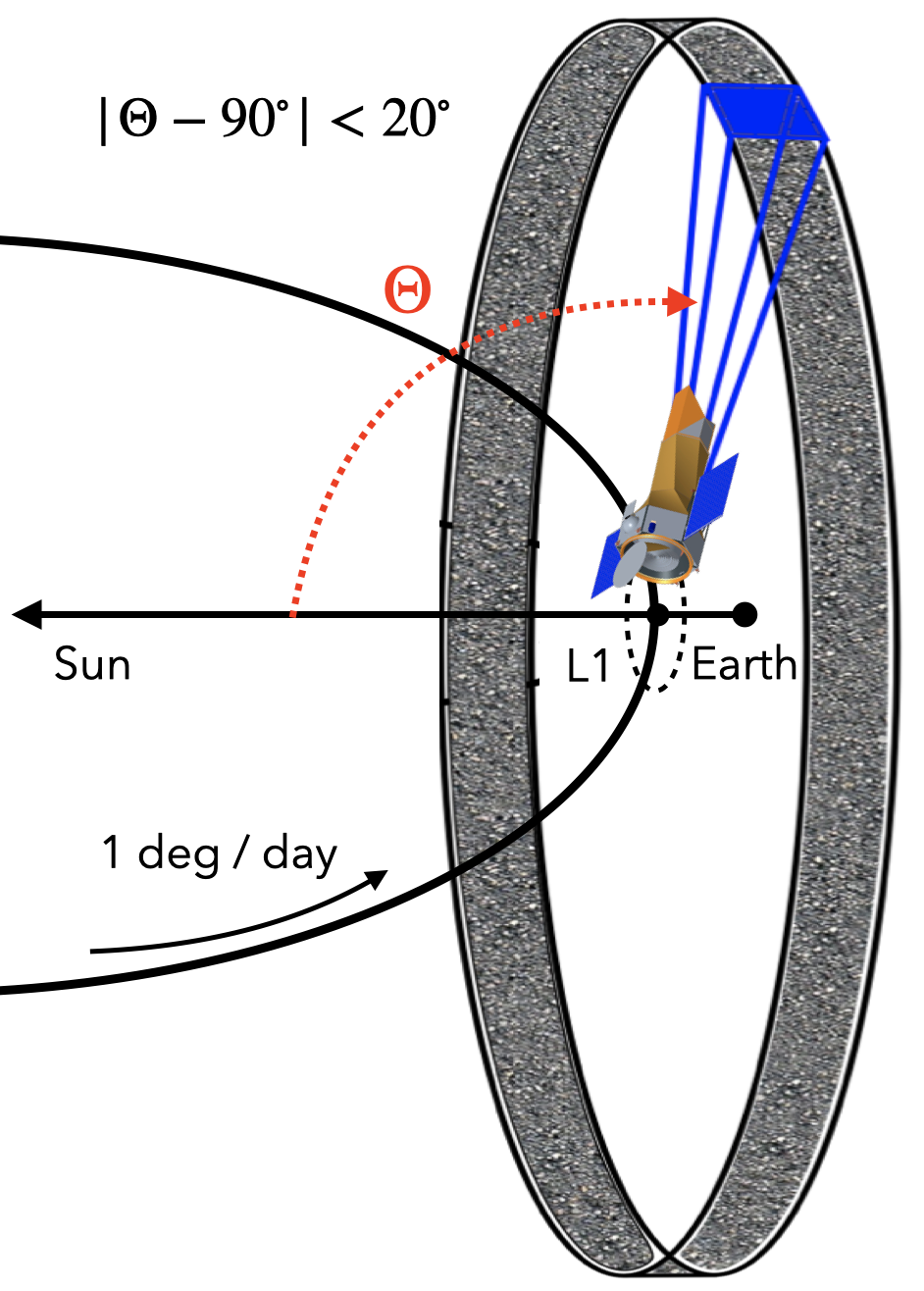}
    %\hspace{3cm}
        \includegraphics[angle=0, trim = 5cm 0cm 5cm 0cm,width=0.9\columnwidth]{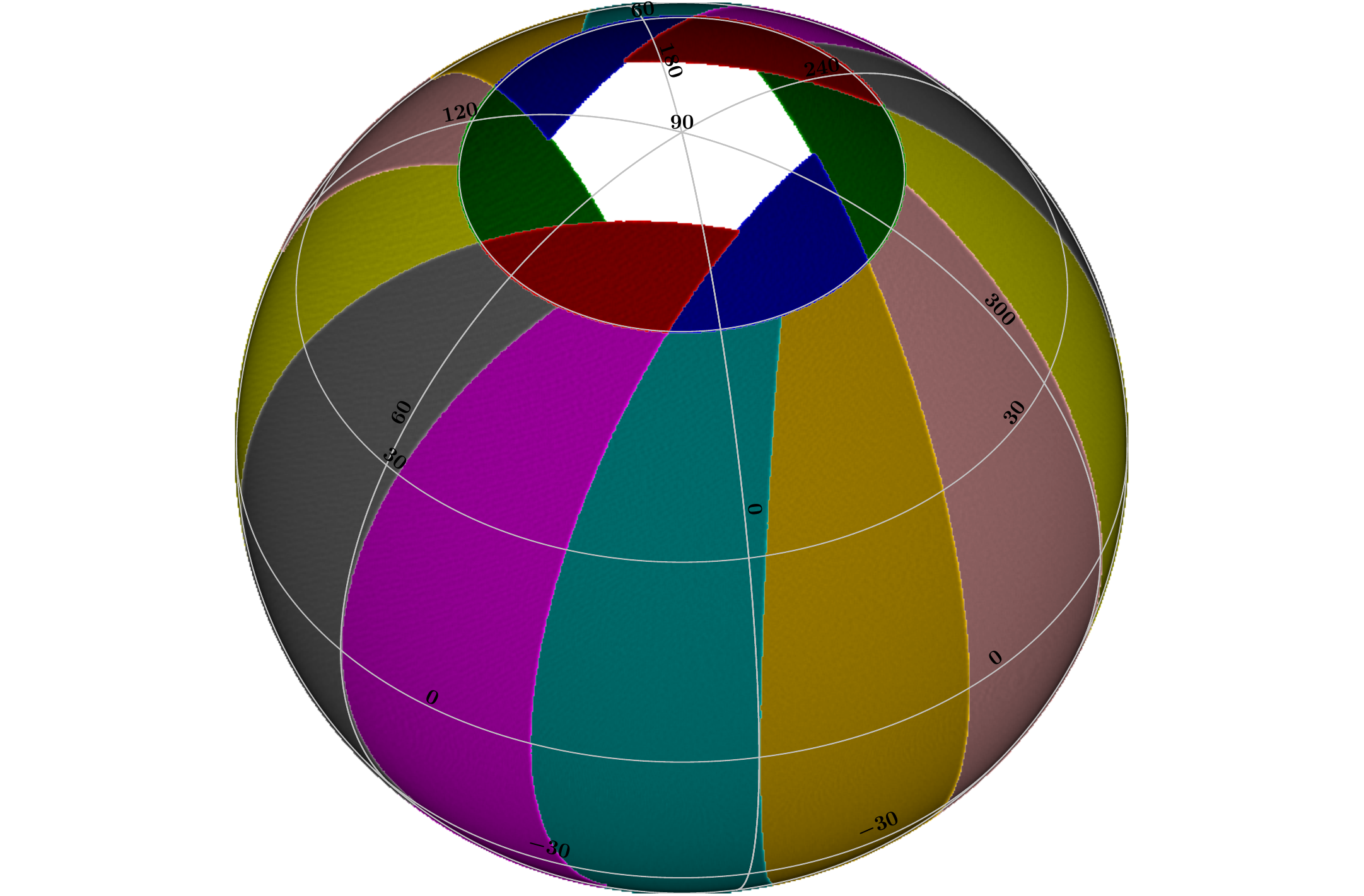}
        \caption{
                \textbf{Top.} An illustration of the sky visibility constraint for a spacecraft on an orbit around L1 point of the Sun-Earth system. 
        \textbf{Bottom.}
An example of the tiling strategy for an all-sky survey satisfying the Solar aspect angle constraint $|\Theta-90^\circ|<20^\circ$ with some margin for scheduling flexibility.  $\Theta$ is the angle between the telescope axis and the direction towards the Sun.  
        The white areas surrounding Ecliptic Poles are visible to the telescope over the whole year, while each of the twelve equatorial sectors and the twelve tiles at $|$ecliptic latitude$|>70^\circ$ can be fully scanned in one continuous episode.}
    \label{fig:lass-tiles}
\end{figure}

According to the provisional \lem~Time Allocation plan, $\sim10\%$ (16 Ms) of the mission guaranteed 5-yr active time will be devoted to the \textcolor{blue}{LASS}, corresponding to a uniform exposure time of 100 seconds for each point at the end of the mission.
%Because of that, 
{
%\color{red}
The allocated time is driven by the requirement to achieve the accuracy necessary to study diffuse emission and, in particular, to separate the SWCX foreground emission from the emission from the LHB and Milky Way CGM. 

A key component of \textcolor{blue}{LASS} is to monitor and study the effect of SWCX over $\sim$half a solar cycle with 10\% accuracy. Previous rocket data have shown that SWCX may change by up to 10\% over a 10 deg scale\cite{2014Natur.512..171G}. Similarly, this is also the scale at which we observe significant structure in the CGM\cite{2013ApJ...773...92H,2020NatAs...4.1072K,2022ApJ...936...72B}(but also \cite{2023A&A...670A..99P}). To keep the uncertainty in SWCX analysis/removal to less than 10\%, the maximum field of view used should be $\rm 10\times10~deg^2$. 

To produce a good characterization of SWCX, \textcolor{blue}{LASS} will cover roughly half a solar cycle, or $\sim$5 years and, assuming a SWCX surface brightness at 0.5 keV of 1~Line Unit (1~photon/s/cm$^2$/steradian) and the \LEM~resolution, 10 s per pointing is required to keep the statistical uncertainty of SWCX below 10\%. The SWCX contribution at any given sky location depends on the Earth's position in its orbit and varies with solar wind conditions. To maximize our constraints on the SWCX behaviour, we would like to revisit each location at different times of the year, and distribute those visits over as much of the Solar cycle as possible.

With the planned location of \lem\ at the L1 Lagrange point, the only limitation on observing direction is the Solar angle constraint due to spacecraft operational considerations. To provide substantial margins, the solar angle constraint has been set to be 70 to 110 degrees, allowing two $40^\circ$ sectors in the ecliptic plane to be observed at any given time, and larger sectors at higher latitudes.  The sky above  $70^\circ$ can be observed at any time.  Spacecraft scan rates are more than adequate to raster-scan arbitrary blocks with the desired 10 s per half-degree with little overhead as long as the scans are at least  $90^\circ$ long.  Data can be taken while scanning with no loss of angular resolution.  An example of a sky tiling strategy satisfying solar aspect constraints and minimizing overlapping coverage of the same areas is shown in Figure \ref{fig:lass-tiles}.

The 100 s exposure time for a given spot on the sky is far shorter than that of the SRG/eROSITA survey, but will give measurements of the stronger lines on areas as small as a square degree.  Integrating over any $\rm 10^\circ\times10^\circ$ region, or a narrow arc or arbitrary shape of similar area, will give an equivalent 40 ks exposure, resulting in a high-resolution spectrum with over 6000 counts per line unit (LU), so accurate measurements of even very faint lines can be made.  If better energy resolution is needed, only the 6\% of the photons from the central array can be selected, giving $\sim 1 \rm eV$ resolution.  The high-statistics eROSITA map can be used to identify spatial features, and \textcolor{blue}{LASS} data integrated over the same area can be used to study the spectrum and resolve ambiguities, such as multiple emitting regions along a chosen line of sight. Very small structures can be cut out and stacked to improve statistics.

The instrument and mission design will also provide low vignetting ($\sim 10\%$) and ability of mapping the sky in large patches with near-uniform exposure. 
Learning from \textit{SRG}/eROSITA, a substantial (e.g. 100 deg$^2$ in size) patch of the sky is planned to be observed to the final depth early in the mission, allowing the first look at the \textcolor{blue}{LASS} data to be obtained and exploited for deep pointed observations, similar to eROSITA Final Equatorial Depth Survey (eFEDS\cite{2022A&A...661A...1B}).% (eFEDS) field. 

\section{GALACTIC CGM}
\label{sec:cgm}
%
%------------------------
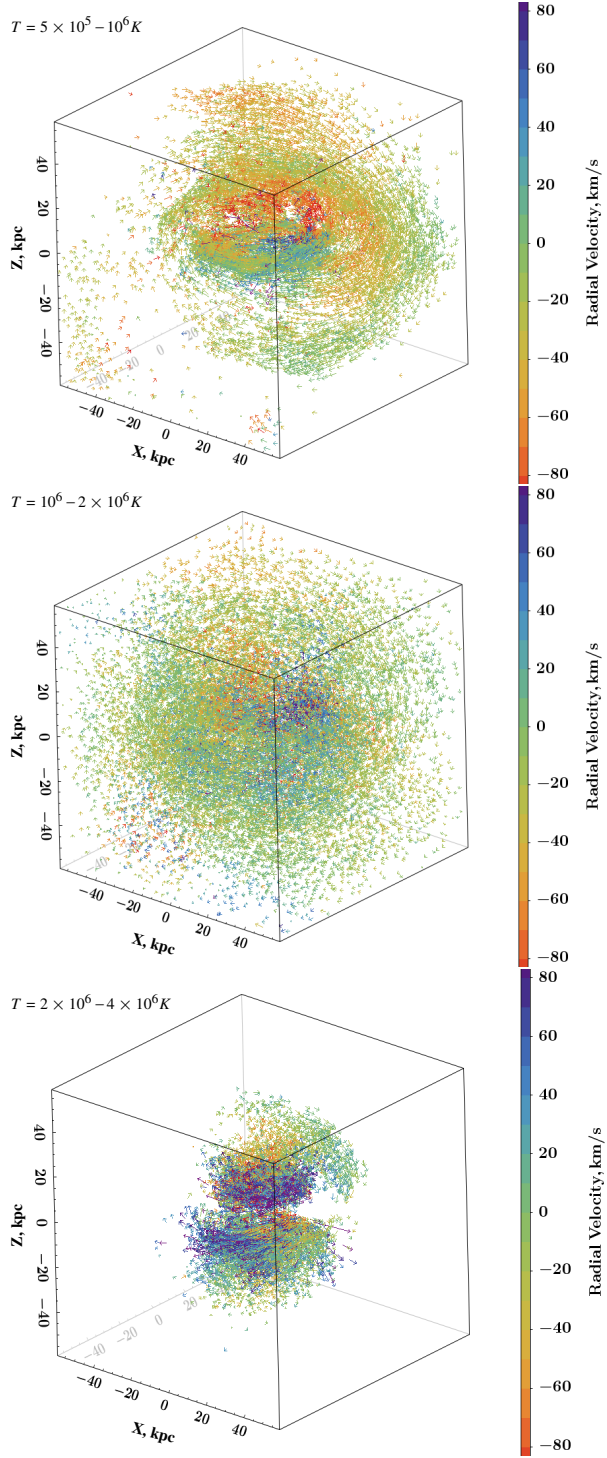
\begin{figure}
\centering
\includegraphics[clip=true,trim=1cm 3.cm 0cm 3cm,angle=0, width=0.96\columnwidth]{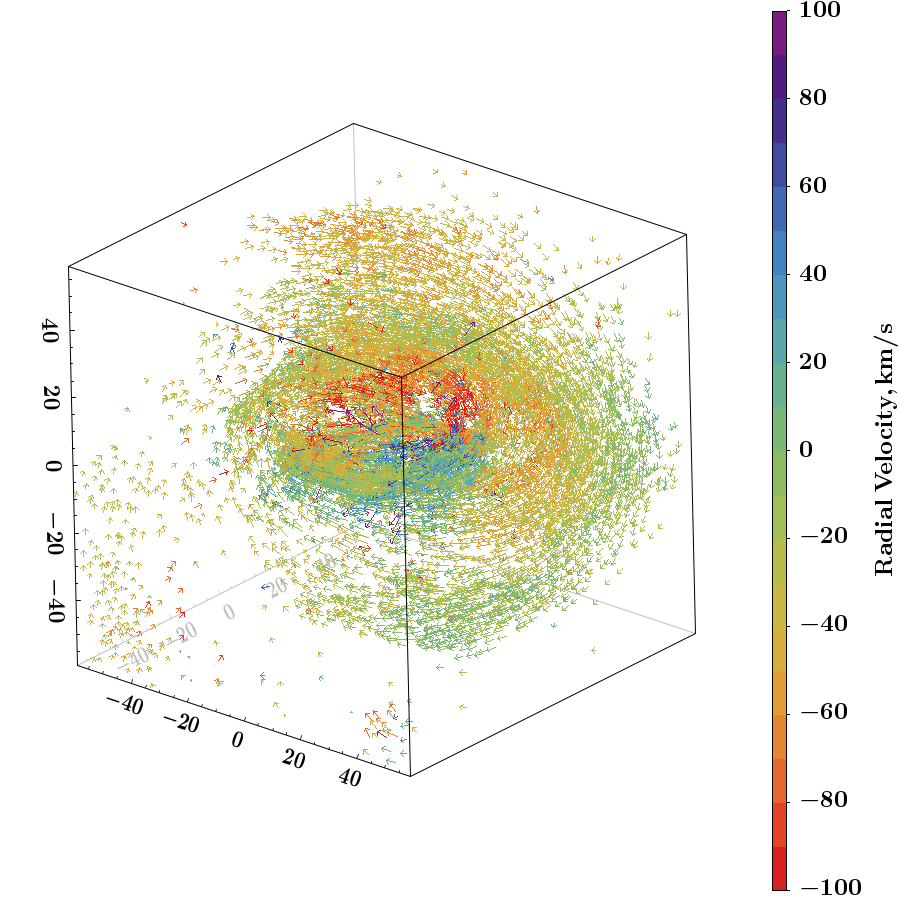}
\includegraphics[clip=true,trim=1cm 3.cm 0cm 3cm,angle=0, width=0.96\columnwidth]{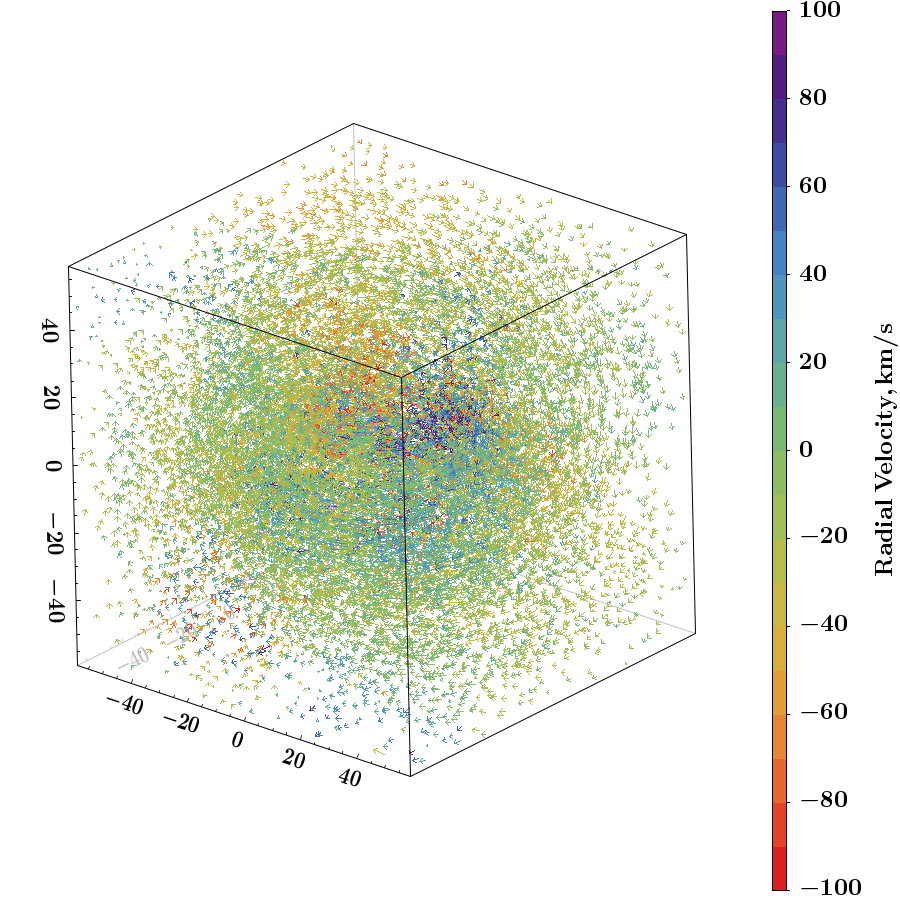}
\includegraphics[clip=true,trim=1cm 3.cm 0cm 3cm,angle=0, width=0.97\columnwidth]{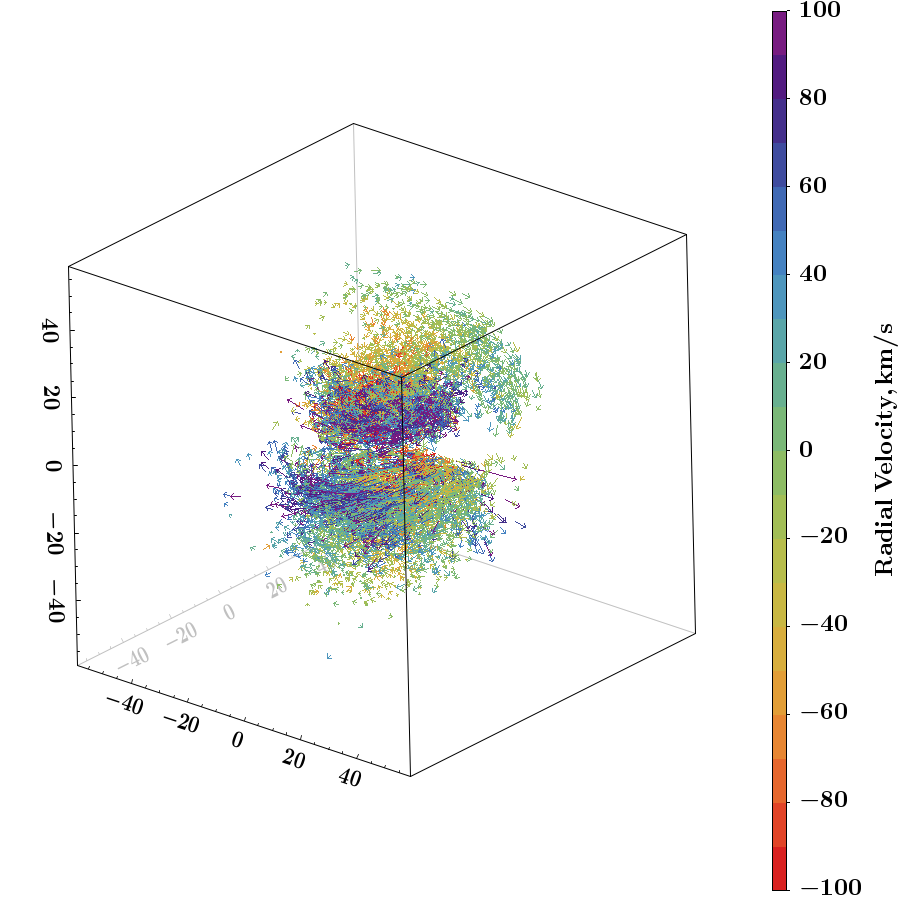}
\caption{Illustration of the velocity patterns within a (60~kpc)$^3$ cut-out region of a Milky Way-type galaxy simulation (Valentini et al.\cite{2023MNRAS.518.1128V}) for gas with temperature between 0.5 and 1 MK (top panel), 1 and 2 MK (central panel), 2 and 4 MK (bottom panel). Colour coding shows the radial velocity of the gas particles (red - inflow, blue - outflow), while direction and length of the arrows show full velocity. The dominance of the inflowing and outflowing patterns is obvious for the warm and hot gas, respectively.}
\label{fig:cgm_phase_vfield}
\begin{picture}(0,0)
\put(-70,155){\rotatebox{-15}{\tiny \bf X, kpc}}
\put(-115,225){\rotatebox{90}{\tiny \bf Z, kpc}}
\put(-70,340){\rotatebox{-15}{\tiny \bf X, kpc}}
\put(-115,410){\rotatebox{90}{\tiny \bf Z, kpc}}
\put(-70,525){\rotatebox{-15}{\tiny \bf X, kpc}}
\put(-115,595){\rotatebox{90}{\tiny \bf Z, kpc}}
\put(-115,685){\rotatebox{0}{\tiny \bf \rm $T=5\times 10^{5}-10^6 K$}}
\put(-115,505){\rotatebox{0}{\tiny \bf \rm $T=10^{6}-2\times 10^6 K$}}
\put(-115,315){\rotatebox{0}{\tiny \bf \rm $T=2\times 10^{6}-4\times 10^6 K$}}

%\put(90,250){\colorbox{white}{\large B}}
%\put(-150,200){\colorbox{white}{Declination [deg]}}
%\put(-520,180){\rotatebox{90}{\large Declination [deg]}}
%\put(-248.,263.5){\circle{20}}
\end{picture}
\end{figure}
%-------------------------

\begin{comment}
    
%------------------------
\begin{figure}
\centering
\includegraphics[trim=10cm 0cm 10cm 0cm, angle=0, width=0.99\columnwidth]{fig/lem_allsky_slice_vert.pdf}
\caption{}
\label{fig:cgm_outer_cut}
\end{figure}
%-------------------------
\end{comment}

%------------------------
\begin{figure}
\centering
\includegraphics[clip=true,trim=4.9cm 0cm 4.9cm 0cm,angle=0, width=0.99\columnwidth]{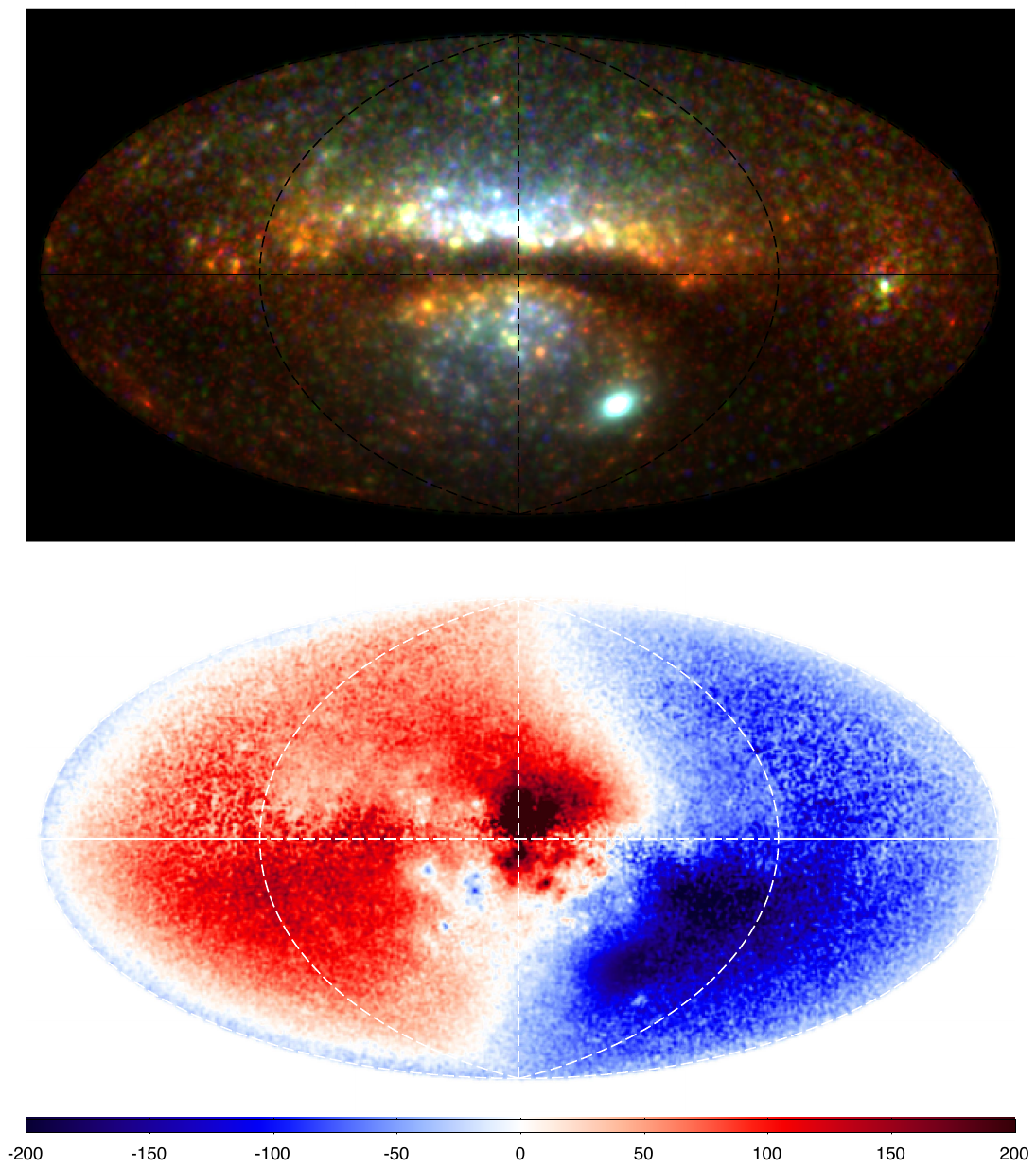}
\caption{An example of the simulated all-sky X-ray map {(Galactic coordinates, Aitoff projection)} obtained by locating a virtual observer inside a simulated Milky Way-like galaxy\cite{2023MNRAS.518.1128V} 8 kpc away from its center. \textbf{Top panel:} an RGB map showing surface brightness in 0.2-0.4 (red), 0.4-0.7(green) and 0.7-1.3 keV (blue) bands. Cold gas in the galactic disc was excluded from simulation, resulting in the Galactic plane "shadow". The predicted average surface brightness in these bands agrees well with the currently available measurements, including their global trends, while the small-scale grainy appearance of the image results from resolution limitations of the simulation (namely, contributions of gas particles in the relative vicinity of the observer).  \textbf{Bottom panel:} Predicted line-of-sight velocity (in km/s) of the gas weighted by its O~VII emissivity and taking into account the observer's rotation with the stellar disc of the galaxy. A clear dipole structure is visible, reflecting the difference of the rotation in X-ray emitting gas and the stellar disc, as well as the outflowing gas structure in the central part of the galaxy. For the Galactic plane, the real velocity map will be different because interstellar absorption (not accounted for in this simulation) results in dominance of the nearby gas in the observed line emission.   }
\label{fig:cgm_map}
\begin{picture}(0,0)
\put(-110,375){\colorbox{white}{\tiny -200}}
\put(-87.5,375){\colorbox{white}{\tiny -150}}
\put(-64.,375){\colorbox{white}{\tiny  -100}}
\put(-37.5,375){\colorbox{white}{\tiny -50}}
\put(-7.5,375){\colorbox{white}{\tiny 0}}
\put(15.5,375){\colorbox{white}{\tiny 50}}
\put(40.5,375){\colorbox{white}{\tiny 100}}
\put(65.5,375){\colorbox{white}{\tiny 150}}
\put(85.5,375){\colorbox{white}{\tiny 200}}
\put(50.5,390){\rotatebox{0}{\small $\rm V_{los}(km/s)$}}
\put(-80,615){\rotatebox{0}{\small \textcolor{red}{0.2-0.4 keV}, \textcolor{green}{0.4-0.7 keV}, \textcolor{blue}{0.7-1.3 keV}}}

%\,\,\,-150     -100   -50   0   50   100   150  200
\end{picture}
\end{figure}
%-------------------------

%------------------------
\begin{figure*}
\centering
\hspace{0.3cm}
\includegraphics[clip=true,trim=0cm 0cm 1.8cm 0cm,angle=0, width=0.95\columnwidth]{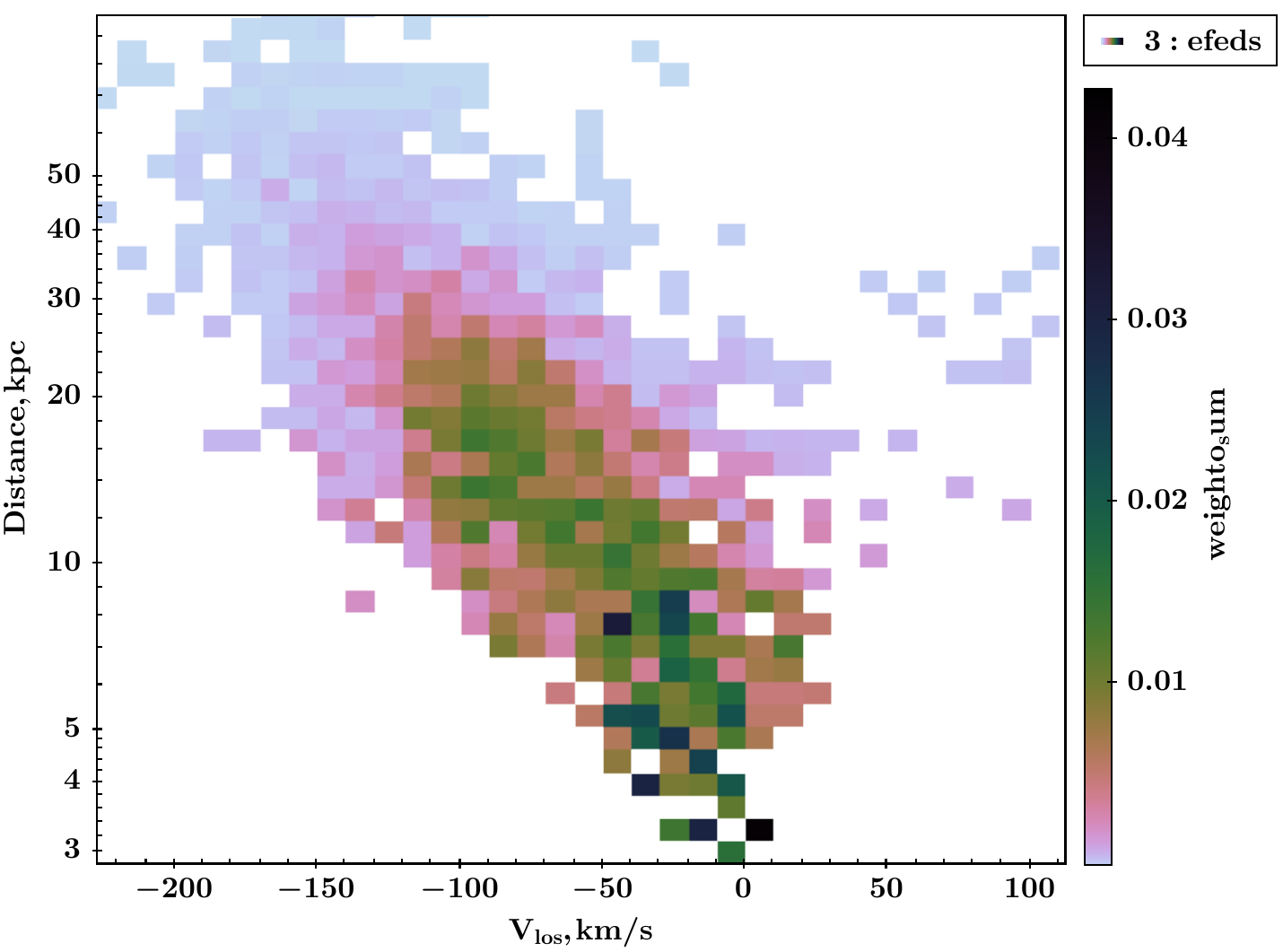}
%\hspace{0.3cm}
\includegraphics[clip=true,trim=0cm 0cm 1.8cm 0cm,angle=0, width=0.95\columnwidth]{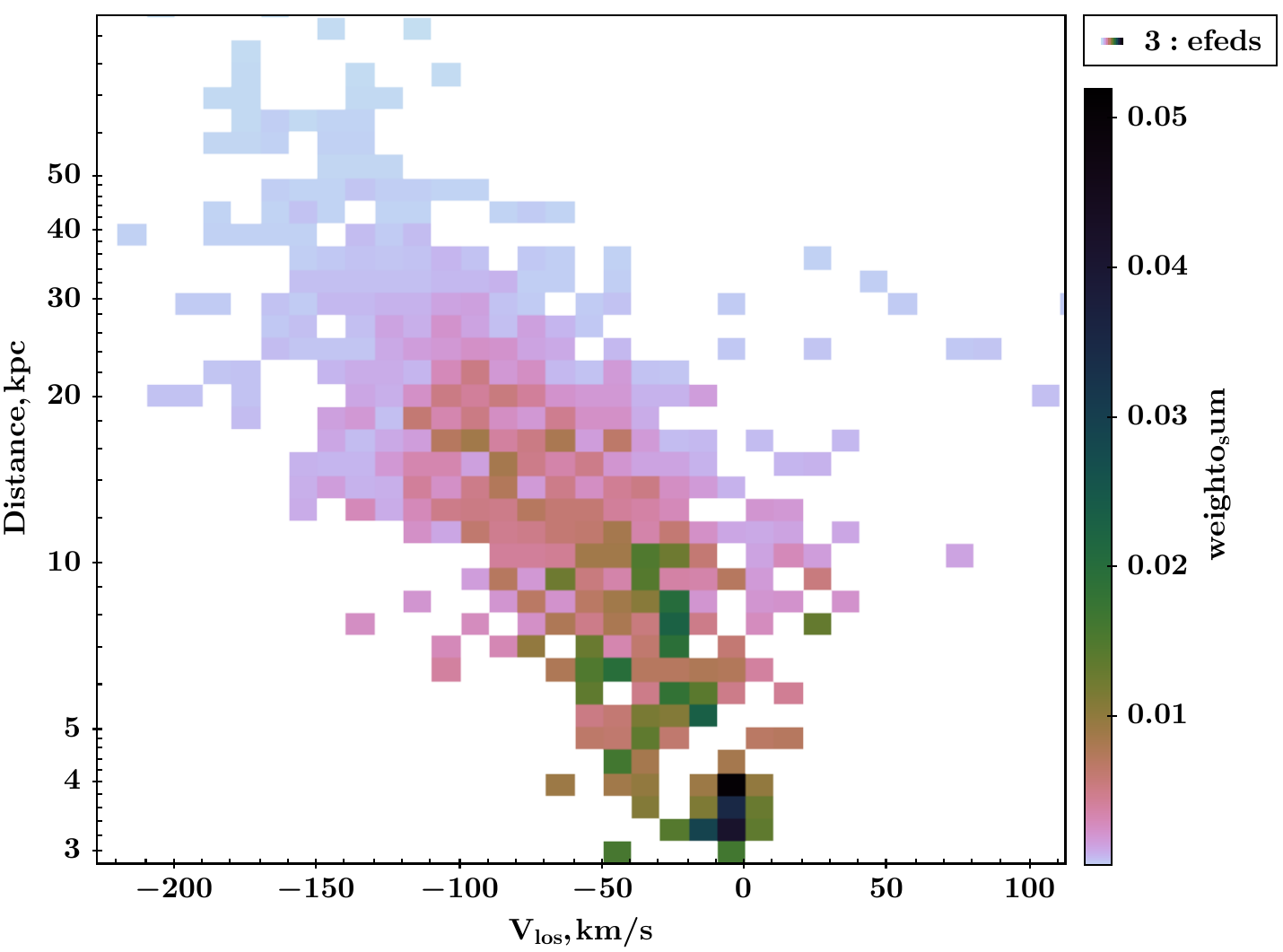}
%\hspace{0.3cm}
\includegraphics[clip=true,trim=0cm 0cm 1.8cm 0cm,angle=0, width=0.95\columnwidth]{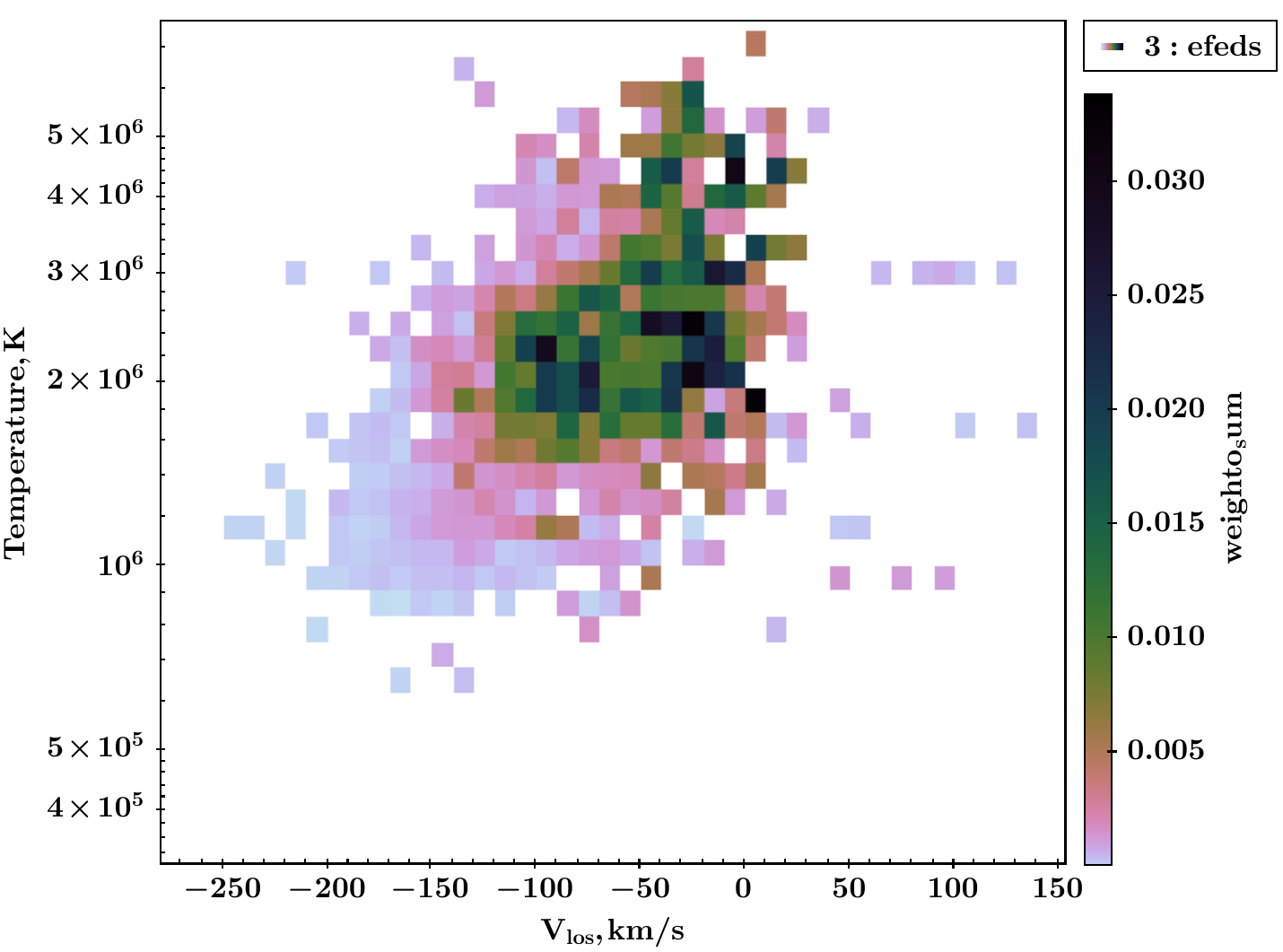}
\includegraphics[clip=true,trim=0cm 0cm 1.8cm 0cm,angle=0, width=0.95\columnwidth]{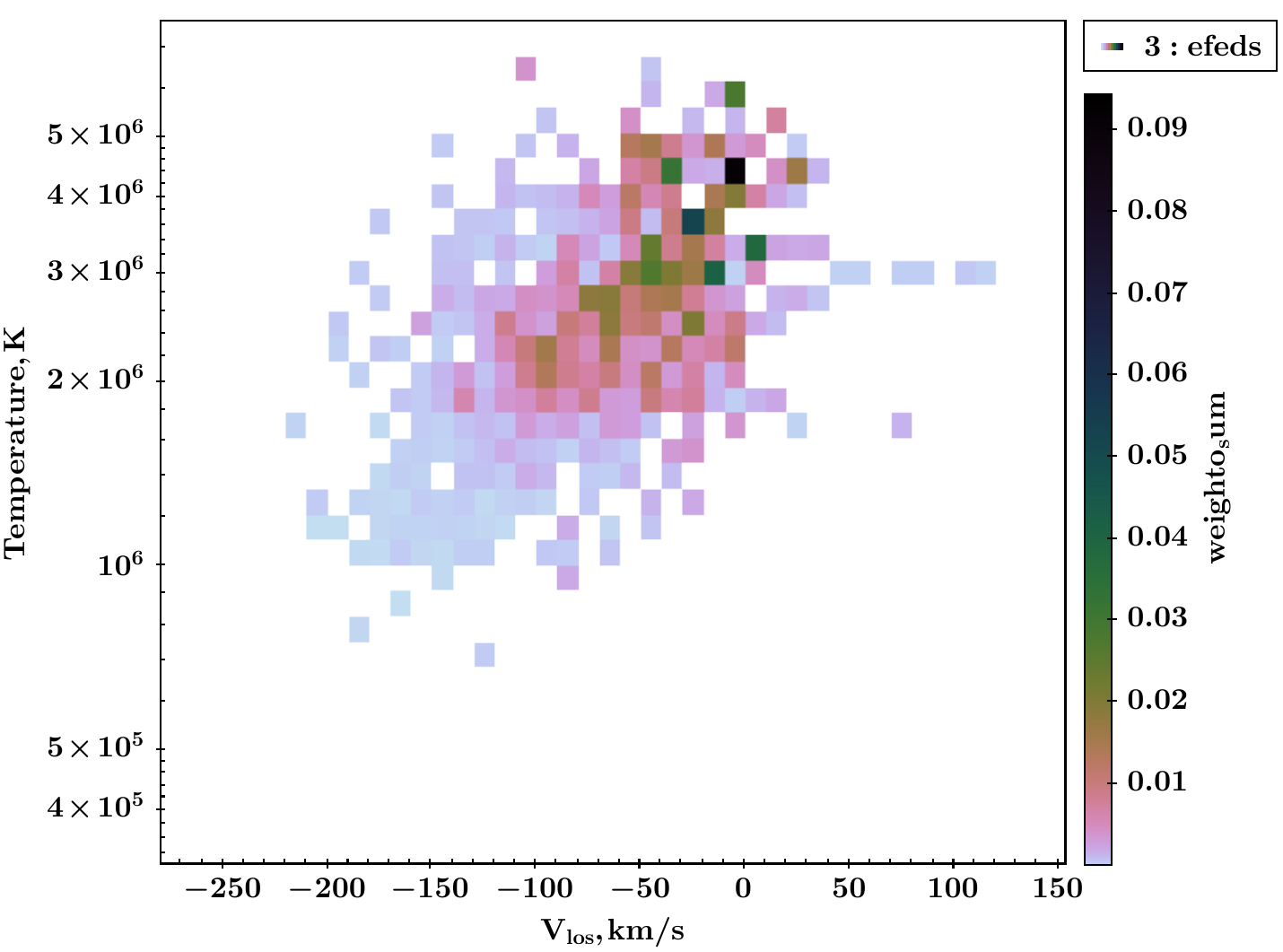}
%\includegraphics[clip=true,trim=0cm 0cm 1.8cm 0cm,angle=0, width=0.99\columnwidth]{fig/fe17_v_vs_d_efeds.pdf}
%\hspace{0.3cm}
%\includegraphics[clip=true,trim=0.0cm 0cm 1.8cm 0cm,angle=0, width=0.99\columnwidth]{fig/fe17_v_vs_t_efeds.pdf}
\caption{The complex structure of the X-ray emitting gas along a single line-of-sight for an observer located inside a simulated Milky Way-like galaxy\cite{2023MNRAS.518.1128V} is revealed by the distribution of simulated O~VII (left) and O~VIII (right) photons with respect to the line-of-sight velocity, distance to the observer(top panels), and temperature of the gas from which they originated (bottom panels). Clear differences in spatial and kinematic signatures of the resulting line emission are visible.
}
\label{fig:cgm_vstruct}
\end{figure*}
%-------------------------

%------------------------
\begin{figure}
\centering
\includegraphics[clip=true,trim=1.5cm 5.5cm 2.5cm 3.5cm,angle=0, width=1.\columnwidth]{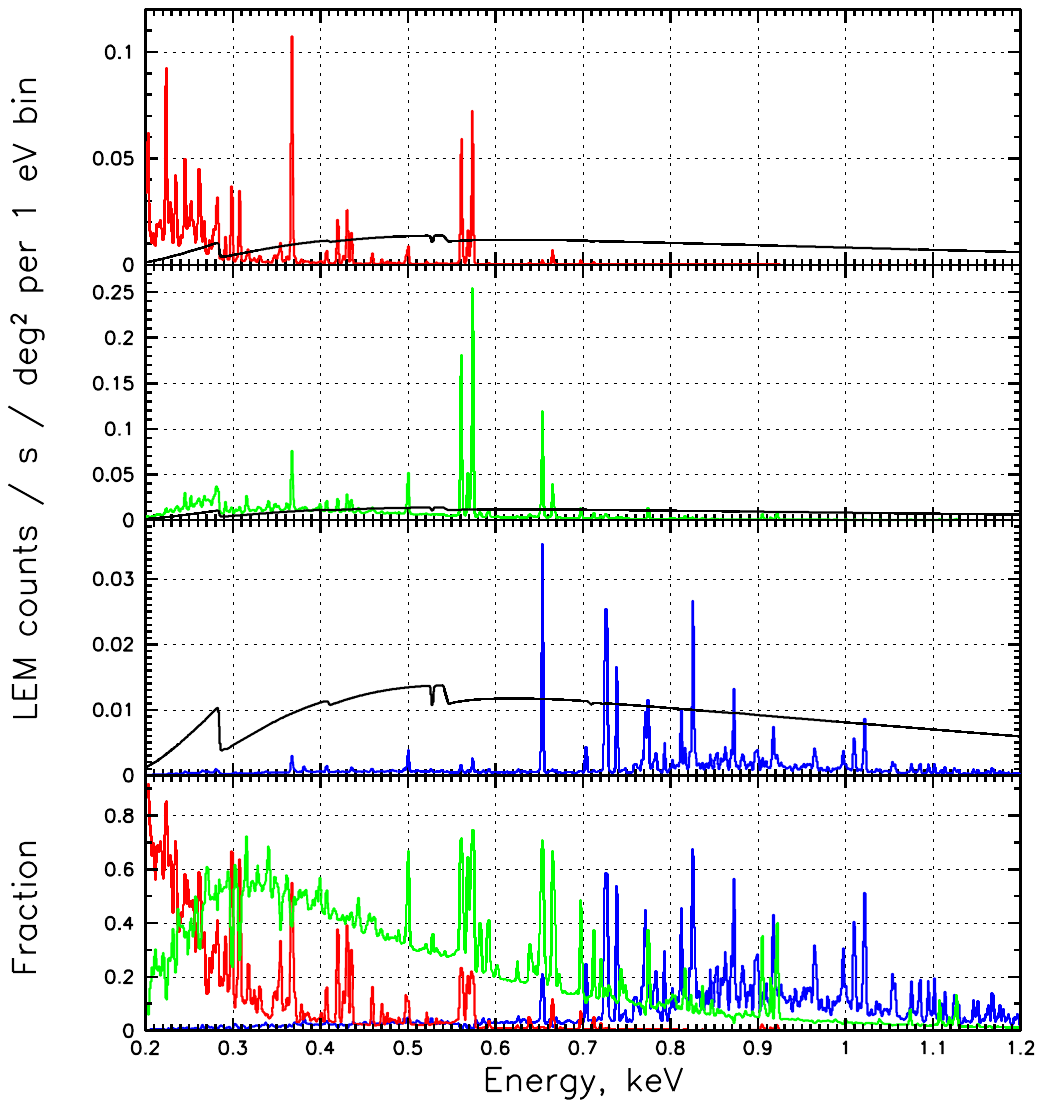}
\caption{Predicted \lem\ count rates for four background components inferred from the \textit{SRG}/eROSITA observations in the eFEDS field\cite{2023A&A...674A.195P}, including the Local Hot Bubble (red), the hot Galactic disk (blue), the Galactic Halo (green), and the Cosmic X-ray Background (black), as well as their relative contribution to the total flux (bottom panel).}
\label{fig:lem_efeds}
\end{figure}
%-------------------------

%------------------------
\begin{figure}
\centering
\includegraphics[clip=true,trim=1.5cm 1.5cm 2cm 2.5cm,angle=0, width=1.\columnwidth]{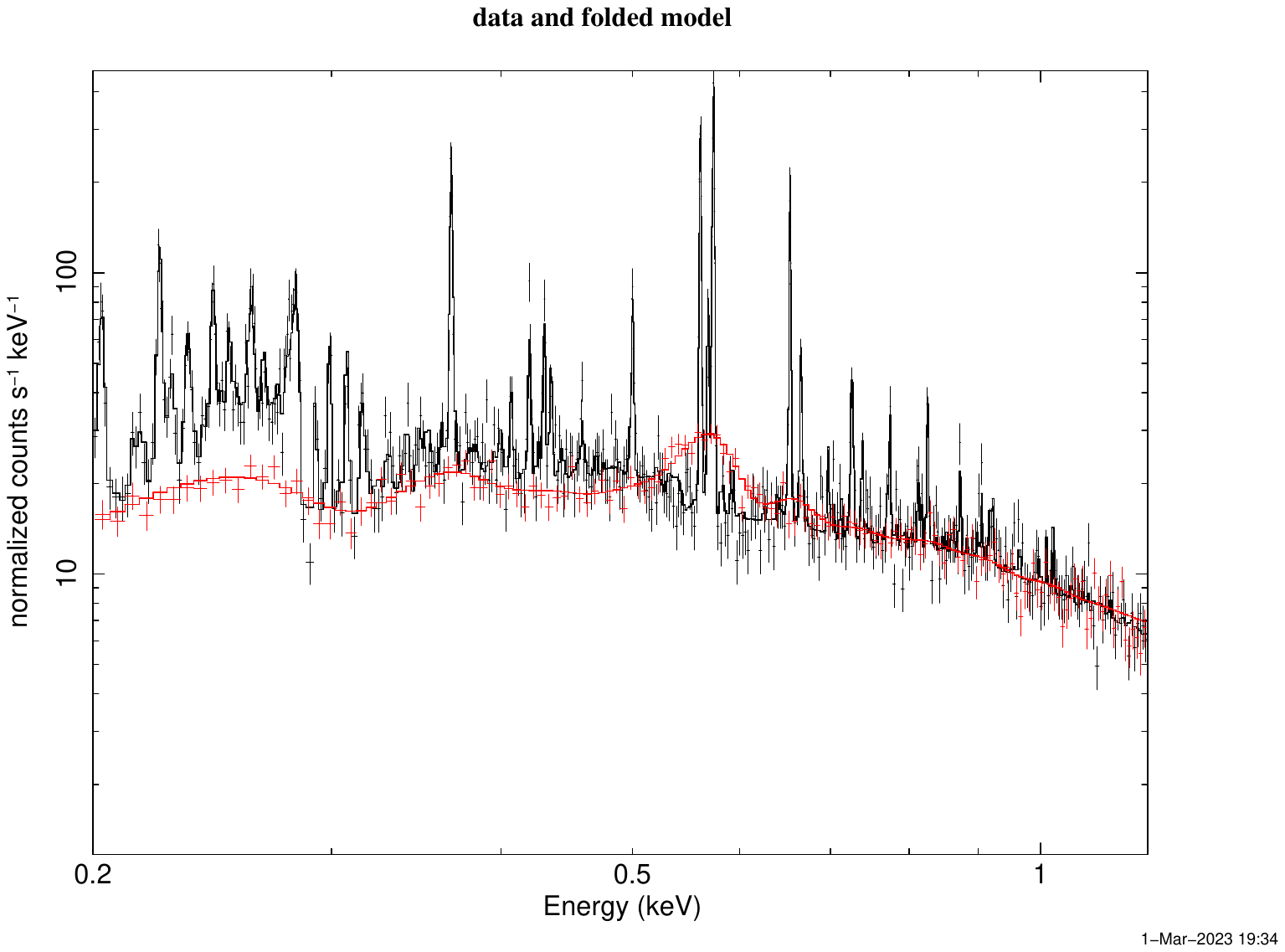}
\caption{A simulated \lem\ spectrum (black) for a 100s exposure of a 10 deg$^2$ sky patch with the spectral model inferred from the \textit{SRG}/eROSITA data in eFEDS field \cite{2023A&A...674A.195P}. The modelled \textit{SRG}/eROSITA spectrum for the same area and exposure time is shown in red, demonstrating richness of the spectral information to be uncovered by the \textcolor{blue}{LASS} data.}
\label{fig:lem_efeds_fake}
\begin{picture}(0,0)
\put(-60,130){\textcolor{red}{eROSITA}}
\put(-90,130){\textcolor{black}{LEM}}
\end{picture}

\end{figure}
%-------------------------

%------------------------
\begin{figure}
\centering
\includegraphics[clip=true,trim=0cm 0cm 3cm 0cm,angle=0, width=1.\columnwidth]{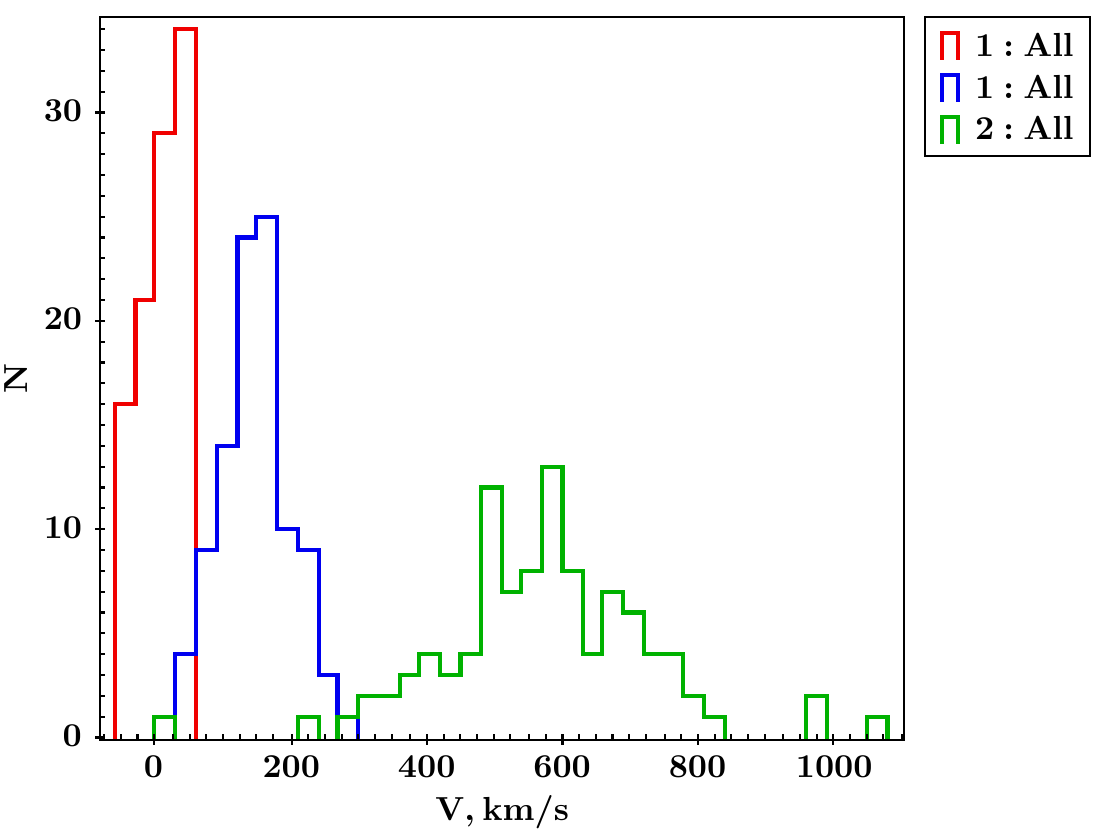}
\caption{The reconstructed line-of-sight velocities for the four-component background emission model consisting of Local Hot Bubble (LHB, red), hot Galactic disk (green), and Galactic Halo (blue) with eFEDS-like\cite{2023A&A...674A.195P} parameters and normalisations, assuming that they have Doppler shifts corresponding to 0, 100, and 600 km/s, respectively, for a 100s long exposure of a 1 deg$^2$ region. In our fitting procedure, the LHB velocity was allowed to vary only within $\pm50$ km/s.}
\label{fig:cgm_vmeasurement}
\begin{picture}(0,0)
\put(40,330){\textcolor{red}{ LHB}}
\put(40,315){\textcolor{blue}{ Halo}}
\put(40,300){\textcolor{green}{ Hot outflow}}

%\put(-150,200){\colorbox{white}{Declination [deg]}}
%\put(-520,180){\rotatebox{90}{\large Declination [deg]}}
%\put(-248.,263.5){\circle{20}}
\end{picture}
\end{figure}
%-------------------------

One of the main driving aspects of such a survey will be study of the Milky Way's Circum-Galactic Medium (CGM). \lem\, will be able to study the details of the bright inner parts of the CGM at the scale-height of a few kpc, which are important as the actual region of contact where the exchange of mass, energy and metals with the outer CGM and beyond takes place. \lem\ will map the velocities of the inner regions of the Milky Way CGM and, in particular, the expansion of the Fermi/eROSITA bubbles, believed to be evidence of feedback from either the SMBH or the star-forming regions in the Galactic Center. \lem\  will map the temperature structure of the inner CGM {\it both} across the sky {\it and} along the line of sight using the lines of the various ion species - something only a calorimeter can do, in the presence of multiple temperature components along each line of sight, and the solar wind charge exchange emission between lines of sight. The MW observations will complement the studies of the CGM in distant galaxies (a critical science driver of \lem\cite{2023ApJ...953...42B,2023MNRAS.522.3665N,2023arXiv230701259S,2023MNRAS.525.1976T,2023arXiv230701269Z}) where \lem\ will map the outer halos but have limited insight into the interface between the disk and the halo. 

\subsection{OUTER GALAXY:\\ DISK - HALO INTERFACE}

Milky Way mass galaxies are relatively efficient in making stars, and consequently producing metals and dispersing them. The depletion of the cold gas and the quenching of star formation is believed to be avoided by the inflow of gas into the disc, some part of which is recycled from the matter previously expelled from the disc by energetic outflows\cite{2005ARA&A..43..769V} or ``fountains''\cite{2023arXiv230917246S,2023ApJ...943...61F}. The dynamics and kinematics of the warm-hot gas taking part in these fountain flows are key but poorly understood ingredients in the current paradigm of Milky Way-like galaxies evolution\cite{2016ApJ...822...21H,2016ApJ...818..112M,2020ApJ...894..142Q}.

In particular, the exact way by which the cold star forming disk of the Milky Way is fed by freshly accreted gas is unclear\cite{2012ARA&A..50..491P,2017ARA&A..55..389T,2023ARA&A..61..131F}, although the CGM gas (replenished by cosmological inflows\cite{2017MNRAS.470.3167V}) likely provides a sufficient baryonic reservoir\cite{2012ApJ...756L...8G,2018ApJ...862....3B}. Moreover, settling of this gas into the cold ISM might proceed either via precipitation-like condensations losing pressure support and falling back on to the disk, via steady radial inflow, or (partially) rotationally-supported  inflow of the hot gas \cite{2018MNRAS.480.2963O,2022MNRAS.514.5056H,2023arXiv230600092S}.  All these scenarios predict drastically different kinematic properties of the X-ray emitting gas in the few tens of kpc above the Galactic disc (see illustration of complex kinematic structure in the multiphase inner CGM in a simulated MW-like galaxy\cite{2023MNRAS.518.1128V} in Figure \ref{fig:cgm_phase_vfield}). Microcalorimetric X-ray observations of the full sky by \LEM~offer a unique opportunity to reconstruct the kinematics of the inner CGM region, since radial velocity measurement down to 50 km/s can be achieved via line centroid measurements. 

A key aspect of \textcolor{blue}{LASS} is the possibility to map emission features with angular resolution  down to a degree scale, which is crucial because line-of-sight-absorption is known to vary at such scales. Such mapping is necessary  to decompose the total X-ray emission for given lines of sight into individual spatial-spectral components. Finally, several epochs of observations of the same sky regions are required to control the time-variable SWCX emission, which might not only change line intensities and their ratios, but also introduce substantial biases in the centroid-based kinematic measurements.

Depending on the level of co-rotation of the X-ray emitting gas with the stellar disk of the Milky Way (along with which we as Sun-attached observers are co-rotating), one can readily predict the expected all-sky kinematic pattern. Figure \ref{fig:cgm_map} shows a simulation-based example provided byX-ray post-processing of the simulation by Valentini et al.\cite{2023MNRAS.518.1128V} (via the PHOX package\cite{2012MNRAS.420.3545B} assuming an optically thin collisional ionization equilibrium (CIE) plasma radiation for all the X-ray-emitting gas particles) . Indeed, if the X-ray emitting CGM gas is rotating only slowly compared to our circular velocity, v$_c\sim$200 km/s, one expects a clear dipole velocity structure, with the largest velocity amplitude at $\ell=90^\circ$ and $270^\circ$. In the nearly co-rotating case, one expects very small velocities across the sky, dominated by the radial inflow components. For the radial inflow/outflow kinematics, the ideal directions are offered by the Galactic poles, $b=\pm90^\circ$, where rotation is not expected to play a significant role. For a similar perspective in the context of observations of the CGM in distant galaxies, see \cite{2023arXiv230701269Z}.  

Measuring the gas velocity, its fluctuations and correlation lengths, as well as the azimuthal and polar variations allows one to distinguish models of smooth and highly fragmented inflows/outflows. Moreover, $\sim 15 '' $ angular resolution retained during the survey enables the extraction of spectra from oddly shaped regions, enabling the study of possible interactions of the hot gas with the cold High Velocity Clouds (HVC), hence probing not only physics of processes like charge exchange and reconnection\cite{1999A&A...342..213K,2012ApJ...751..120S,2014ApJ...791...41H}, but also multiphase kinematics, thanks to the exquisite 21cm velocity maps of the HVCs\cite{2009ApJ...698.1485H,2012ARA&A..50..491P}.  

A clear prediction of the numerical simulations is that multiple temperature components contribute to the soft X-ray signal at any given direction of the sky. To disentangle or model them, one must exploit the spectral information encoded in the brightest emission lines. The line ratios in the 0.3-1.1 keV band provide temperature sensitivity (see Appendix A), while the velocity “stratification” of the separate emission components (as illustrated for an example sky region in the outer galaxy in Figure \ref{fig:cgm_phase_vfield}), can be tracked by line centroid shifts at the level of 50 km/s.

Recently, \textit{SRG}/eROSITA measured the intensity of four main components of the X-ray background in the direction of the outer Galaxy with very high precision, exploiting the Performance Verification and All-sky Survey data in eFEDS field\cite{2023A&A...674A.195P}. In Figure \ref{fig:lem_efeds} we show predicted \LEM~count rates for all of these components and their relative contribution to the total signal. Although the depth of the \textit{SRG}/eROSITA data allows mapping of the parameters at the sub-degree scale, the ultimate limitation is set by the spectral resolution of the CCD detector, precluding more realistic multi-temperature models from being tested. \textcolor{blue}{LASS} data will offer a dramatic improvement (see Figure \ref{fig:lem_efeds_fake})! Using the sky brightness of these models we can predict the precision to which velocity can be measured for a given angular scale and exposure time. A simulation of the velocity reconstruction exercise based on 100 realisations of the eFEDS-like spectrum to be collected by \lem\ in the 100 s-deep survey within 1 deg$^2$ is shown in Figure \ref{fig:cgm_vmeasurement}, demonstrating the \lem\ capabilities even for such a small integration area. 

\subsection{INNER GALAXY:\\
AGN BUBBLES IN OUR BACKYARD? }

The central region of the Galaxy and the inner CGM is a natural place to look for the signatures of the feedback from our supermassive black hole, as indicated by the presence of such large-scale energetic perturbations as the Fermi and eROSITA bubbles\cite{2020Natur.588..227P}. While the former require efficient acceleration of gamma-ray emitting particles and their presence at least ~8 kpc away from the Galactic Center, the latter likely  demonstrates CGM gas compressed and heated by a shock wave at  distances of at least ~15 kpc from the disk\cite{2020Natur.588..227P,2023NatAs...7..799G}. 
This collisionless shock wave in the tenuous and magnetized CGM plasma is not only valuable for studying the shock microphysics in a  spatially resolved manner, but it also provides stringent constraints on the energetics of the original event produced it. In particular, direct measurement of the gas velocity behind the shock is crucial for determining the “age” of the bubbles (the shock velocity might be estimated indirectly from the shock Mach number as well, exploiting Rankine-Hugoniot jump conditions for the gas heating and compression\cite{2020Natur.588..227P}, but this relies on accurate spectral modeling and deprojection of the X-ray emission), and hence the average luminosity required to inflate them.

Figure~\ref{fig:bubble_velocities} illustrates the \lem\ capabilities in probing the velocity structures of a simulated eROSITA-like bubble. In this illustrative exercise, we create a \lem\ mock observation of a simulated MW-like galaxy taken from the TNG50 simulation\cite{2021MNRAS.508.4667P}. The simulated X-ray map exhibits morphological features that resemble the observed eROSITA bubble. We show that the velocity structures of the bubble leave distinct imprints on the emission profile of prominent lines such as Fe~XVII (826 eV). Notably, when the line of sight is through the bubble, the expanding shock front causes the line profile to split, producing a characteristic double-horned shape (Panel A of Fig.~\ref{fig:bubble_velocities}). This is in contrast to the case when the sight line cuts near the bubble edge, in which case the line profile exhibits only a single peak (Panel B of Fig.~\ref{fig:bubble_velocities}). With $\sim$1.3 eV spectral resolution of the inner array, \lem\ would be able to resolve the ``double horn'' feature, paving a way to further investigations of the bubble velocity structures. This measurement of bubble kinematics will help to distinguish between a relatively short and luminous AGN burst and more gentle and prolonged episodes of enhanced star formation in the Central Molecular Zone, accompanied by massive Galactic winds.

%------------------------
\begin{figure*}
\centering
\includegraphics[angle=0, width=2.0\columnwidth]{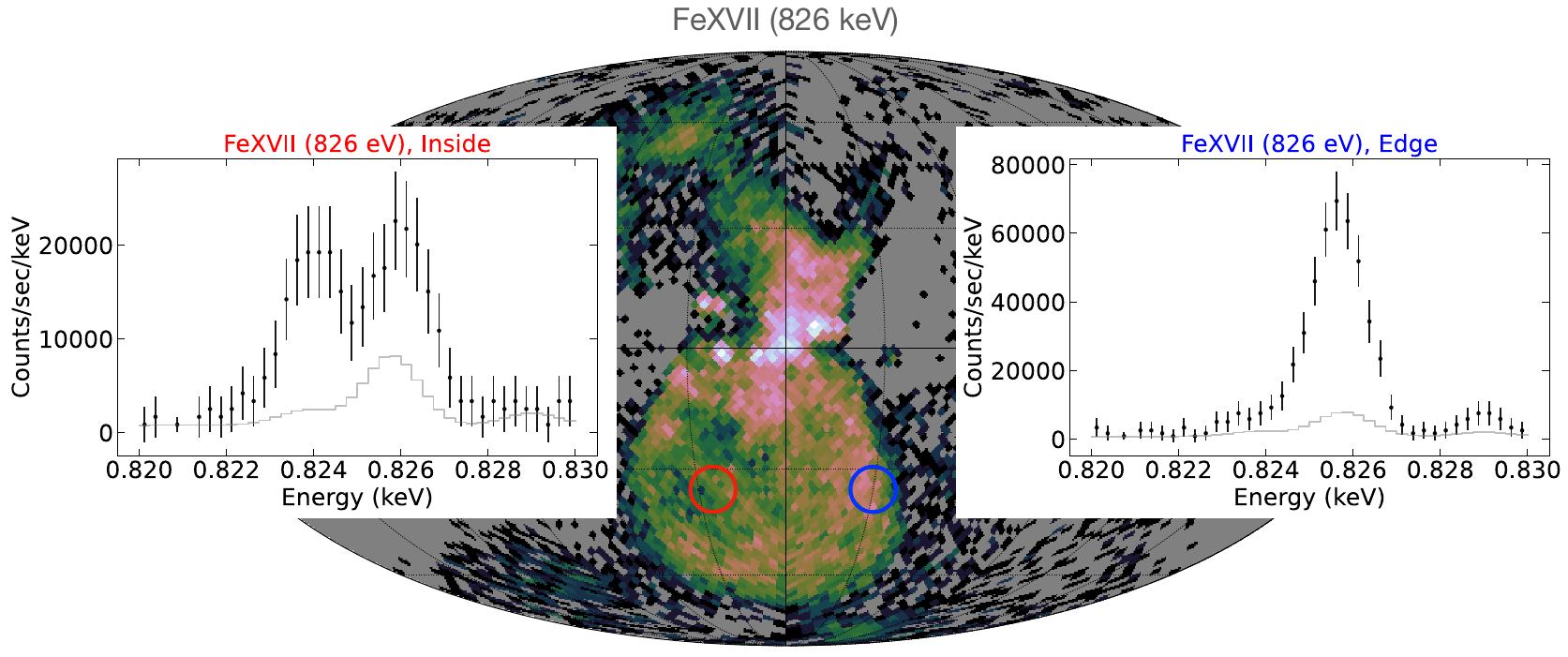}
\caption{\lem\ capabilities for probing an eROSITA-like bubble's velocity structures. \textbf{Main panel:} the all-sky surface brightness map seen in Fe~XVII (826 eV) of a simulated Milky Way-like galaxy from TNG~50\cite{2021MNRAS.508.4667P}. The map is created based on an mock \textcolor{blue}{LASS} observation assuming that a virtual observer located in the Galactic disk at a chosen distance away from the Galactic Center so that the simulated bubble covers a similar region on the sky as the eROSITA bubble. The flux map is calibrated to be consistent with the eROSITA bubble's averaged flux of ${\rm 5.75\times 10^{-12}\ erg/s/cm^2}$ in the [0.6-1.0] keV band\cite{2020Natur.588..227P}. \textbf{Inset subpanels:} the Fe~XVII emission profiles extracted from an inside-bubble region (inset A, red circle) and from an edge region (inset B, blue circle). The gray solid lines show representative background signal, including particle and cosmic X-ray backgrounds, Local Hot Bubble and the Milky-Way halo emission. }
\label{fig:bubble_velocities}
\begin{picture}(0,0)
\put(-190,250){\colorbox{white}{\large A}}
\put(90,250){\colorbox{white}{\large B}}
%\put(-150,200){\colorbox{white}{Declination [deg]}}
%\put(-520,180){\rotatebox{90}{\large Declination [deg]}}
%\put(-248.,263.5){\circle{20}}
\end{picture}
\end{figure*}
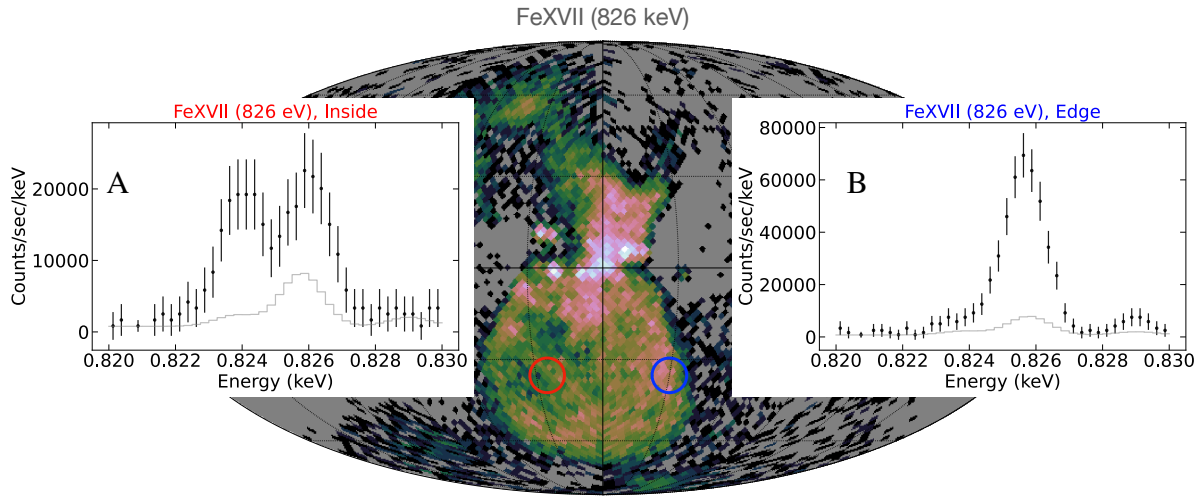
%-------------------------

The enrichment of the CGM gas by outflows is also of great interest, so measurement of the metal abundance patterns can also reveal massive winds transporting significant amounts of iron-rich medium into the CGM. In contrast, the AGN jet driven perturbations would result only in the compression and heating of the CGM gas, with no significant change in its chemical composition.  Sensitive mapping of the relative abundances of oxygen and iron across the whole inner Galactic region is thus one of the most crucial measurements to be enabled by the \textcolor{blue}{LASS}.

\vspace*{-1mm}
\section{LOCAL HOT BUBBLE}
\vspace*{-1mm}

The Solar neighbourhood is the closest and most easily studied sample of the Galactic interstellar medium, an understanding of which is essential for models of star formation and galaxy evolution. 
Observations of an unexpectedly intense diffuse flux of easily absorbed \textonequarter{}~keV X-rays, coupled with the discovery that interstellar space within about a hundred parsecs of the Sun is almost completely devoid of cool absorbing gas, led to a picture of a ``Local Cavity'' filled with X-ray-emitting hot gas, dubbed the Local Hot Bubble (LHB).

The LHB is a key component of the local ecosystem. It is likely responsible for most of the star formation within $\sim100$~pc from the Sun and is a key driver of the evolution of our Galactic neighbourhood\cite{2022Natur.601..334Z}. 
Due to our unique inside viewpoint the LHB also provides a unique opportunity to understand how superbubbles form and evolve; the soft X-ray emission from other similar superbubbles is quickly absorbed by neutral gas in the Galaxy. 

Despite the fact that the solar system resides within this Bubble, very little is known about its nature and physical characteristics --- to the point that its very existence has been challenged multiple times \cite{2005A&A...436..615W,2009ApJ...696.1517K,2021ApJ...920...75L}.
Studying the properties of the LHB is made particularly hard by its relatively low temperature ($\sim0.09$~keV, which emits primarily in the \textonequarter{}~keV X-ray band), the abundance and complexity of line emissions in that energy range, and contamination from SWCX. Current missions have had poor performance in the \textonequarter{}~keV band, so little work has been done on the LHB with modern instruments.

%------------------------
\begin{figure*}
\centering
\includegraphics[angle=0, width=2.0\columnwidth]{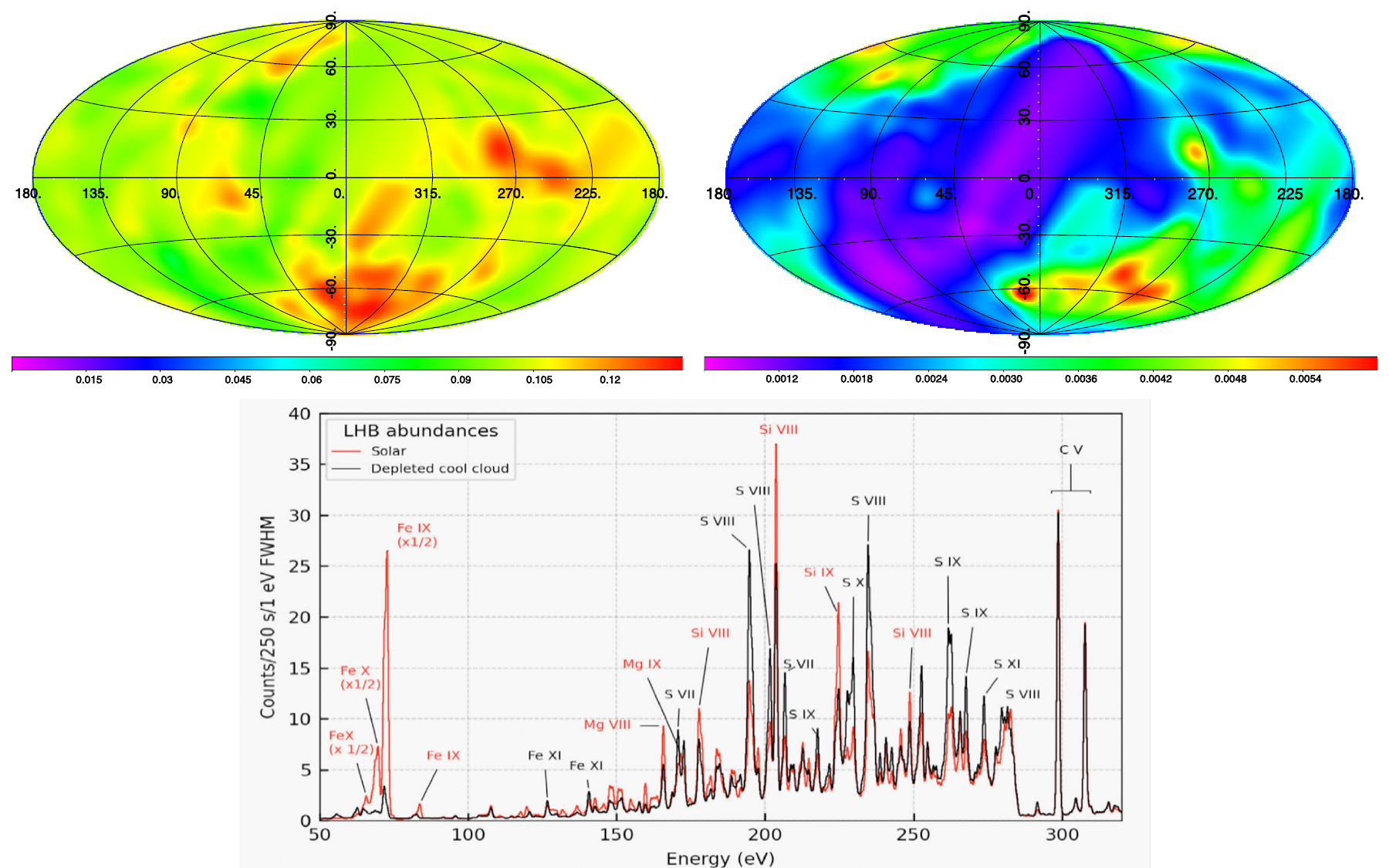}
\caption{\textbf{Top panel:} The LHB temperature (left) in $\rm keV$ and Emission Measure (right) in $\rm cm^{-6} pc$, obtained by combining ROSAT and DXL data (Liu et al.\cite{2017ApJ...834...33L}) assuming thermal equilibrium. \textbf{Bottom panel:} Predicted \textcolor{blue}{LASS} spectrum of the LHB for different metal abundances: Solar and depleted\cite{1996ARA&A..34..279S}. The \textcolor{blue}{LASS} will be able to measure individual lines to identify different temperature components and track the relative abundance within each component to study the origin and evolution of the LHB (see Appendix A).}
\label{fig:lhb}
\end{figure*}
%-------------------------

The \textcolor{blue}{LASS} will be able to identify suitable emission lines of  elements such as Si, S, Mg, C, N, Ar, O, and Fe and isolate them from the surrounding weak and poorly-understood  lines in this very crowded spectral region (Fig.~\ref{fig:lhb}). The lines can then be used to map the LHB temperature, emission measure, ionization state, and relative abundances while identifying the time-variable SWCX contribution  (Appendix A). 

The \textcolor{blue}{LASS} will provide spatially and temporally resolved, high energy resolution spectra over the whole sky. The temporal separation of the spectra (the same area will be covered in $\sim10$~s depth intervals spread over the mission lifetime), combined with the high energy resolution, will allow a separation of LHB emission from SWCX. The spatial resolution will be used to remove contamination from point sources and other bright objects (using complementary data from other surveys such as ROSAT and eROSITA), and will also enable the use of high latitude soft X-ray shadows as a direct probe of the LHB emission. The complete sky coverage will allow a study of the global LHB structure. 

Combined with detailed surveys of the Milky Way structure by Gaia and other source population surveys (e.g., Euclid, 4MOST, Rubin), the \textcolor{blue}{LASS} will address the interaction of the LHB with the Local Cavity and address the LHB origin, physical state, and evolution within the local Galaxy. 

At the end of the survey, \LEM~will be able to  \\
(a) quantify the temperature distribution,\\
(b) infer the emission measure, \\
(c) determine the ionization state, and \\
(d) measure the relative element abundances 
as well as the extent and scales on which they vary.\\
This information will be used, in turn, to determine the origin and evolution of the LHB and the structure of the surrounding Galactic interstellar medium and will lead to resolving the fundamental questions:\\
(a) How did the LHB form and evolve? \\
(b) How does it affect the local interstellar medium and is affected by it? \\
(c) How does it compare to other superbubbles observed in the Milky Way and nearby galaxies?

\begin{comment}
\begin{figure}
    \centering
    \includegraphics[width=\columnwidth]{fig/lem_gp_draco_deg2_t1000.pdf}
    \caption{A simulation of a Draco-like shadow observation in LASS.}
    \label{fig:lem-draco}
\end{figure}
\end{comment}

\vspace*{-1mm}
\section{HELIOSPHERE}
\vspace*{-1mm}

The heliosphere is the Sun's own astrosphere, a bubble-like region blown into space and filled with solar wind (SW) plasma. Initially emitted at supersonic speeds (300-800 km/s), the SW slows down abruptly at the termination shock ($\sim$ 80 AU) due to the pressure of the interstellar medium (ISM). The outer heliospheric limit is the heliopause, at first order defined by the pressure equilibrium between the SW and IS plasmas, and situated approximately $\sim$130 AU from the Sun \citep{2015ApJS..220...32I}. 

Solar wind charge exchange (SWCX) X-rays are emitted by highly charged SW ions when they capture electrons from neutral atoms and molecules. In the case of heliospheric SWCX, the neutral targets are ISM atoms flowing through the heliospheric interface and into the solar system \citep{2005Sci...307.1447L}. This ISM wind, composed mainly by hydrogen (H) and $\sim$15\% helium (He), appears to flow at $\sim$ 25 km/s from the direction $(\ell,b)=(3.2^{\circ},15.5^{\circ})$,
%(255$^{\circ}$, 5$^{\circ}$) in ecliptic coordinates, 
due to the relative motion of the Sun with respect to the local ISM. 

%------------------------
\begin{figure}
\centering
\includegraphics[angle=0, width=1.0\columnwidth]{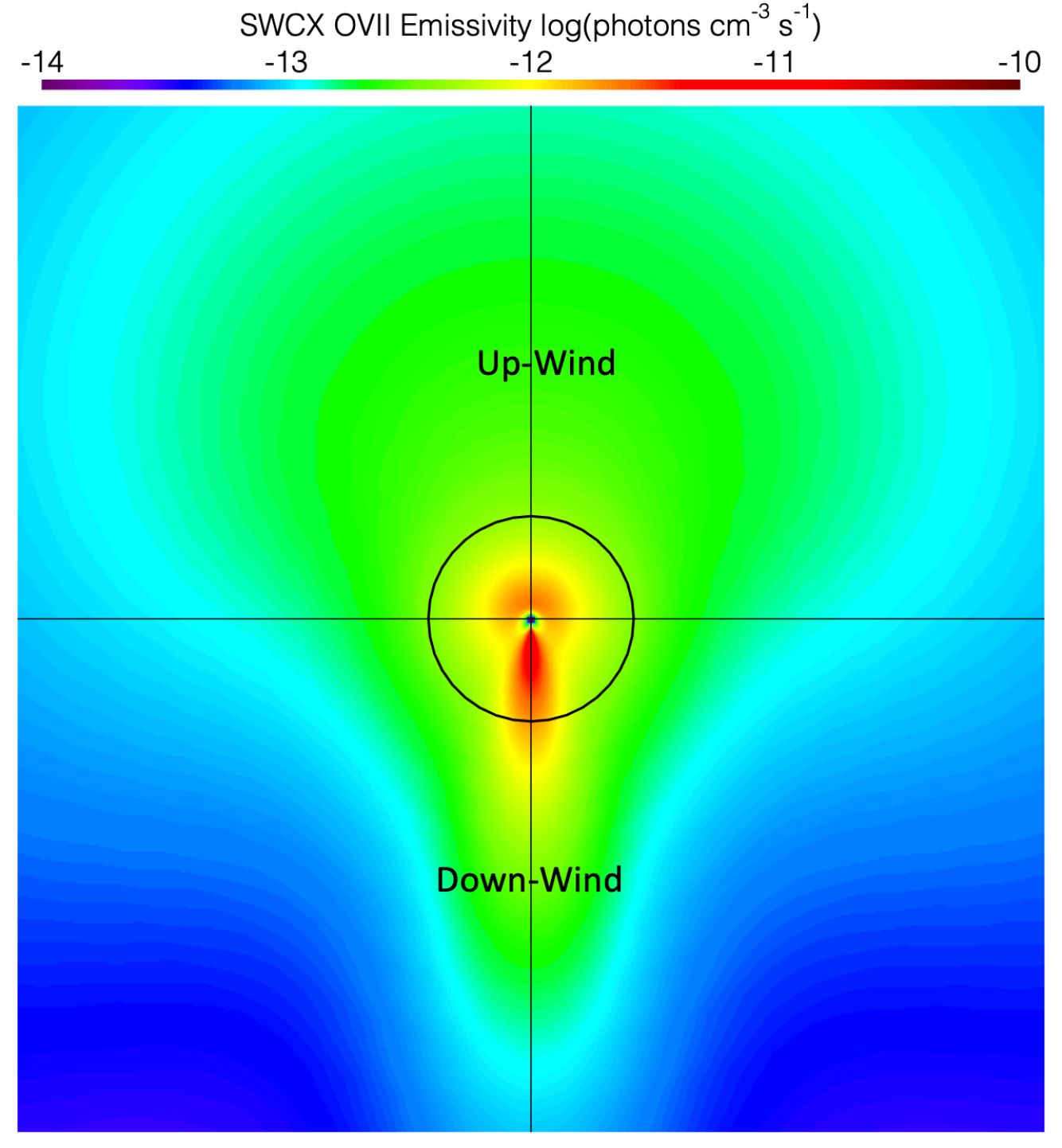}
\caption{Ecliptic plane cut of the total O~VII Solar Wind Charge eXchange volume emissivity in log units of photons/cm$^3$/s for steady-state slow solar wind in solar maximum conditions. The Sun is in the center and the black circle represents roughly the Earth/\LEM~position throughout the year. The Up-Wind (Down-Wind) direction corresponds to early June (December) on the orbit, and points roughly near the Galactic Center (anti-center).}
\label{fig:SWCX_emissivity}
\end{figure}
%-------------------------

To first order, the neutral atom distribution in interplanetary space, and hence the SWCX volume emissivity, is axi-symmetric around the ISM flow vector \citep{2012AN....333..341K}, as shown in Figure \ref{fig:SWCX_emissivity}. SWCX with Hydrogen atoms is the strongest towards the incoming flow direction (upwind, or towards the Galactic Center $\sim$ 2-3 AU from the Sun, represented by the green area in Figure \ref{fig:SWCX_emissivity}), as it is strongly ionised by charge-exchange with solar protons near the Sun. SWCX with He atoms dominates the inner solar system (<1 AU; orange/red area in Figure \ref{fig:SWCX_emissivity}) and especially a cone-like structure in the downwind direction where He atoms are focused due to the Sun's gravity (H atoms cannot be focused like He, because they are subjected to solar radiation pressure). Therefore, for an observatory like \lem , orbiting the Sun, an all-sky map of SWCX emission will appear very different at different times of year, since the map traces different neutral column densities due to parallax effects \citep{2006A&A...460..289K}. In Figure \ref{fig:swcx_lass} we present average all-sky maps of the SWCX OVII triplet emission, constructed over periods of 6 months (the time it will take \LEM~for a complete scan of the sky) starting at different times of the year, highlighting the heliospheric parallax effects. These maps are calculated for steady-state slow solar wind parameters \citep{Koutroumpa2023}, roughly corresponding to maximum solar activity.

\begin{figure}
    \centering
    \includegraphics[width=\columnwidth]{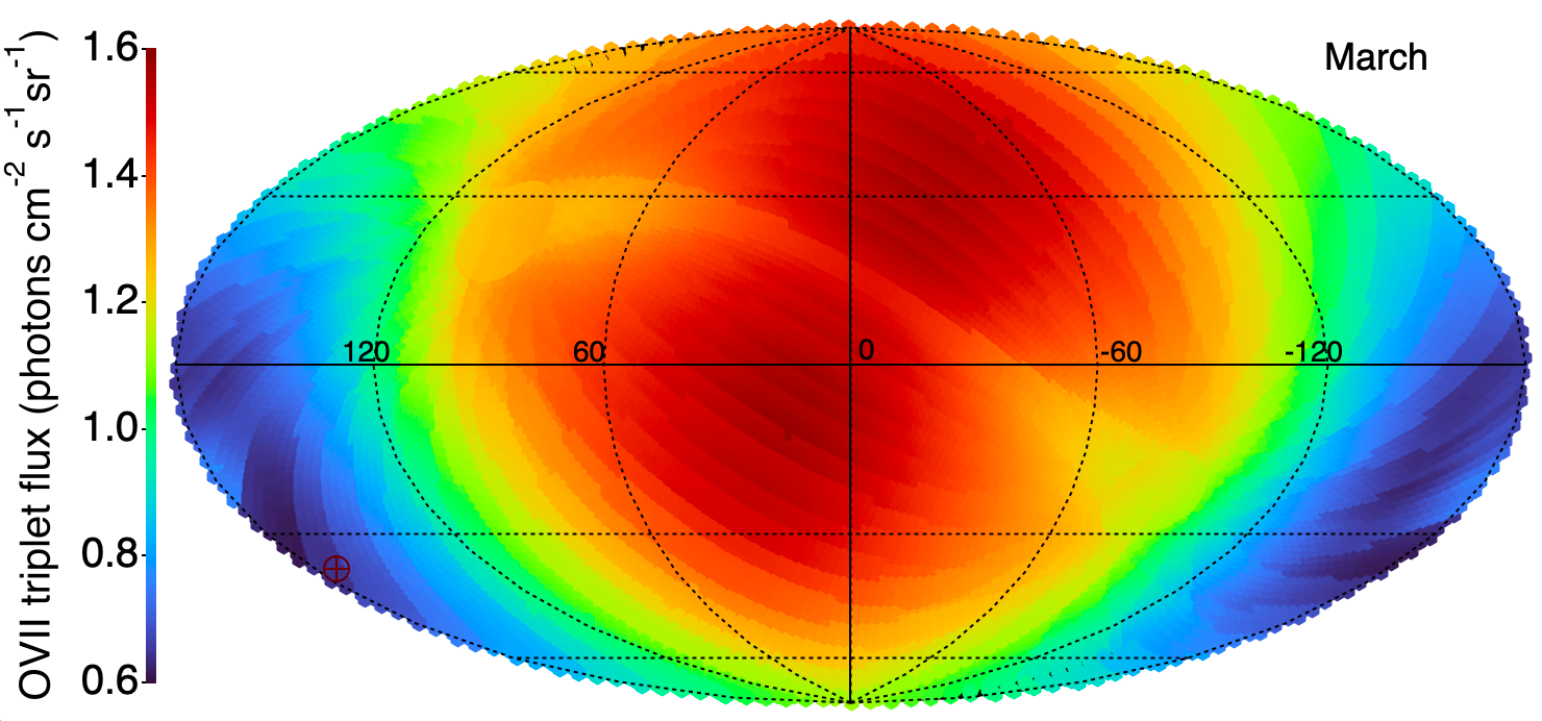}
    \includegraphics[width=\columnwidth]{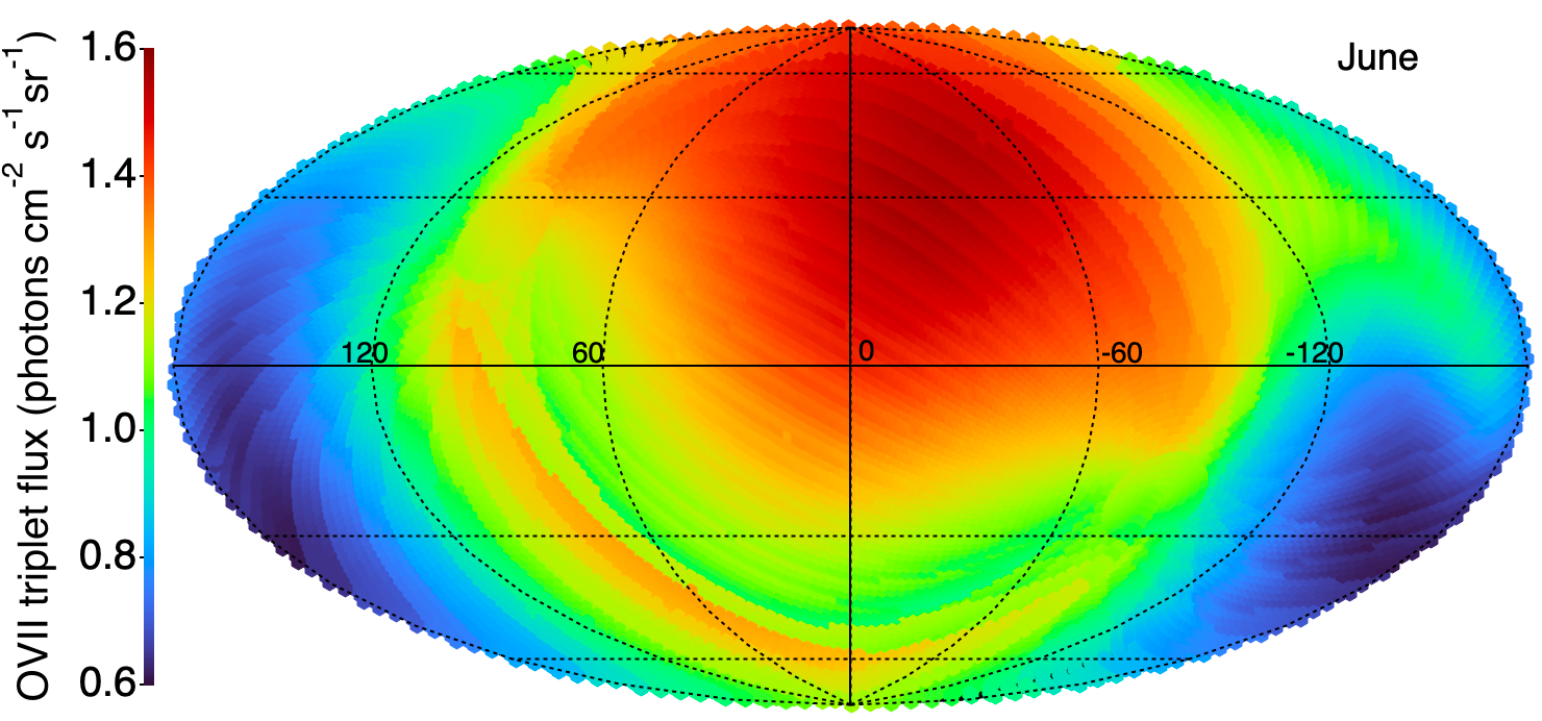}
    \includegraphics[width=\columnwidth]{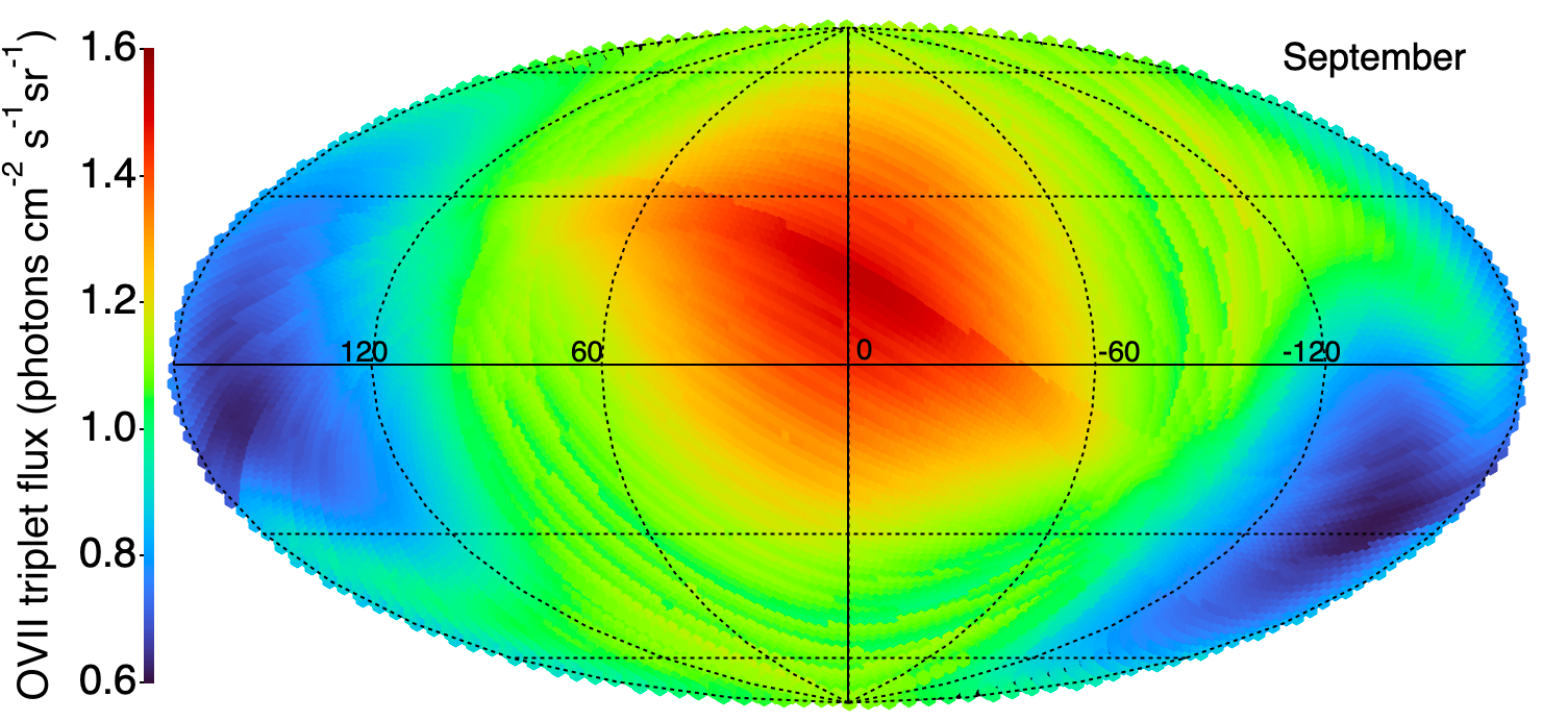}
    \includegraphics[width=\columnwidth]{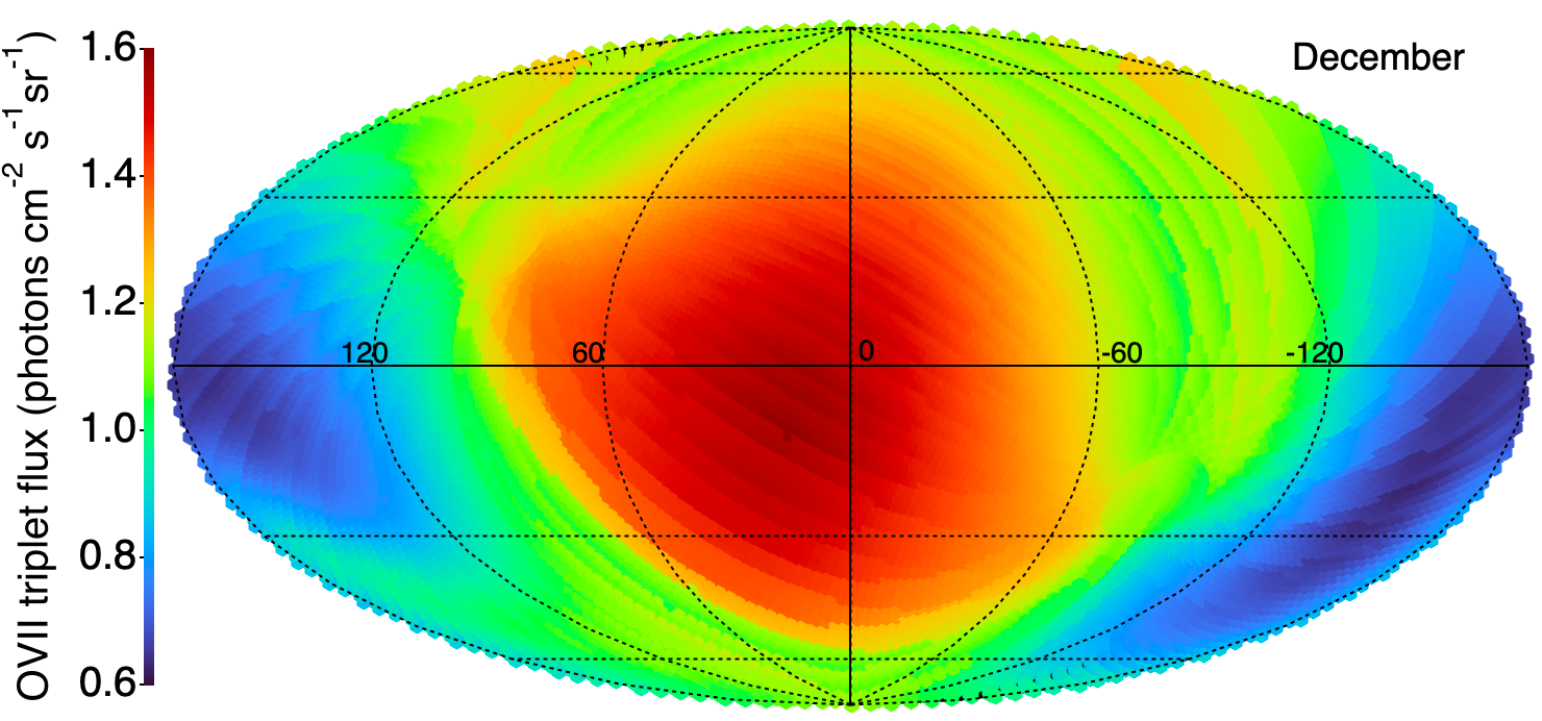}
    \caption{Average \lem\ All-Sky Survey O VII SWCX maps in units of photons/cm$^2$/s/sr for steady-state slow solar wind conditions (solar maximum). The maps are in zero-centered Galactic Aitoff coordinates with longitude increasing to the left. From top to bottom, the maps are calculated using a continuous 6-month period on \lem 's orbit starting in March, June, September and December, and respecting the Solar aspect angle constraints. The discontinuities are due to the change of observer position for each tile scan.}
    \label{fig:swcx_lass}
\end{figure}

In reality, the SWCX emission will have more complex characteristics, involving temporal variations due to the SW intrinsic variability, in addition to the spatial variations of the neutral distributions. The \textcolor{blue}{LASS}, in high resolution spectroscopic mode, will provide innovative diagnostics for the heliospheric SWCX science. A few key features that may be studied are discussed in the next few paragraphs.

Latitudinal distribution of the solar wind (SW): the solar wind has a sharp bimodal configuration during solar minimum, with a narrow zone of high density, high temperature, slow ($\sim$ 390 km/s) SW around the ecliptic/solar equator, while at high ecliptic latitudes large coronal holes emit low density, low temperature, fast (> 600 km/s) SW (e.g. Ulysses data in Figure 1 of~ \cite{2008GeoRL..3518103M}). The two SW types have considerably different ion composition that imprints on the SWCX spectral signal (Figure~\ref{fig:swcx_SWtype}). In the slow SW the high-state ions (e.g. C$^{6+}$, N$^{7+}$, O$^{7, 8+}$, Ne$^{9+}$), that produce higher energy spectral lines, have higher relative abundances, while in the fast SW the high-state ions abundances are lower. 
On the other hand, the lower charge states (e.g. C$^{5+}$, O$^{6+}$) have opposite behaviour, with increased relative abundances in the fast SW. 

The composition changes between slow and fast SW result in a change of hardness ratio in the SWCX spectra. By comparing lines-of-sight near the ecliptic that are completely embedded in the slow solar wind, and high latitude lines-of-sight that cross mainly coronal hole type solar wind with only a small contribution of equatorial slow SW, we may provide a complementary analysis of the latitudinal distribution of the SW, in particular for the extended mission that will encompass a large period of the solar cycle covering the maximum and minimum activity periods. 

Identification of previously unresolved lines: The bulk of the SWCX emission (and the Local Bubble) is produced below 0.3 keV, that current instruments cannot access due to limited sensitivity/grasp at those energies. \LEM's efficiency and spectral resolution will allow for the first time to identify the key spectral lines and measure their relative strengths (Figure~\ref{fig:swcx_SWtype}). 
It may also be possible to measure solar abundances for very minor species that are not measured with solar {\it in situ} instruments (e.g. Ar, P) which will provide new insights into the processes that make the abundances in the solar wind different from those of the surface of the Sun.

Localized enhancements/Doppler measurements due to ICMEs (Interplanetary Coronal Mass Ejections): ICMEs can have both very high velocities, and significant changes in SW ion composition \cite{2004JGRA..109.9104R}. Detecting transient SWCX events, through localized enhancements and/or Doppler shift measurements, in repeated \textcolor{blue}{LASS} pointings of the same regions will put constraints on the SW propagation models used by the heliophysics community for space weather predictions at Earth.

Doppler measurements of the inner heliosphere and the outer heliosheath regions: the heliospheric interface is a complex region of great interest for the physics of the solar wind interaction with the interstellar medium (cf Voyagers’ studies of the termination shock and heliosheath \cite{2005Sci...309.2017S}, and future plans of an Interstellar Probe mission \cite{2023SSRv..219...18B}). At the termination shock the supersonic SW (400-800 km/s) decelerates abruptly to speeds <50 km/s, while it is heated from a few thousands K to nearly 2 MK (Figure~\ref{fig:swcx_MgXI} - top). In the meanwhile the interstellar hydrogen population undergoes charge exchange with protons piled up in the heliosheath (region between the termination shock and the heliopause), resulting in a filtration and modification of its distribution (density, velocity and temperature). The SWCX emission from the heliosheath will be near the rest-frame of the spectral line energies, while the inner heliosphere SWCX emission will be red-shifted by $\sim$ 2eV, especially for the higher energy lines (e.g. Mg~XI, Figure \ref{fig:swcx_MgXI} - bottom). Measurements with \LEM~(Figure \ref{fig:swcx_MgXI_lem}) near the nose of the heliosphere $(\ell,b) \sim (3.2, 15.5)$, where the velocity change and thus the Doppler shift is the largest, with accumulated (but not necessarily continuous) exposures during the All-Sky Survey, will allow detection of the heliosheath SWCX emission and will constrain heliospheric interaction models.

%------------------------
\begin{figure*}
\centering
\includegraphics[angle=0, width=2.0\columnwidth]{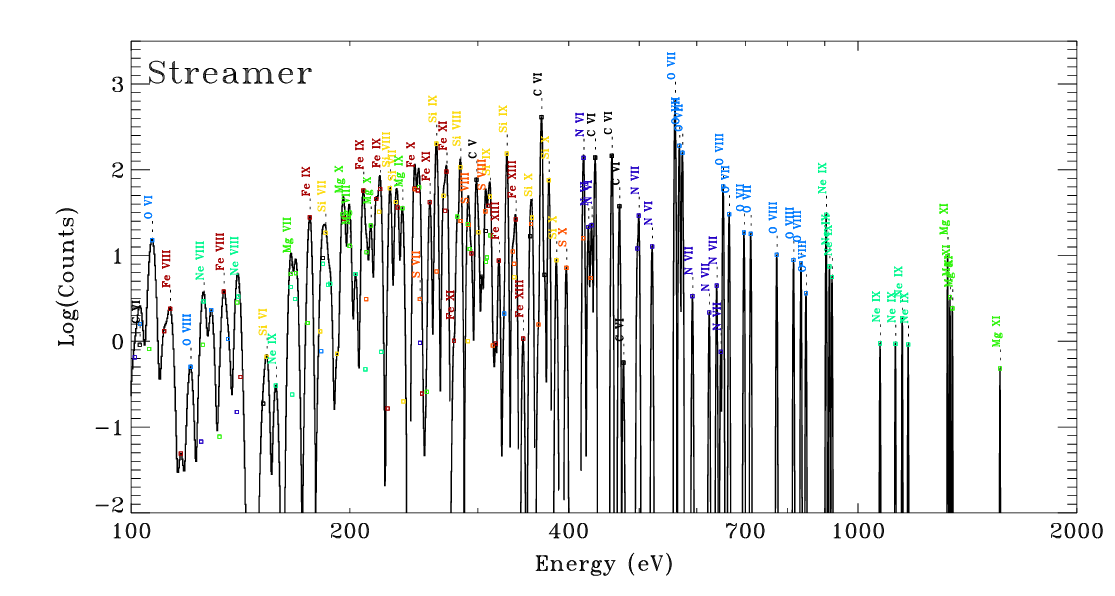}
\includegraphics[angle=0, width=2.0\columnwidth]{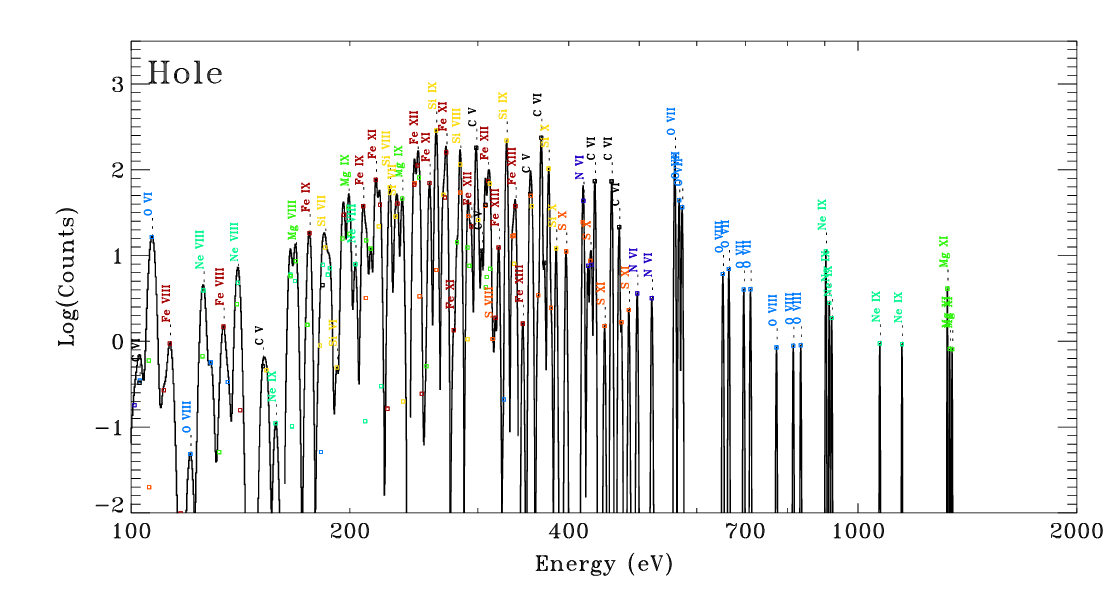}
\caption{\textbf{Top}: Theoretical heliospheric SWCX spectrum for streamer (slow) type of SW in counts for a 100 s observation of a $10^\circ$ by $10^\circ$ region. \textbf{Bottom}: Same as the top panel, only for coronal hole (fast) SW type (see text). The \lem\ response and line spread function has been applied. Major lines have been labelled. \textcolor{blue}{LASS} will e able to detect and identify not only a few bright lines, but an abundance of weaker lines as well. The boxes indicate the strength of an individual line; if it is not at the observed peak, then the observed line is strongly contaminated with other lines. The two SW types have considerably different ion composition which can be readily seen in the spectra. Thus, \lem\ will not only detect the SWCX emission, but will also be able to determine what type of solar wind is producing it. These spectra also demonstrate how the variety of accessible species, and thus our capability to diagnose the physical state of a plasma, increases sharply as the exposure time increases.}
\label{fig:swcx_SWtype}
\end{figure*}
%-------------------------

%------------------------
\begin{figure}
\centering
\includegraphics[angle=0, width=1.0\columnwidth]{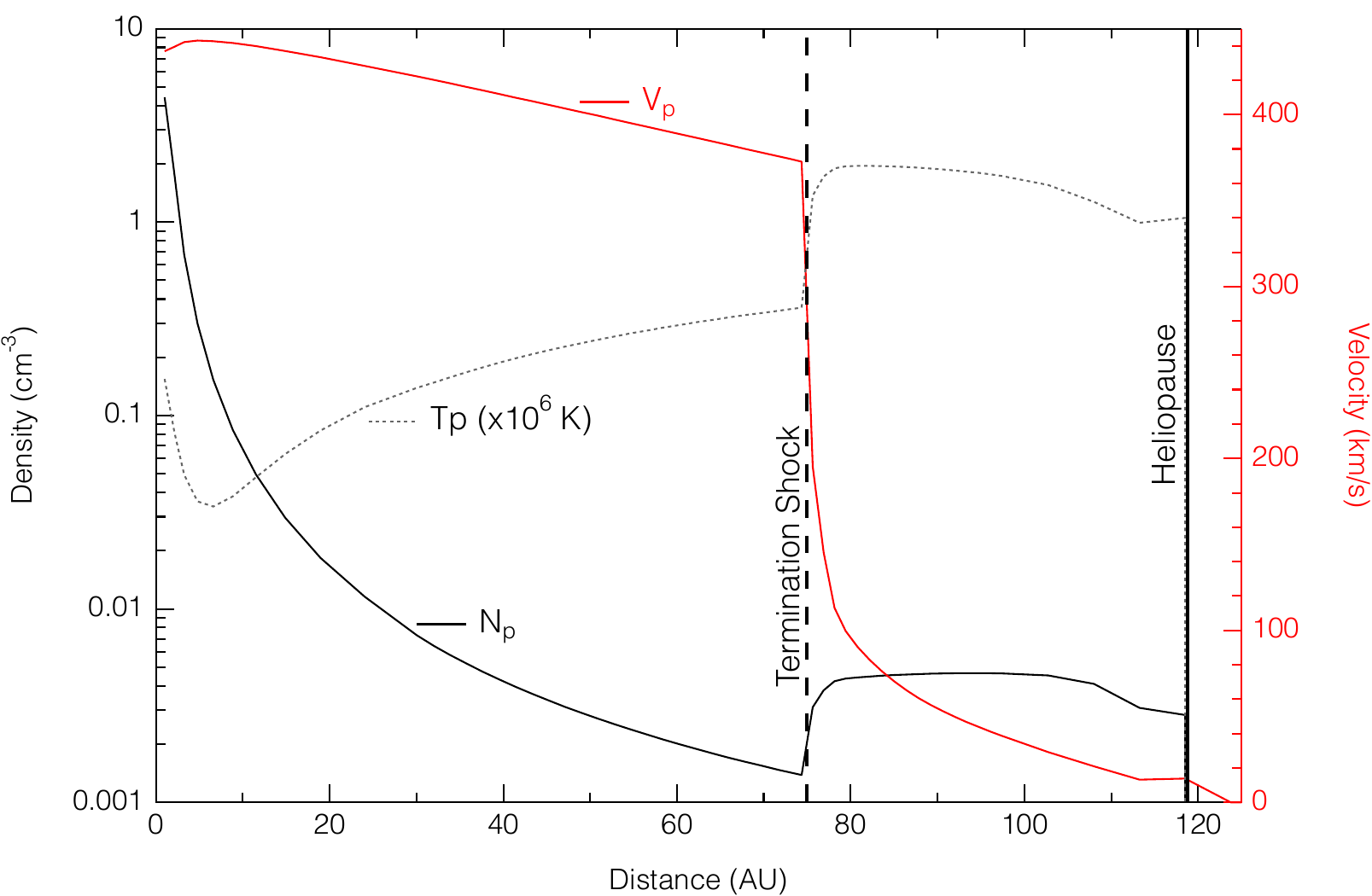}
\includegraphics[angle=0, width=1.0\columnwidth]{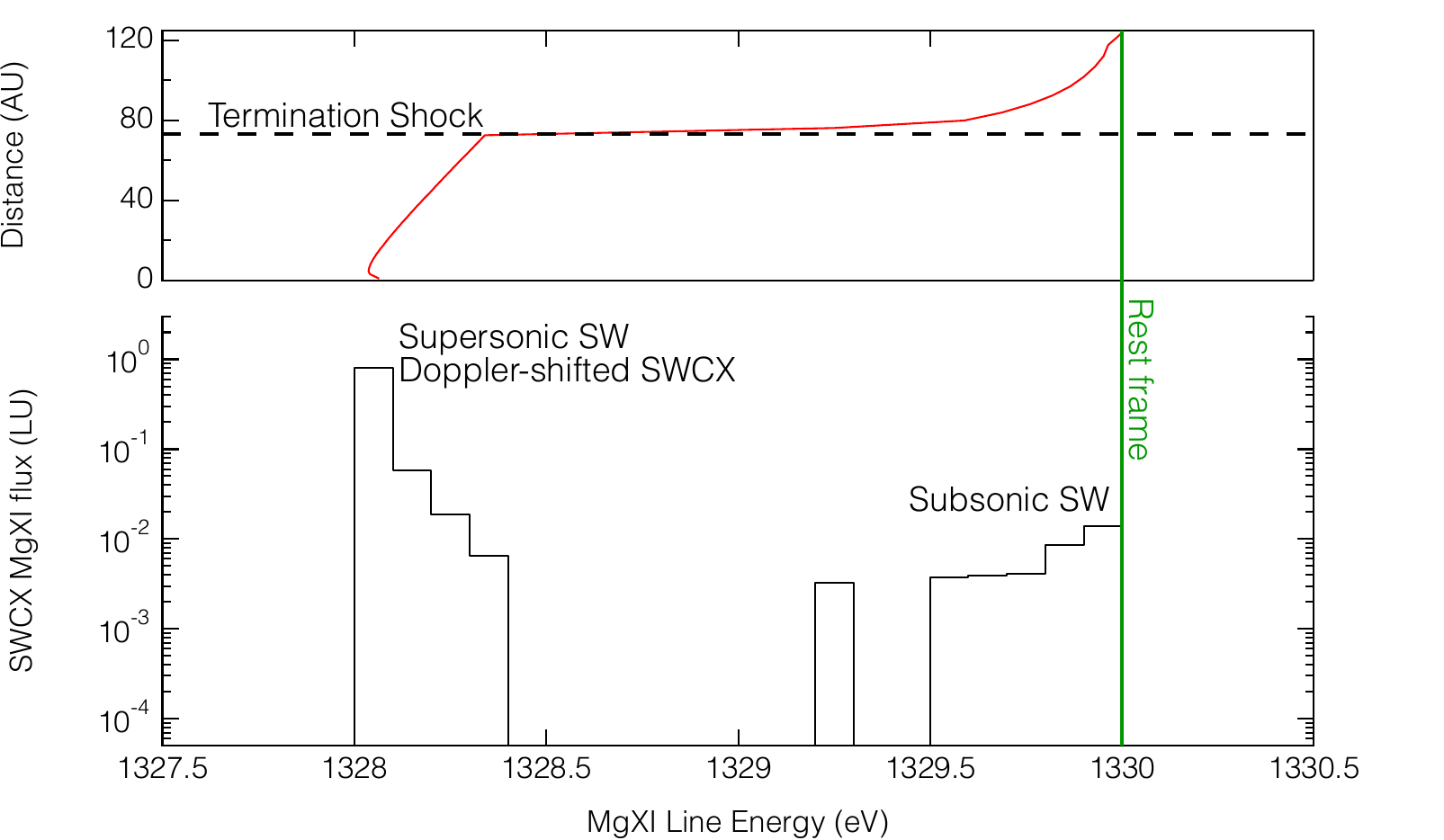}
\caption{\textbf{Top}: Solar wind proton density (full black curve), temperature (dotted black curve), and velocity (full red curve; right axis) in the direction of the nose of the heliosphere, based on the model from \cite{2015ApJS..220...32I}. The vertical lines represent the termination shock (dashed) and heliopause (full). \textbf{Bottom}: SWCX flux (in photons/cm$^2$/s/sr) of the Mg XI line at 1.33 keV, binned in 0.1 eV bins. The line is separated in two components, one Doppler-shifted, produced by supersonic solar wind ions up to $\sim$ 80 AU (top sub-panel), and one near the rest frame, produced by subsonic solar wind ions after they cross the termination shock (see text for details). 
}
\label{fig:swcx_MgXI}
\end{figure}
%-------------------------

%------------------------
\begin{figure}
\centering
\includegraphics[angle=0, trim=1cm 12cm 1cm 2.5cm, width=1.0\columnwidth]{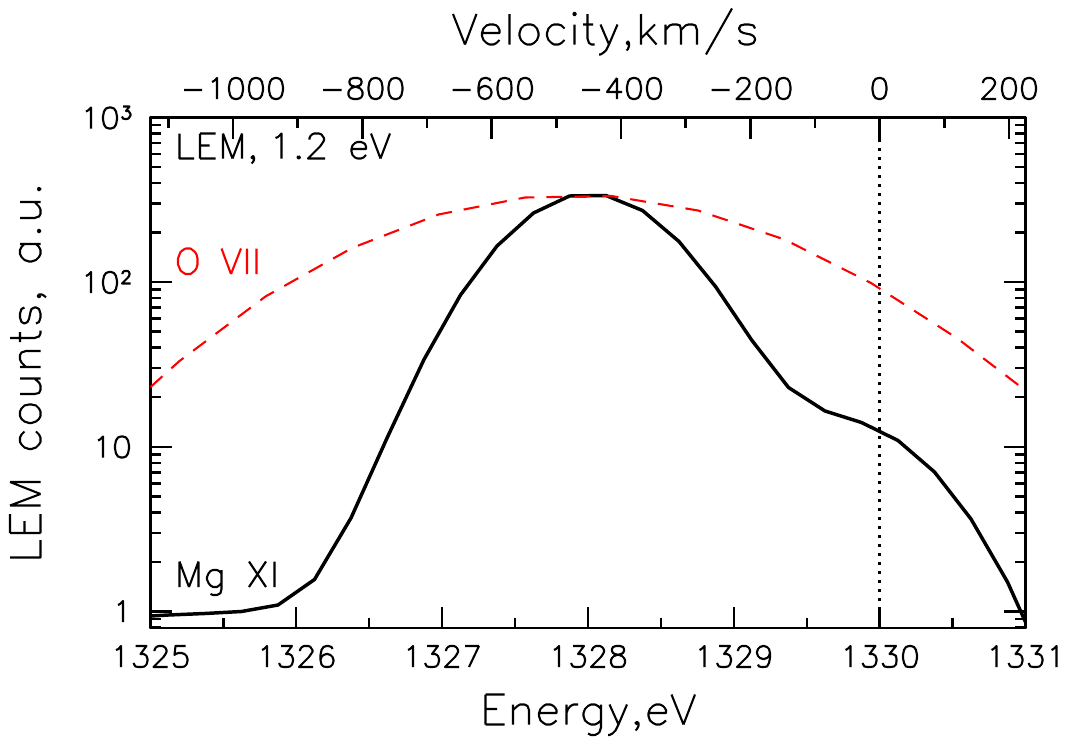}
\caption{ Convolution of the predicted velocity profile (shown in Figure \ref{fig:swcx_MgXI}) for the Solar Wind Charge Exchange (SWCX) Mg~XI line with \lem\, response function for the inner array (FWHM=1.2 eV). For comparison, the red dashed line shows the convolved SWCX velocity profile for O~VII,  demonstrating the unique power of the Mg~XI line for resolving the subsonic Solar wind component.}
\label{fig:swcx_MgXI_lem}
\end{figure}
%-------------------------

%\clearpage
\vspace*{-1mm}
\section{STACKING \& CROSS-CORRELATING}
\vspace*{-1mm}

%------------------------
\begin{figure}
\centering
\includegraphics[angle=0,trim=0cm 5cm 0cm 2cm, width=1.0\columnwidth]{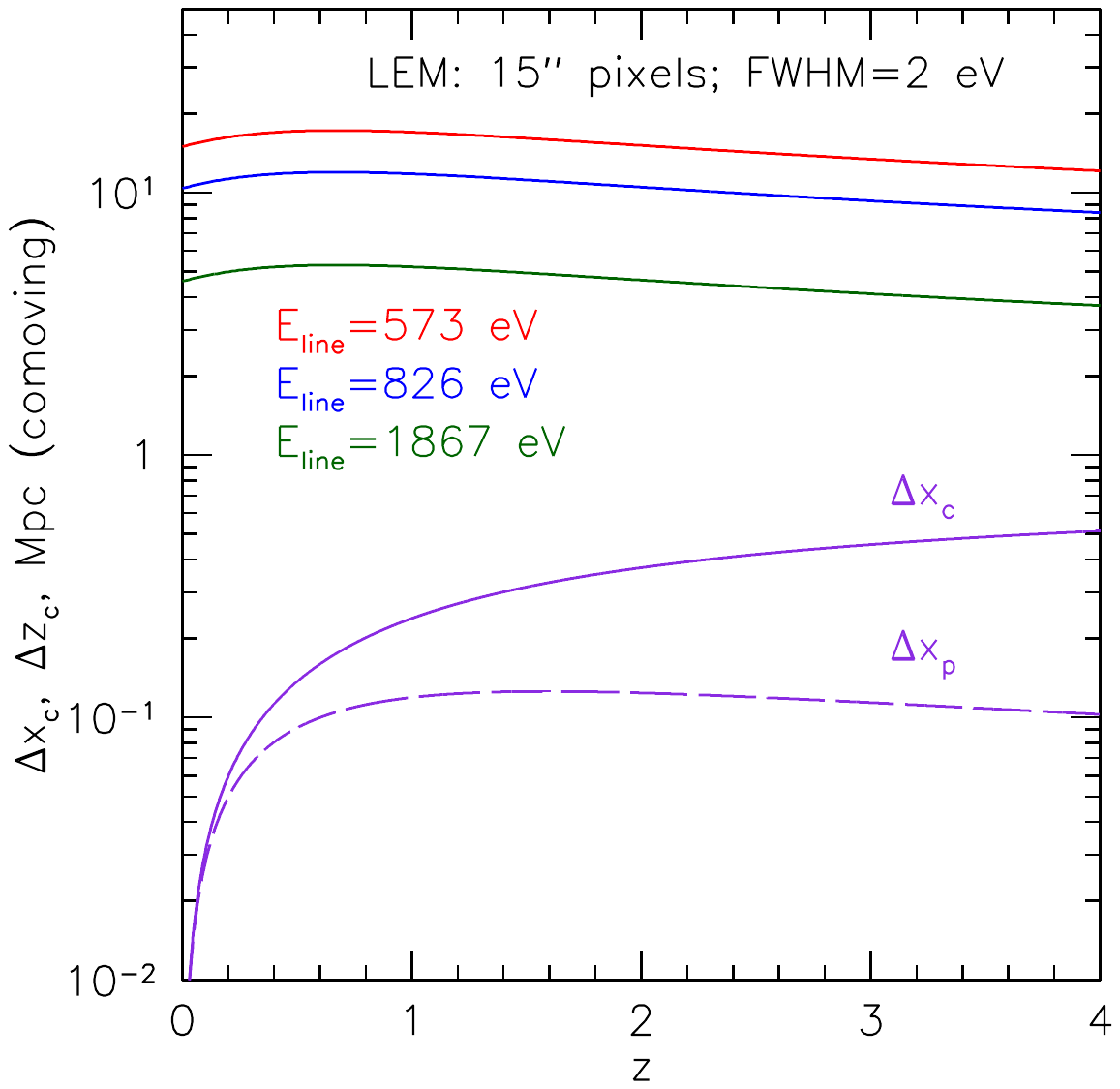}
\caption{\lem~ "3D" spatial resolution as a function of redshift. The violet lines show the comoving ($\Delta x_c$, the solid line) and physical sizes ($\Delta x_p$, the dashed line) corresponding to $15''$ pixels. The other three lines show the effective resolution along the lines of sight, corresponding to the comoving distance between objects in the Hubble flow.}
\label{fig:lem_res}
\end{figure}
%-------------------------

Compared to pointed observations, the main feature of the \textcolor{blue}{LASS} data is the access to multiple objects of the same class, e.g. galaxies, clusters, filaments, etc., that can be combined in a way that alleviates the problem of variance, associated with peculiarities of individual objects and selection biases.  For all objects that have a mean surface number density on the sky larger than 1 per 0.25 deg$^2$ FOV (or 160,000 on the sky), a comparable S/N ratio will be achieved when summing the signal from multiple objects in the \textcolor{blue}{LASS} data and doing the same for pointed observation in a random direction with a similar combined exposure. For less abundant objects, the effective stacked exposure in the \textcolor{blue}{LASS} data will be $t\approx t_{LASS} \times N_{obj}/1.6\times10^5\approx N_{obj}\times 100 \,{\rm s}$. In other words, for any class of objects with more than $10^4$ members, the \textcolor{blue}{LASS} will automatically provide 1~Ms worth of data. Of course, pointed observations of selected or truly unique objects, e.g. the nearest ones (or the most prominent) will be indispensable\cite{2023ApJ...953...42B,2023arXiv231004499M,2023MNRAS.525.1976T,2023arXiv230701259S,2023arXiv231002225Z}. Still, if the cosmic variance is an issue, the \textcolor{blue}{LASS} represents an ultimate dataset.

\subsection{Galaxy clusters and groups}

A comparison between observations and simulations of galaxy cluster density and temperature profiles is only meaningful when averaging over many objects, revealing trends free from stochastic variations. Related to averaging is another appealing feature of \textcolor{blue}{LASS}, namely an effectively unlimited FoV that allows one to map extended objects in their entirety. The \textcolor{blue}{LASS} offers an opportunity to reach the faintest possible levels in terms of the mean surface brightness, being limited only by the photon counting noise, since background/foreground variations are suppressed by a factor $1/\sqrt{N_{obj}}$ (see an example\cite{2023MNRAS.525..898L} demonstrating that by stacking some 40 clusters it is possible to reach radii $\sim 3 R_{500c}$, where the X-ray surface brightness of cluster emission drops below 1\% of the sky background). With the \textcolor{blue}{LASS}, one can measure ``averaged'' properties of the transition region between hot virialized gas and infalling matter, and use emission lines to infer the spread in temperature and the degree of clumpiness. Without the information from soft emission lines, it is nearly impossible. The same approach applies also to groups and isolated elliptical galaxies.

Cluster distant outskirts, all the way up to the turn-around radius $r_{ta}\sim 10\,{\rm Mpc}$,  are the natural targets for \LEM~in the \textcolor{blue}{LASS}. This is a highly dynamic, very irregular, and very multiphase region where the local overdensity changes from 50-100 down to 2-3. Stacked data of known clusters near $r_{ta}$ would provide a fair sample of the Universe X-ray emissivity in a trans-nonlinear regime.

The same technique can be applied to protoclusters - the most massive halos beyond redshift $z\sim 1.5-2$ (e.g., the Spiderweb cluster\cite{2023Natur.615..809D} at $z\sim 2.16$). This is the epoch of rapid growth of SMBHs and their X-ray emission can outshine the thermal emission of the diffuse gas. In the case of the Spiderweb, inverse Compton scattering of the CMB photons by AGN jets dominates\cite{2022A&A...667A.134T} the core of the cluster. \LEM~can resolve lines from the ICM on top of the non-thermal emission and measure the gas properties there. Stacking various protocluster proxies (e.g., Spiderweb was discovered\cite{1994A&AS..108...79R} as an overdensity around an ultra-steep spectrum radio galaxy), coupled with the \lem\ spectral resolution is the promising route to characterise their population.   

\subsection{Diffuse IGM}

Stacking of objects and/or cross-correlation are actively utilised in observational cosmology, e.g. by using WMAP or Planck maps in combination with other data. These studies mainly exploit spatial information. \LEM’s excellent energy resolution adds additional capabilities to this procedure. Indeed, the $15''$ angular resolution of \LEM~corresponds to a physical distance of less than 100 kpc at any redshift, while 2~eV energy resolution allows \lem\ to distinguish lines for objects that are separated in the Hubble flow by $\sim 10$ Mpc (see Fig.~\ref{fig:lem_res}).
This would make a cross-correlation of \LEM’s data with the massive redshift surveys of galaxies 
%(e.g. Euclid Wide Survey) 
extremely powerful and convert the usual position-based stacking to full 3D stacking/cross-correlation. The possibility using the redshift ``dimension''  increases the amount of information one can get from X-ray data. Moreover, since the expected \lem\ signal sensitively depends on the density, temperature (see Fig.\ref{fig:igm_3inst}), and metallicity of the gas, the cross-correlation analysis can be tuned to emphasise a particular group of temperature-sensitive lines, adding yet another dimension to the problem. While the distribution of galaxies is a biased tracer, as is the X-ray signal, one can further use \LEM's angular resolution to selectively exclude halos of a given mass from the cross-correlation analysis, effectively disentangling the contributions of different halos. 

Because of the strong Galactic foreground emission, the stacking analysis will be limited to particular "redshift windows", preventing the IGM emission lines (primarily O~VII and O~VIII) from being overwhelmed by their Galactic counterparts. An example of such a window is the redshift range from 0.05 to 0.1. This volume of the Universe is actually already relatively well-studied thanks to the 2MASS Redshift Survey\cite{2012ApJS..199...26H}, and many of its global mass distribution properties are well understood and can be reliably modelled via constrained cosmological simulations (see an example of such reconstruction in Figure \ref{fig:lem_constrained_igm} based on the SLOW\cite{2023A&A...677A.169D} simulation).

\begin{comment}

%------------------------
\begin{figure}
\centering
\includegraphics[angle=0, trim=1cm 5cm 0cm 2cm,width=1.0\columnwidth]{fig/lem_spec_whim_r1}
\caption{Placeholder}
\label{fig:lem_whim}
\end{figure}
%-------------------------
\end{comment}

\subsection{Stars}

While the effective area of \lem\, is unprecedented for a high-resolution X-ray spectrometer, there are still key parts of high-energy stellar physics that will remain out of reach within reasonable pointed observation exposure times. In some of these cases, stacking analysis based on the known positions of stars from deep surveys such as Gaia can provide key insights into the X-ray characteristics of groups of similar types of stars, such as plasma temperature distributions, densities, and chemical compositions. We cite three examples here, each of which is relevant for exoplanet science as well as intrinsic astrophysical interest. 

High-mass stars with masses $M>6 M_{\odot}$ and spectral types O to early B generate copious X-rays through instabilities and resulting shocks in their radiatively driven winds\cite{2001ApJ...554L..55C,2001A&A...365L.312K}. Low-mass stars with masses $M<1.5 M_{\odot}$ generate X-rays through convection and rotation-driven magnetic activity\cite{2023hxga.book..132D}. The intermediate-mass later B- and A-type stars are X-ray dark\cite{2006ApJ...636..426P,2008AJ....136.1810A}. Little is known about the transitions between these regimes. Scant detections of stars at the lower mass transition point suggest coronae become cool as well as faint\cite{2012ApJ...750...78G}. At the higher mass transition, the situation is less clear, with detections at types mid-to-late B suggesting a possible transition to, or hybrid of, wind and coronal X-ray mechanisms\cite{1997ApJ...487..867C}. Stacking analysis of intermediate-mass stars will probe the physics and plasma conditions of these boundaries. 

Little is known about magnetic activity at the end of the main sequence. Both X-ray and chromospheric H$\alpha$ emission undergo a quite precipitate drop in mid-M to L field dwarfs\cite{2006A&A...448..293S,2010ApJ...709..332B} but radio emission remains strong\cite{2023hxga.book..132D}. In the case of brown dwarfs the situation is complicated by evolutionary effects, with young objects resembling M dwarfs or late T Tauri stars. The very cool, neutral photospheres of late M dwarfs and brown dwarfs should be largely incapable of sustaining magnetic stresses, yet flaring on these objects appears to be quite frequent. Generally too X-ray faint for detailed study,  these objects are sufficiently numerous in the solar vicinity that stacking of \LEM~spectra of field late M dwarfs and brown dwarfs can be key for revealing the details of their plasma physics. 

Radiative losses from stellar coronae are dominated by metal lines. One outstanding problem in coronal physics is whether metal-poor stars sustain X-ray emitting coronae, and if so what is the effect of metal paucity on the coronal energy balance \cite{1996ApJ...472L.101F,2023hxga.book..132D}. Population II stars are metal-poor, but they are also old and expected to be of low magnetic activity and faint in X-rays. Stacked \lem\, spectra of Population II stars will provide a window into this astrophysical regime that will be generally inaccessible through pointed observations.  
 
\subsection{AGN}

For directed, Deep Survey fields up to 1 Msec, \lem\, will reach 0.2-2 keV point source flux levels of $\ge2\times10^{-16}~{\rm erg~cm^{-2}~s^{-1}}$ and individual 6.4 keV rest-frame Fe~K$\alpha$ lines for heavily absorbed AGN may be detectable down to a line flux of $4\times 10^{-17}$ erg ${\rm cm^{-2}~s^{-1}}$ above redshifts of 3 for 1 Ms, depending upon the iron abundances within these AGN populations.\ A more comprehensive all-sky survey will provide even more source photons from background sources. 
Serendipitous science from \textcolor{blue}{LASS} provides a further rich opportunity for AGN studies. 

Cross-comparisons with multi wavelength catalogs from other missions will allow us to stack photons from different source populations. Following previous studies\cite{2012A&A...537A..86I,2011A&A...530A..42C}, we will stack spectra according to AGN type and probe intrinsic absorbing column densities, accretion rates, Eddington ratios (given estimates for the black hole mass), and the evolution of the relationship between X-ray reflection strength and intrinsic AGN source luminosity. JWST has provided a detailed study of AGN hosts from z$\sim3-5$ \cite{2023ApJ...946L..14K} (Yang et al., in press). \lem~is perfectly suited to provide complementary observations and high-energy X-ray insight into this population of host galaxies, allowing for more detailed study of the accretion properties of AGN that reside inside them. 

There are also significant efforts to increase the populations of detected high-z galaxies and AGN with current and future observatories for statistically significant studies. For example, COSMOS-Web \cite{2023ApJ...954...31C} is expected to unearth nearly 1,000 galaxies with 6.5<z< 7.5 with JWST\cite{2022MNRAS.516.1047M}. These studies will be further advanced in the advent of Roman Space Telescope and Euclid\cite{2011arXiv1110.3193L}, which will offer high sensitivity over wider fields of view, where yields of hundreds of quasars at z$\sim$6-8 are expected to be discovered over the course of their survey periods\cite{2019A&A...631A..85E,2020MNRAS.499.3819M}. 
All these samples are excellent candidates for \lem~stacking studies.

\subsection{CGM} 
%NT

A key objective of the \LEM~science program is to explore the circumgalactic medium surrounding individual galaxies. Within its directed program, \LEM~will investigate a sample of 30 individual galaxies, encompassing a spectrum of stellar masses, star formation rates, and black hole masses. These observations will probe how gas and metals flow into, through, and out of galaxies. Concurrently, the \textcolor{blue}{LASS} will enable the study of circumgalactic medium properties by leveraging data from thousands to tens of thousands of galaxies in the local Universe. Although the X-ray emission from individual galaxies will be undetected, co-adding (i.e.\ stacking) the X-ray photons from a large sample of galaxies will yield a statistically significant signal. With a large galaxy sample and the application of a stacking analysis, it will be possible to explore the galaxy parameter space, covering properties such as morphology, stellar mass, star formation rate, and black hole mass. This analysis, in turn, will probe how the large-scale CGM properties depend on these parameters, allowing for a clear separation of various effects.

While similar studies utilizing {\it ROSAT}, and more recently {\it Chandra}, {\it XMM-Newton} and {\it SRG/}eROSITA CCD data have been carried out \cite{2013ApJ...762..106A,2015ApJ...804...72B,2022ApJ...936L..15C,2022A&A...666A.156C}, the stacking analysis will be dramatically improved by the \textcolor{blue}{LASS} data. The 1-2 eV spectral resolution of \lem\, will allow an investigation of the emission lines of major ion species, such as O~VII, O~VIII, or Fe~XVII. By isolating these emission lines from the strong Milky Way foreground emission lines, the signal-to-noise ratios of the stacked sample will be significantly boosted. Given that the stacking analysis involves galaxies at varying redshifts (e.g., $z=0.01-0.05$), it is crucial to accurately determine the redshifts of galaxies to correct for their individual redshifts. This will be aided by redshift surveys, such as 2MASS Redshift Survey\cite{2012ApJS..199...26H} (see Figure~\ref{fig:lem_constrained_igm}). 

The stacking analysis, conducted with 1-2 eV \LEM~spectral resolution, will have several major advantages over previous studies:\\
(1) it will enable probing of the X-ray-emitting CGM to significantly larger radii ($R_{500}$) compared to earlier studies, \\
(2) it will allow exploring of the physical properties of galaxies in finer bins, thereby leading to a more thorough exploration of the parameter space, and \\
(3) it will allow deducing essential physical properties, like the temperature, of the CGM by using the ratio of O~VII to O~VIII. 

Overall, the results obtained in the stacking analysis will complement those obtained from the deep observations of individual objects.

%\subsection{IGM}

%------------------------
\begin{figure*}
\centering
\includegraphics[angle=0,trim=1cm 5cm 0cm 2cm,width=0.66\columnwidth]{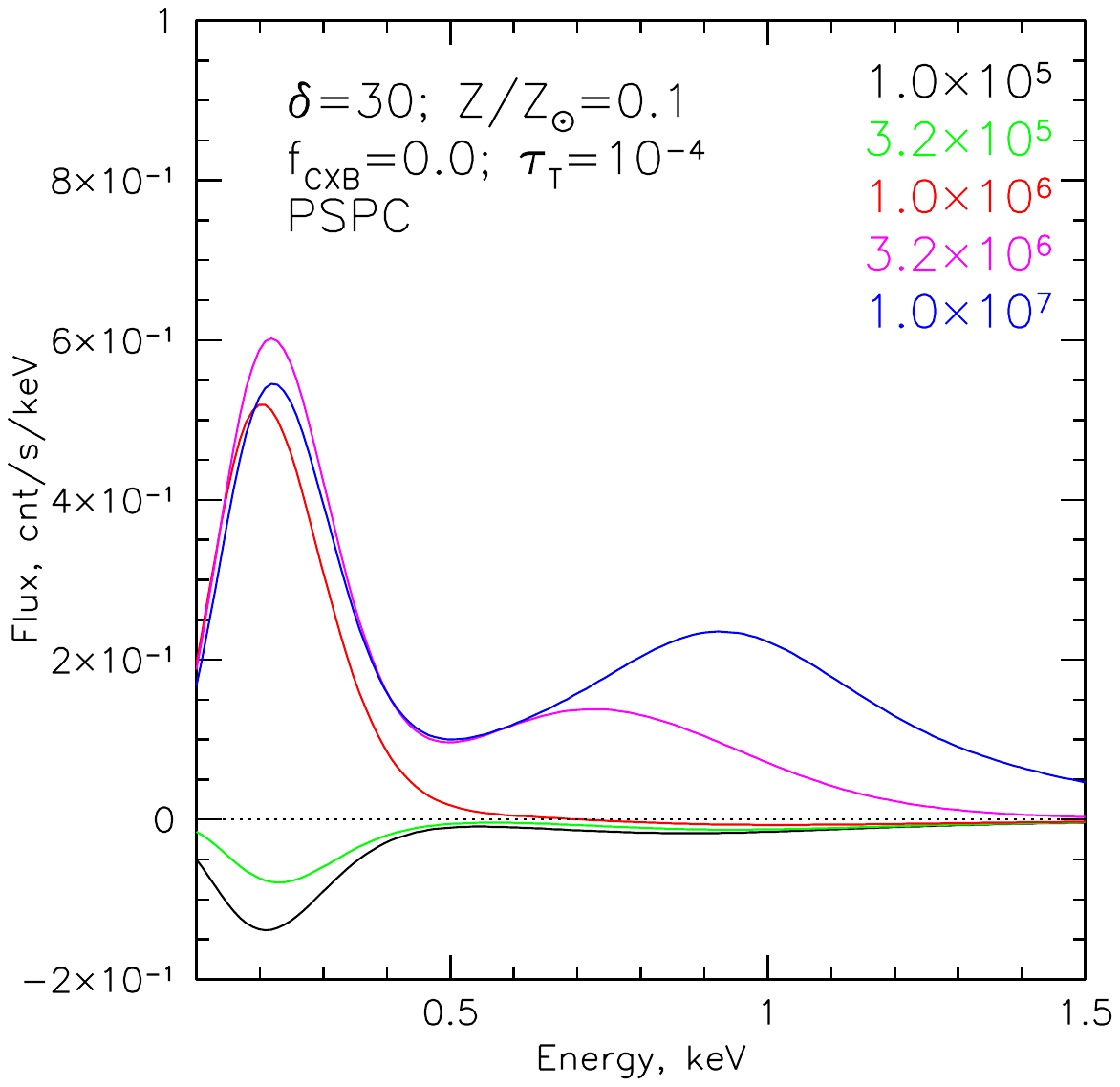}
\includegraphics[angle=0,trim=1cm 5cm 0cm 2cm,width=0.66\columnwidth]{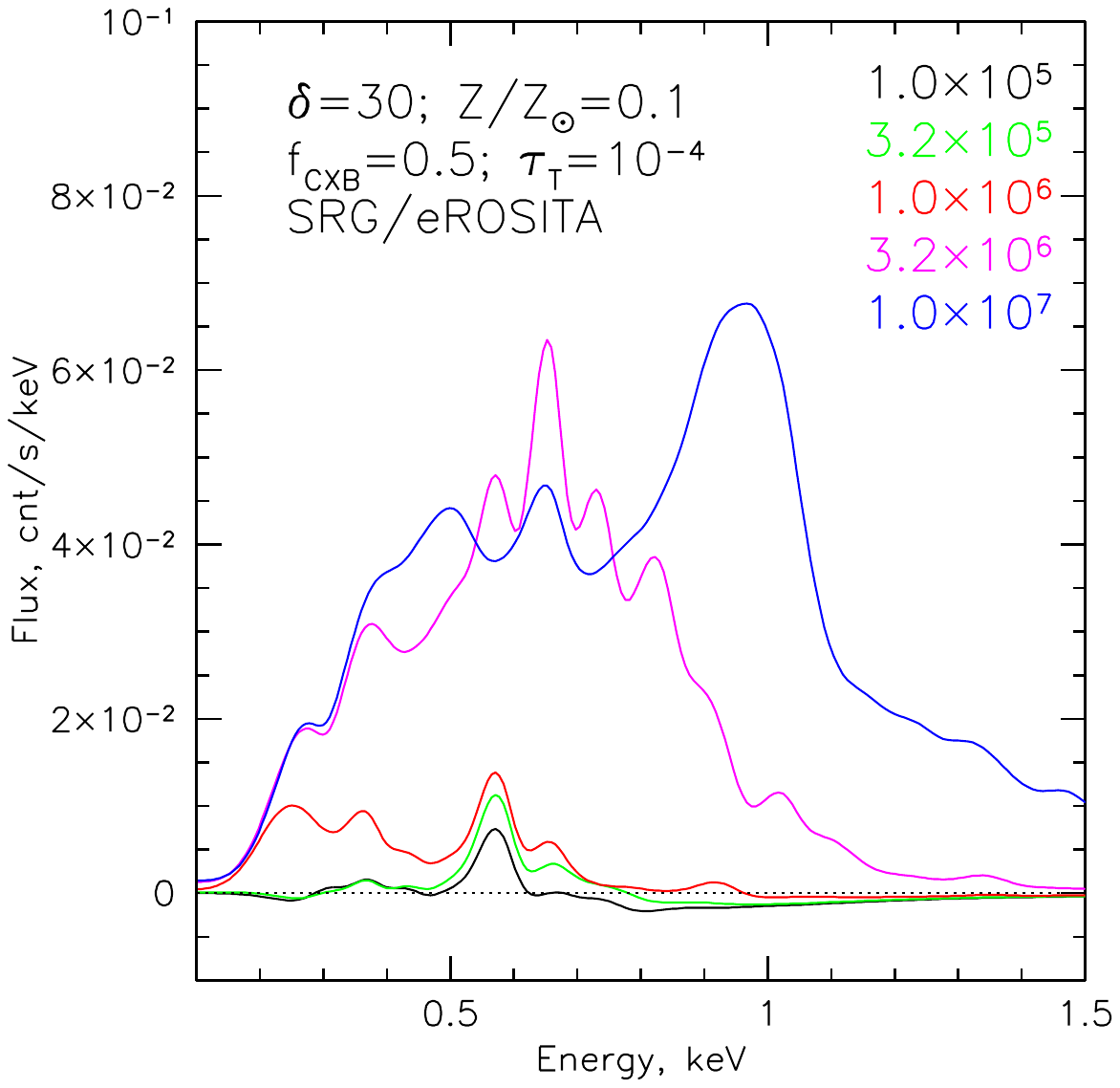}
\includegraphics[angle=0,trim=1cm 5cm 0cm 2cm,width=0.66\columnwidth]{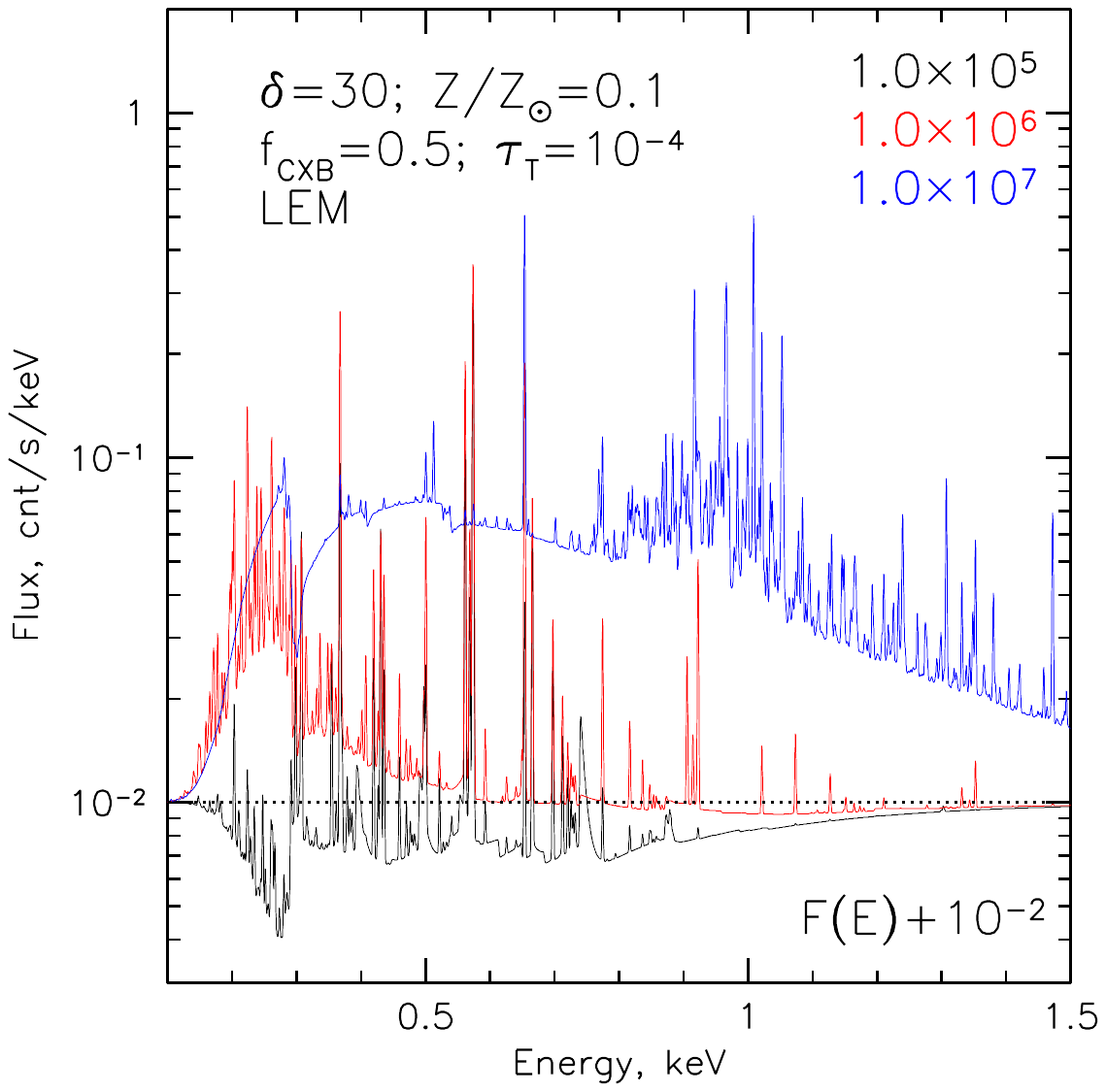}
\caption{\lem~prospects for Warm-Hot Intergalactic Medium detection. Simulated "on-filament" minus "off-filament" spectra for \textit{ROSAT}/PSPC, \textit{SRG/}eROSITA, and \lem\, missions for several values of the gas temperature in the range from $10^5$ to $10^7$~K. This is essentially a signature of a diffuse IGM imprinted in the observed spectra. The spectra have been convolved with the default responses of these three instruments, and it is assumed that the IGM subtends the entire FoV of each instrument.  %For the adopted overdensity, the transition between the two regimes (at low and high temperatures) is very clear. 
For \lem, a constant ($10^{-2}$) was added to the $T=10^5$~K spectrum (black line), to show the negative parts of the spectrum on the log scale of the y-axis. The resolved fraction of the Cosmic X-ray Background is set to $f_{\rm CXB}$=0 for \textit{ROSAT}/PSPC and $f_{\rm CXB}$=0.5 for \textit{SRG/}eROSITA and \lem. }  
\label{fig:igm_3inst}
\end{figure*}
%-------------------------

%------------------------
\begin{figure*}
\centering
\includegraphics[clip=true, angle=0,trim=1.5cm 8cm 1cm 7cm,width=1.0\textwidth]{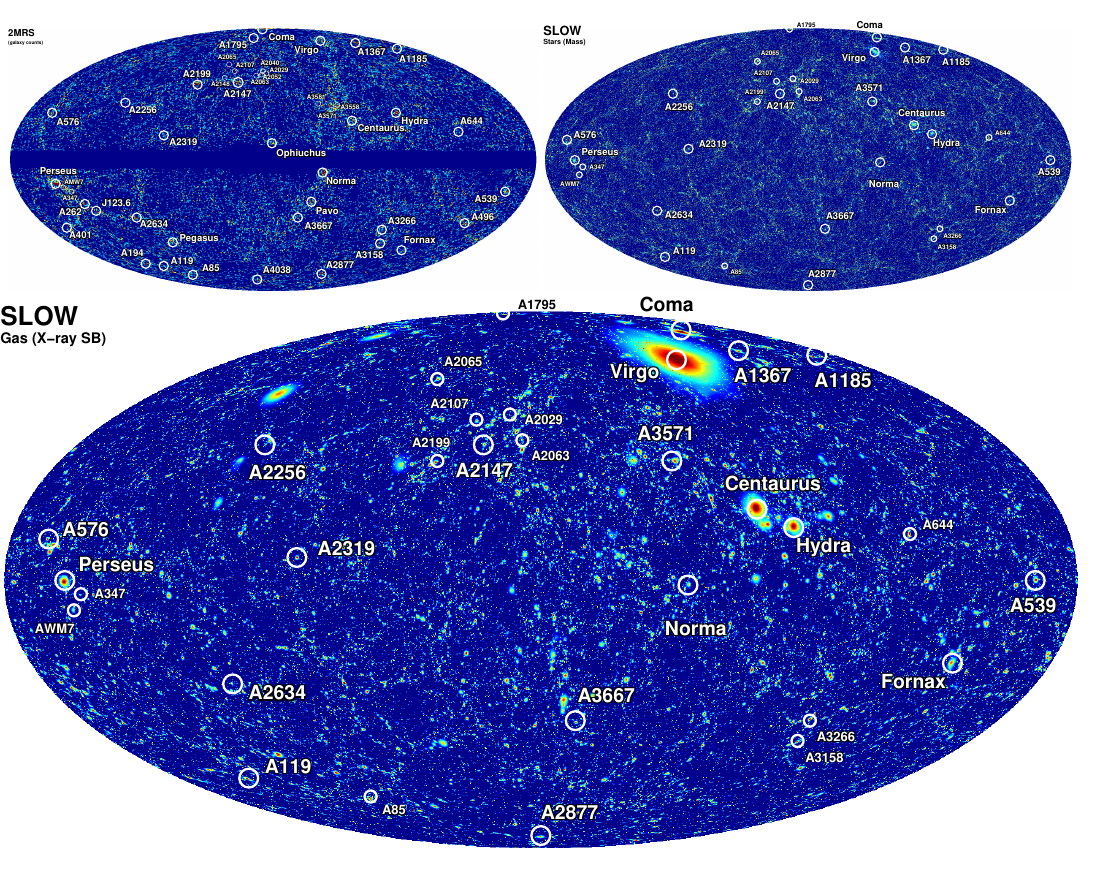}
\caption{Predicted all-sky map of soft X-ray emission (bottom panel) from all gas particles in the constrained cosmological hydrodynamical simulation SLOW \citep{2023A&A...677A.169D}, which is capable of reproducing the global galaxy distribution (top right panel) similar to that observed (e.g. in the 2MASS Redshift Survey\cite{2012ApJS..199...26H}, top left panel). Future spectroscopic surveys will provide us with galaxy density maps in the optimal ``redshift windows'' for WHIM detection, enabling the possiblity of global stacking to detect very faint signals.   
}
\label{fig:lem_constrained_igm}
\end{figure*}
%-------------------------

%-------------------------
\vspace*{-1mm}
\section{GALACTIC DIFFUSE OBJECTS}
\vspace*{-1mm}
%-------------------------

Soft X-ray emission in the Galactic plane is heavily absorbed by ubiquitous atomic gas and molecular clouds. Nonetheless, very bright and relatively nearby objects, located in the Galactic plane, will dominate soft X-ray maps as individual objects. These are supernova remnants and regions of massive star formation. Once again, compared to such objects in external galaxies, that might be less affected by the absorption, the Galactic objects can be studied with unprecedented detail. Many of these objects will benefit from deep targeted observations, while some others are very extended and bright enough to provide excellent statistics just from the \textcolor{blue}{LASS} data. This refers in particular to the brightest (e.g. Cygnus Loop\cite{2008PASJ...60..521M,2011ApJ...730...24K,2014ApJ...787L..31C,2015MNRAS.449.1340R} and Vela\cite{1995Natur.373..587A,2000A&A...362.1083L,2023A&A...676A..68M}) and largest (e.g. Monogem\cite{1996ApJ...463..224P}) supernova remnants, the Orion-Eridanus superbubble\cite{2023ApJ...943...61F,1993ApJ...406...97B}, and Cygnus X star-forming region\cite{1980ApJ...238L..71C,2001A&A...371..675U,2006A&A...458..855S}. For some objects, the \textcolor{blue}{LASS} data will serve as a pathfinder for deeper follow-up observations. The unique possibility of \lem~to single out the dominant diffuse line emission will offer an opportunity to search for soft non-thermal continuum emission for many of them.

\vspace*{-1mm}
\subsection{NORTH POLAR SPUR}
\vspace*{-1mm}

%------------------------
\begin{figure*}
\centering
\includegraphics[angle=0, trim=0cm 0cm 0cm 0cm, width=2.0\columnwidth]{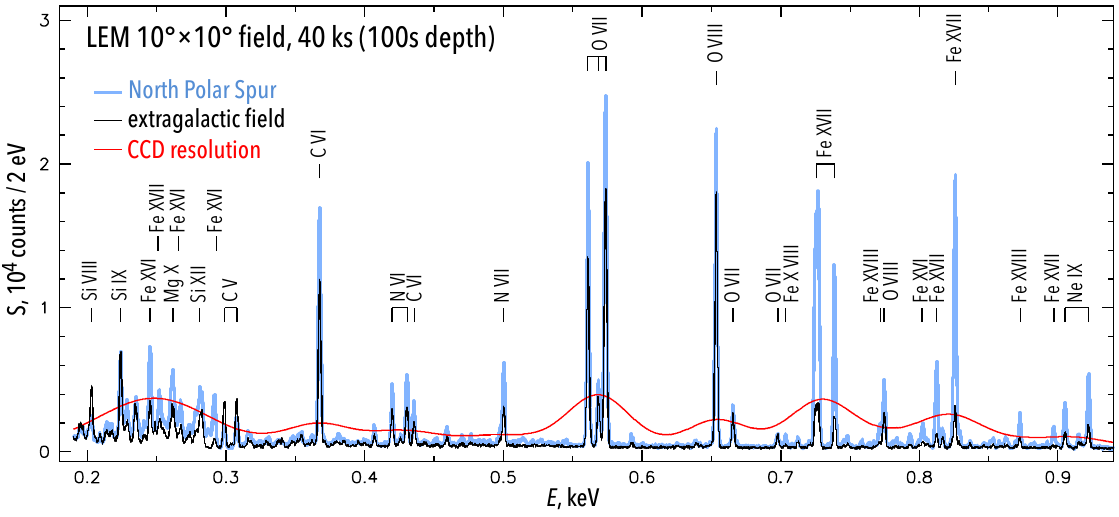}
\caption{Simulated example
spectra for $10^\circ\times10^\circ$ regions and 100s depth are shown for an average high-latitude
region outside of any bright features (“extragalactic field”), and another for the North Polar Spur (NPS) with inclusion of the cosmic X-ray background. Purely thermal CIE plasma models are assumed. Red curve shows the NPS spectrum at CCD
resolution (e.g., \textit{SRG}/eROSITA). \LEM~will resolve the forest of lines and gain access to line diagnostics for temperature, non-equilibrium and charge exchange processes.}
\label{fig:lass-nps}
\end{figure*}
%-------------------------

Although the North Polar Spur (NPS) is a very prominent feature of the X-ray and radio sky, its nature is still under debate\cite{2020ApJ...904...54L}, with the two main models connecting it either to a local superbubble, or to the Galactic Center structures related with the Fermi and eROSITA bubbles\cite{2022arXiv220301312L}.  Thanks to the brightness of its emission above 0.7 keV, \lem~ will be able not only to obtain high quality spectra of NPS (cf. Figure \ref{fig:lass-nps}), but also map its properties in great detail. This will reveal physical conditions within it, as well as the measurement of the absorbing column density (crucial for the distance measurements), and possible signatures of the hot gas interaction with the colder surrounding medium (e.g. via characteristic CX line ratios\cite{2016A&A...594A..78G}).

\begin{comment}
    
\vspace*{-1mm}
\subsection{GALACTIC CENTER}
\vspace*{-1mm}
Chimneys and roots of the Fermi/eROSITA bubbles.

\end{comment}

\vspace*{-1mm}
\subsection{FERMI BUBBLES}
\vspace*{-1mm}

After more than 10 years since their discovery\cite{2010ApJ...724.1044S}, the nature of the Fermi Bubbles remains unclear\cite{2018Galax...6...29Y,2022ApJ...927..225N}. Extending above and below the Galactic plane, the bubbles are aligned nearly symmetrically with respect to the Galactic Center and shine brightly in the GeV gamma-ray band. The gamma-rays are believed to be the inverse Compton emission produced by the population of cosmic ray electrons that also generates the synchrotron ``microwave haze'' observed by \textit{WMAP} and \textit{Planck}\cite{2014ApJ...793...64A,2022NatAs...6..584Y}. In some models, the most energetic of those cosmic ray electrons may generate detectable synchrotron emission in soft X-rays, with an intensity strongly dependent on the cosmic ray energy spectrum above the limit directly probed by \textit{Fermi} and on the magnetic field strength inside the bubbles\cite{2022MNRAS.510.5834O,2022MNRAS.516.1539O}. 
        
Thanks to its ability to resolve the bright emission lines from the ISM and CGM of the Milky Way that dominate the soft X-ray sky, \lem~ may be able to detect the X-ray continuum synchrotron emission from those cosmic ray electrons\cite{2013ApJ...779...57K}. \LEM~spectra, collected over the entire Fermi Bubbles in the course of the \textcolor{blue}{LASS}, will provide constraints on, or even positive detection of, this non-thermal emission, delivering essential insights into the energetics of the electron population and the magnetic field strength within the bubbles, and in particular in the direct vicinity of the bubbles ``working surface'' or the Galactic wind's termination shock\cite{2014MNRAS.444L..39L}.

\vspace*{-1mm}
\subsection{GALACTIC CENTER}
\vspace*{-1mm}
%text below is by Gabriele Ponti and IK

{

\begin{figure*}
    \centering
    \includegraphics[clip=true,width=1.8\columnwidth,trim=1cm 8.cm 1cm 6.6cm]{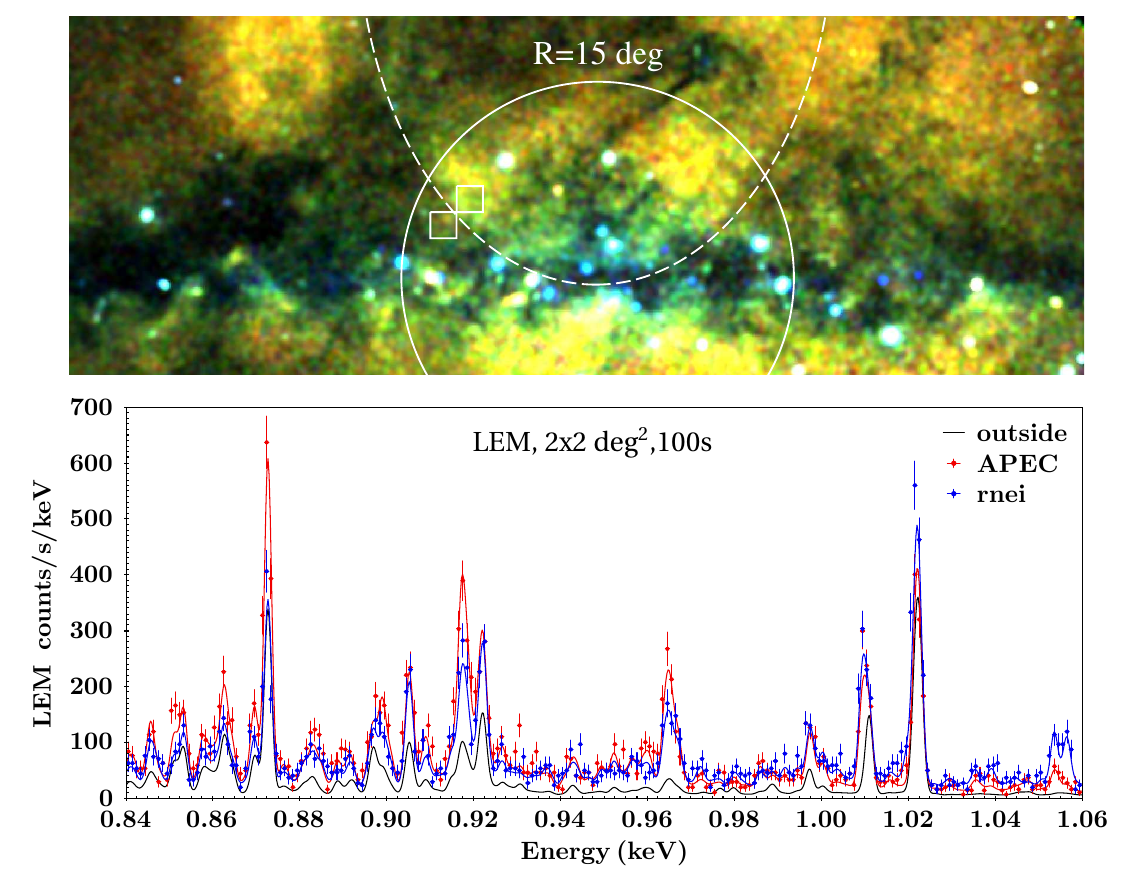}
    \caption{\textbf{Top panel:} \textit{ROSAT} all-sky RGB (R5-R6-R7\cite{1997ApJ...485..125S}) map of the central region of the Milky Way. The dashed ellipse highlights the approximate location of the X-ray ridges associated with the base of the \textit{Fermi} bubbles. The two white squares indicate the two regions of 2x2 deg$^2$ used to simulate the LASS spectra. 
    \textbf{Bottom panel:} \lem~ spectra of two regions at the bottom of the Galactic outflow, one inside and one outside the edges of the \textit{Fermi} bubbles. The red and blue data and model show the expected \LEM~spectra from the interior of the \textit{Fermi} bubbles in case of plasma in thermal equilibrium (APEC) or out of equilibrium (rnei), respectively. The black line shows the best model (obtained by fitting the ROSAT data\cite{1997ApJ...485..125S}) of the 2x2 deg$^2$ region outside the \textit{Fermi} bubbles. The detailed characterisation of the brighter lines will allow to constrain the kinematic of the outflow, while the detection of tens of faint lines will constrain its physical properties.}
    \label{fig:lem_center}
\end{figure*}

Deep scans of the Milky Way center have revealed chimney-like features originating from the central parsecs-to-tens-of-parsecs of the Milky Way, a region characterised by the presence of both Sgr~A* and a burst of star formation within the past few $10^6$ years. The chimneys are observed to rise about a hundred parsecs above and below the Galactic Center. The hot plasma associated with these features emits prominently in the soft X-ray band and the non-thermal component of the outflow is seen clearly in the radio band \cite{2019Natur.567..347P,2019Natur.573..235H,2021A&A...646A..66P,2022ApJ...925..165H}.

On larger scales, of about a kiloparsec, the \textit{ROSAT} soft X-ray images reveal an interesting morphology connecting the edges of the large scale Galactic outflow, traced by the edges of the Fermi bubbles, with the central few hundred parsecs of the Milky Way, which contains about 10\% of the Galactic star formation, the chimneys and Sgr A* at its center \cite{2010ApJ...724.1044S}(see the top panel of Figure~\ref{fig:lem_center} for \textit{ROSAT} soft X-ray image of this region). These edge-brightened features are thought to trace the shock heated plasma at the edges of the Galactic outflow \cite{2010ApJ...724.1044S,2015ApJ...808..107C}.

The all sky coverage of the LASS, will allow  measurements of the physical properties of the hot plasma at the base of the Galactic outflow, connecting the hundred parsec chimney-like features with the kiloparsec edge-brightened structures observed by \textit{ROSAT}. 

The key aspect provided by \textcolor{blue}{LASS} will be the separation of the individual emission lines produced by the outflowing hot plasma. 
By integrating the \textcolor{blue}{LASS} data on 2x2 square degree fields at the base of the Galactic outflow, it will be possible to characterise the basic physical parameter of the X-ray plasma, through the characterisation of tens of the brightest soft X-ray lines (bottom panel of Figure~\ref{fig:lem_center}). 
While eRASS data will allow measuring intensities and ratios of the brightest lines, \textcolor{blue}{LASS} will enable measurement of their shapes and centroid positions, as well as to give access to numerous weaker lines. 
The former will uncover the kinematic structure of the outflowing gas (to be compared with kinematic structures of the colder gas \cite{2017ApJ...834..191B,2018ApJ...855...33D}), the latter will reveal whether the plasma is in thermal equilibrium, establish its temperature, relative abundances of elements, and, possibly, constrain the presence of any non-thermal component. 

These measurements will be critical for establishing how the Galactic Center outflows are plugged into the Galactic disc and to shed light on the primary mechanism powering them: are these structures powered by either current or past jet-like outflow from Sgr~A*?; by winds from Sgr A*’s accretion disc?; or alternatively are they inflated by either hot plasma pressure or cosmic rays from star formation events \cite{2008ApJ...674..258E}? }

%\vspace*{-1mm}
%\subsection{STAR-FORMING REGIONS}
%\vspace*{-1mm}

%E.g. Cygnus X region. 

\vspace*{-1mm}
\section{SERENDIPITOUS SCIENCE}
\vspace*{-1mm}

Observing large sky areas allows one to detect rare and bright transient sources, with the discovery volume for the telescope being proportional to its grasp, making \lem\, a very important transient detection facility as well. In particular, this is relevant for X-ray transients with soft X-ray spectra, e.g. Tidal Disruption Events (TDE) and supernova shock break-outs. While for the former, the typical variability timescale is from months to years, the latter last for just seconds or hours. Given that the \textcolor{blue}{LASS} will be accumulated in several chunks separated by half of a year, there is a possibility to look for TDE-like variability in a very efficient manner, similar to \textit{SRG}/eROSITA\cite{2014MNRAS.437..327K}. 

In fact, the search for transients in the \textcolor{blue}{LASS} will benefit strongly from the upper limits on X-ray flux at any position provided by the \textit{SRG}/eROSITA all-sky survey\cite{2012arXiv1209.3114M,2021A&A...647A...1P}. This upper limit will be at least an order of magnitude below the detection sensitivity for \lem\ in one 10s scan, meaning that transient finding will be primarily determined by the latter. Thanks to \lem's spectral resolution, the X-ray background for a soft spectrum source, dominated by continuum emission (e,g., in the simplest scenarios, both TDEs and supernova shock break-outs are expected to have black body spectra) can be strongly diminished by excluding narrow spectral bands contaning Galactic emission lines. Thus, Poisson statistics will determine source detection significance. 

As an example following\cite{2014MNRAS.437..327K}, in Figure \ref{fig:lem-tde} we show predicted \lem\, count rates for TDE-like sources with a blackbody spectrum and peak bolometric luminosity defined as a function of the disrupting black hole (the former follows from the standard Shakura-Sunyaev prescription, while the latter is fixed at the Eddington limit, assuming that in the early and brightest phase the accretion of the fallback material is Eddington-limited) and redshift of the source. Similarly to the \textit{SRG}/eROSITA case \cite{2014MNRAS.437..327K}, for  \lem\  there is also only weak dependence of the predicted count rate on the black hole mass in the range from $10^6$ to $10^{7}$ M$_\odot$ for $z<0.1$.

%
%\begin{comment}
    
\begin{figure}
    \centering
    \includegraphics[width=\columnwidth]{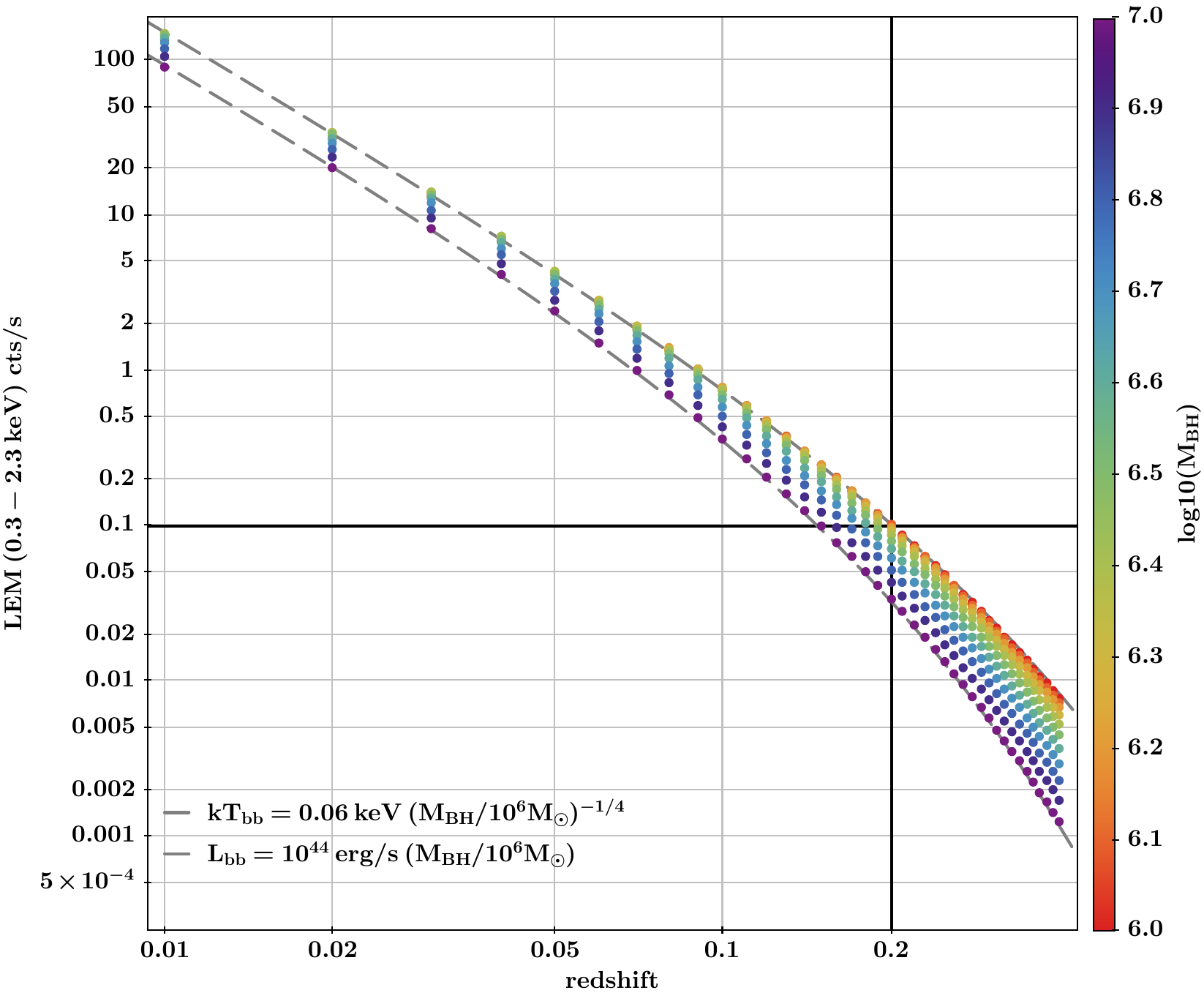}
    \caption{Predicted \lem\, count rate for sources with TDE-like blackbody spectra and bolometric luminosity corresponding to the black hole mass indicated by the color as a function of redshift\cite{2014MNRAS.437..327K}.}
    \label{fig:lem-tde}
\end{figure}
%
%\end{comment}

\vspace*{-1mm}
\section{CONCLUSIONS}
\vspace*{-1mm}

The \textcolor{blue}{Line Emission Mapper All Sky Survey} will fundamentally transform our understanding of the Milky Way at all scales and beyond, from physical processes within the solar system, to gas kinematics in the Circumgalactic Medium. 

The richness of high spectral resolution data will shed new light on the structure and evolution of our Galaxy and the interaction between disk and halo. This includes how the cold star forming disk is fed by the hot, accreted gas in the CGM, and how feedback mechanisms may feed energy back into the CGM in the form of large scale perturbations such as the Fermi and eROSITA bubbles. 

The \textcolor{blue}{LASS} will be highly complementary both to previous X-ray surveys such as ROSAT and eROSITA, by adding a new dimension with its high-resolution spectra. Combined with source population surveys such as Gaia, Euclid, 4MOST, and Rubin, the \textcolor{blue}{LASS} will be able to investigate the interaction between hot gas in superbubbles and the cooler disk surrounding them, and will allow stacking studies of classes of sources too dim to be studied individually.

These are just a few examples of the groundbreaking science provided by the \textcolor{blue}{LASS}. Like its predecessors ROSAT and eROSITA, the science return of the \textcolor{blue}{LASS} will span almost all branches of high energy astrophysics.The new parameter space to be opened by \textcolor{blue}{LASS} will lead to new discoveries and new phenomena that we do not even know exist, and will provide a unique legacy for many years to come: no other mission currently approved or on the drawing board has the combination of energy resolution and grasp to get even close to cover the whole sky at the high energy resolution provided by the \textcolor{blue}{LASS} (while maintaining good angular resolution and imaging capabilities).

%\clearpage
\vspace*{-1mm}
\section{APPENDIX A. SELECTED LINE DIAGNOSTICS}
\vspace*{-1mm}
\label{sec:appendix}
{
%\color{red} 
Analyzing \LEM~data will require updating our methods of analysing and interpreting results. Current plasma models, while quite advanced, are incomplete and the data for many lines will not match the \LEM~spectra. Line data will need to be updated to keep pace with all the lines that will be resolved by \LEM. In the meantime, our analysis will begin with the isolated lines with the best atomic data.

For example, for LHB analysis, for each direction in the sky, all isolated lines for all of the accessible ionization states for an element can be identified. The ratio of lines from the same ionization state can be used to determine the temperature with large uncertainty, but using lines from multiple ionization states produces very small uncertainties. Comparison of the state Q with Q-1 may provide a different ratio than a comparison with Q+1, and different species may indicate different temperatures, all of which provides clues to the extent to which the plasma is out of equilibrium
and can be used to determine the ionization state, an important indicator of the past thermal history, including the possibility of a past photoionization event.  Ratios of lines from different elements in the same component give relative abundances, which can be used, for example, to track the origin of the plasma. Based on the abundance of resolved individual lines in \LEM~spectra (Fig.~\ref{fig:srge_fermi_planck}), this will allow a detailed analysis of LHB nature and structure.

As an example, Figures~\ref{fig:C_T} and \ref{fig:C_err} show the dependence of major Carbon line ratios on temperature (including uncertainty) and the associated uncertainty on the inferred temperature. The plots were generated assuming the \LEM~response, an observing time of 100 s, and a field of view of $10\times10~\rm deg^2$. Under these conditions, the temperature is expected to be recovered with an accuracy of better than 10\%. The major lines associated with C emission are reported in Table~\ref{tab:c}.

\begin{figure}
    \centering
    \includegraphics[clip=true,trim=0cm 0.3cm 0cm 0cm,width=\columnwidth]{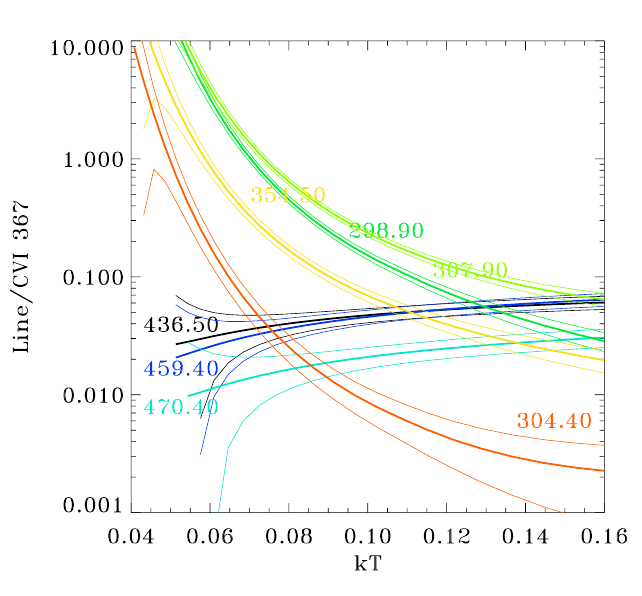}
    \caption{The ratio of various C lines to the C VI 367~eV line. For each emission line the thick line is the ratio, the thin lines are the uncertainties in the ratio.}
    \label{fig:C_T}
\end{figure}

\begin{figure}
    \centering
    \includegraphics[width=\columnwidth]{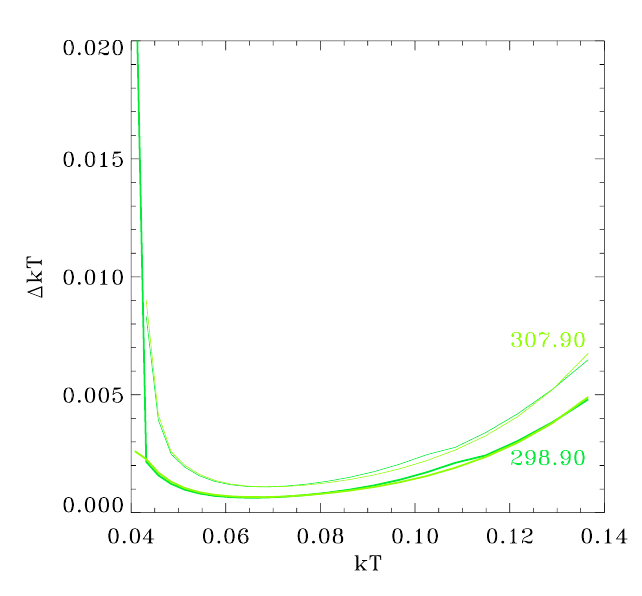}
    \caption{The uncertainty in $kT$ as a function of $kT$ for several different line ratios. The thick lines are the lower $kT$ uncertainties, while the thin lines show the upper uncertainties. This plot needs to be cleaned up.}
    \label{fig:C_err}
\end{figure}

\begin{table}[]
    \caption{Major Carbon lines detectable with \textcolor{blue}{LASS}}
    \centering
    \begin{tabular}{ | c | c | c | }
    \hline
    Ion & Energy & \LEM~Bin \\
    State & (eV) & Number \\
    \hline
    VI & 367.47 \& 367.53 & 3674 \\
    VI & 435.56 \& 435.54 & 4354 \& 4355 \\
    VI & 459.37 & 4593 \\
    VI & 470.40& 4703 \\
    V & 298.96 & 2989 \\
    V & 307.90 & 3078 \\
    V & 354.52 & 3544 \\
    V & 304.40 & 3043 \\
    \hline
    \end{tabular}
    \label{tab:c}
\end{table}
}

The most important temperature diagnostics for hotter gas, above 0.1 keV, come from lines of He-like and H-like oxygen and neon, and L-complex lines of iron (cf. top panel in Figure~\ref{fig:cgm_lines}).  The steep sensitivity of the line ratios to the resonant component of the O~VII triplet (cf. bottom panel in Figure~\ref{fig:cgm_lines}) enables very precise temperature diagnostics in the single temperature case. However, in real situations this is very rarely the case, so sets of line ratios need to be considered, ideally with orthogonal dependence on competing effects (e.g. comparison of the lines of the same ionization state to the ratio of ionization state sensitive ratios reveal deviations from simple CIE case). For the CGM temperature structure analysis, one will have to perform forward modelling of the emission spectra, guided by the more complex mixture models arising from numerical simulations.

%------------------------
\begin{figure*}
\centering
%\hspace{4cm}
\includegraphics[clip=true,trim=0cm 7cm 0.5cm 7cm,angle=0, width=1.15\columnwidth]{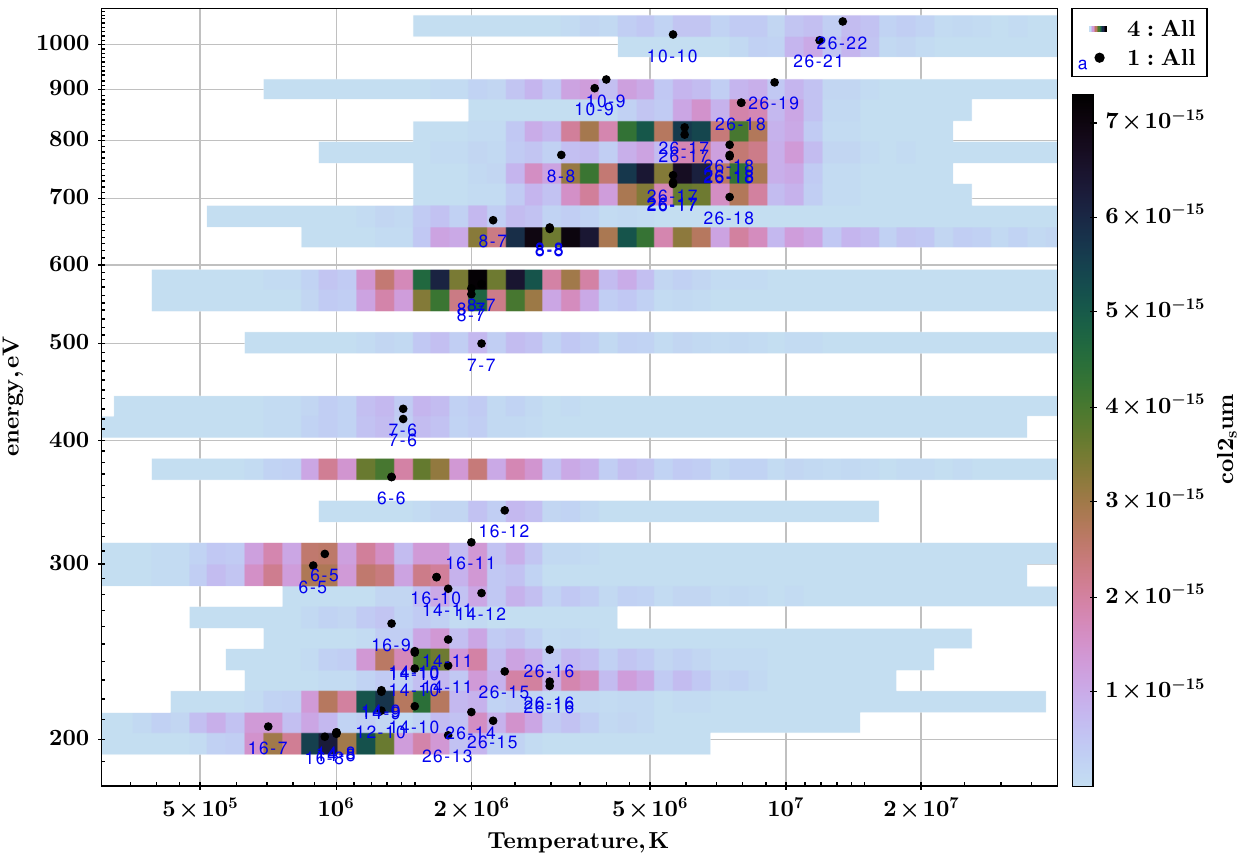}
\includegraphics[clip=true,trim=0cm 0cm 7.1cm 0cm,angle=0, width=0.85\columnwidth]{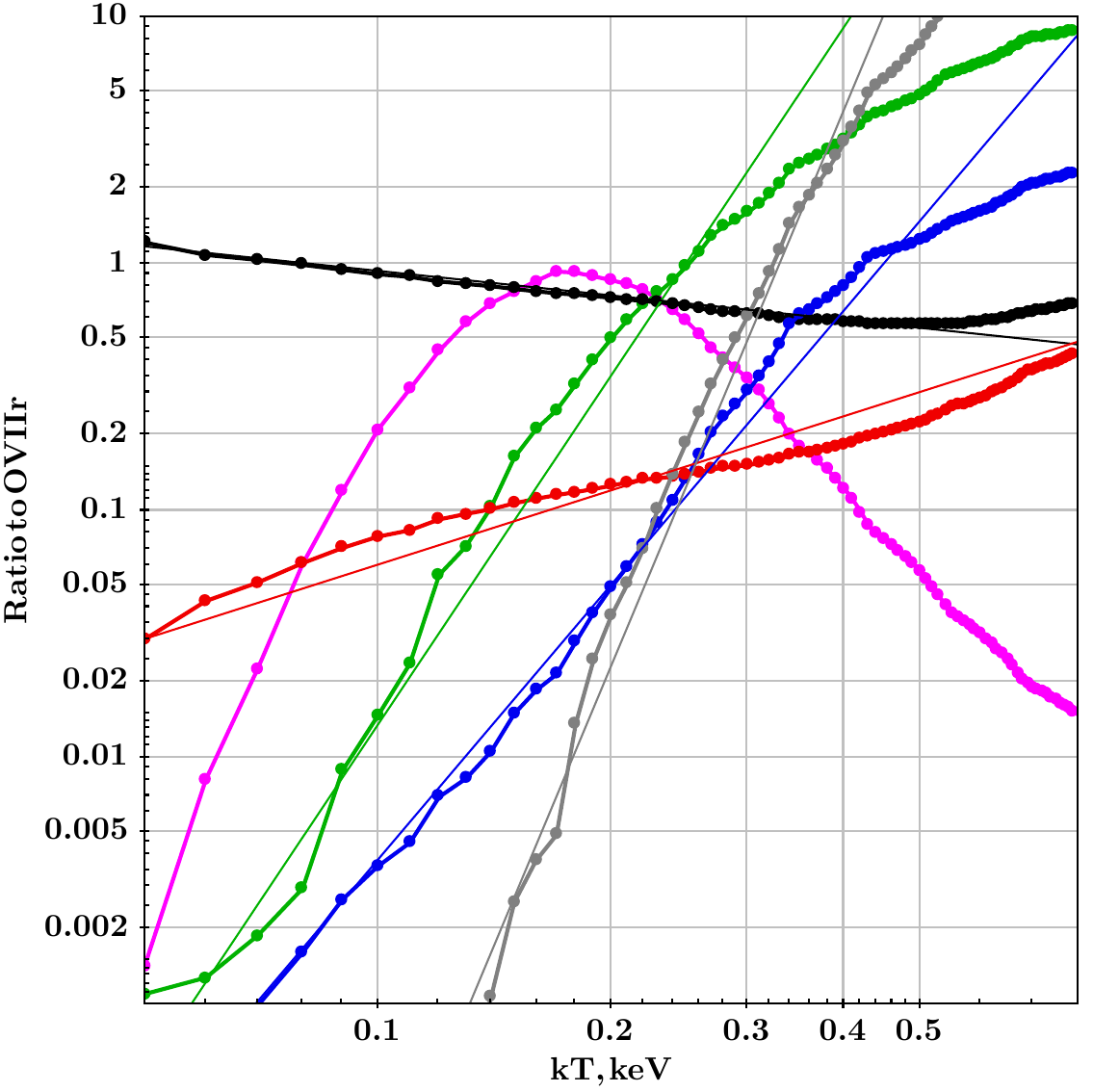}
\caption{\textbf{Left panel:} Emissivity (color-coded) in narrow spectral bins for CIE plasma (APEC) with Solar abundance of metals as a function of gas temperature and spectral bin energy. The dots with blue numbers label central energy and the peak emissivity temperature for the most prominent lines with the first number giving element's nuclear charge (e.g. 8 for Oxygen) and the second number giving spectroscopic symbol (e.g. 7 for O~VII). The behaviour of the peak temperature as a function of bin energy is non-monotonic.
\textbf{Right panel:} Ratios of line emissivity to emissivity in resonant line of O~VII(r) as a function of gas temperature in APEC CIE model: O~VII(f)-black, O~VII~K$\beta$ - red, O~VIII - green, Ne~IX (resonant) - blue, Fe~XVII (826 eV) - grey. Magenta line shows temperature dependence of O~VII(r) line emissivity normalised to the peak one. The straight lines show powerlaw fits to these ratios, with slopes equal to -0.33, 4.7, 3.7, 1.0, and 7.5 for O~VII(f), O~VIII, Ne~IX, O~VII~K$\beta$, and Fe~XVII(826 eV), respectively.     
}
\label{fig:cgm_lines}
\end{figure*}
%-------------------------

%\clearpage
\vspace*{-1mm}
\section{ACKNOWLEDGEMENTS}
\vspace*{-1mm}

IK acknowledges support by the COMPLEX project from the European Research Council (ERC) under the European Union’s Horizon 2020 research and innovation program grant agreement ERC-2019-AdG 882679.
WF acknowledges support from the Smithsonian Institution, the Chandra High Resolution Camera Project through NASA contract NAS8-03060, and NASA Grants 80NSSC19K0116, GO1-22132X, and GO9-20109X. The material is based upon work supported by NASA under award number 80GSFC21M0002.

%\section*{REFERENCES}
\clearpage
%%%%%%%%%%%%%%%%%%%%%%% REFERENCES:
\small


\vspace{-0mm}
\parindent=0cm
\baselineskip=11.5pt

\vspace*{-1mm}
\section{REFERENCES}
\vspace*{-10mm}

\begin{thebibliography}{200}
%ZEncoding:latex%ZLinelength:0\bibitem[%4m(%Y)]{%R} %5.3l\ %Y, %j, %V, %p. \url{https://doi.org/%d} 
\bibitem[Ackermann et al.(2014)]{2014ApJ...793...64A} Ackermann, M., Albert, A., Atwood, W.~B., et al.\ 2014, \apj, 793, 64. \url{https://doi.org/10.1088/0004-637X/793/1/64} 

\bibitem[Anderson et al.(2013)]{2013ApJ...762..106A} Anderson, M. E., Bregman, J. N., Dai, X., et al.\ 2013, \apj, 762, 106. \url{https://doi.org/10.1088/0004-637X/762/2/106}

\bibitem[Aschenbach, Egger, \& Tr{\"u}mper(1995)]{1995Natur.373..587A} Aschenbach, B., Egger, R., \& Tr{\"u}mper, J.\ 1995, \nat, 373, 587. \url{https://doi.org/10.1038/373587a0} 

\bibitem[Ayres(2008)]{2008AJ....136.1810A} Ayres, T.~R.\ 2008, \aj, 136, 1810. \url{https://doi.org/10.1088/0004-6256/136/5/1810} 

\bibitem[Berger et al.(2010)]{2010ApJ...709..332B} Berger, E., Basri, G., Fleming, T.~A., et al.\ 2010, \apj, 709, 332. \url{https://doi.org/10.1088/0004-637X/709/1/332} 

\bibitem[Biffi, Dolag, B{\"o}hringer, \& Lemson(2012)]{2012MNRAS.420.3545B} Biffi, V., Dolag, K., B{\"o}hringer, H., \& Lemson, G.\ 2012, \mnras, 420, 3545. \url{https://doi.org/10.1111/j.1365-2966.2011.20278.x} 

\bibitem[Bluem et al.(2022)]{2022ApJ...936...72B} Bluem, J., Kaaret, P., Kuntz, K.~D., et al.\ 2022, \apj, 936, 72. \url{https://doi.org/10.3847/1538-4357/ac8662} 

\bibitem[Bogdan et al.(2015)]{2015ApJ...804...72B} Bogdan, A., Vogelsberger, M., Kraft, R. P., et al.\ 2015, \apj, 804, 72. \url{https://doi.org/10.1088/0004-637X/804/1/72}

\bibitem[Bogd{\'a}n et al.(2023)]{2023ApJ...953...42B} Bogd{\'a}n, {\'A}., Khabibullin, I., Kov{\'a}cs, O.~E., et al.\ 2023, \apj, 953, 42. \url{https://doi.org/10.3847/1538-4357/acdeec} 

\bibitem[Bordoloi et al.(2017)]{2017ApJ...834..191B} Bordoloi, R., Fox, A.~J., Lockman, F.~J., et al.\ 2017, \apj, 834, 191. \url{https://doi.org/10.3847/1538-4357/834/2/191} 

\bibitem[Brandt et al.(2023)]{2023SSRv..219...18B} Brandt, P.~C., Provornikova, E., Bale, S.~D., et al.\ 2023, \ssr, 219, 18. \url{https://doi.org/10.1007/s11214-022-00943-x}

\bibitem[Bregman et al.(2018)]{2018ApJ...862....3B} Bregman, J.~N., Anderson, M.~E., Miller, M.~J., et al.\ 2018, \apj, 862, 3. \url{https://doi.org/10.3847/1538-4357/aacafe} 



\bibitem[Brunner et al.(2022)]{2022A&A...661A...1B} Brunner, H., Liu, T., Lamer, G., et al.\ 2022, \aap, 661, A1. \url{https://doi.org/10.1051/0004-6361/202141266} 



\bibitem[Burrows et al.(1993)]{1993ApJ...406...97B} Burrows, D.~N., Singh, K.~P., Nousek, J.~A., Garmire, G.~P., \& Good, J.\ 1993, \apj, 406, 97. \url{https://doi.org/10.1086/172423} 



\bibitem[Casey et al.(2023)]{2023ApJ...954...31C} Casey, C.~M., Kartaltepe, J.~S., Drakos, N.~E., et al.\ 2023, \apj, 954, 31. \url{https://doi.org/10.3847/1538-4357/acc2bc}

\bibitem[Cash et al.(1980)]{1980ApJ...238L..71C} Cash, W., Charles, P., Bowyer, S., et al.\ 1980, \apjl, 238, L71. \url{https://doi.org/10.1086/183261} 



\bibitem[Cassinelli et al.(2001)]{2001ApJ...554L..55C} Cassinelli, J.~P., Miller, N.~A., Waldron, W.~L., MacFarlane, J.~J., \& Cohen, D.~H.\ 2001, \apjl, 554, L55. \url{https://doi.org/10.1086/320916} 

\bibitem[Chadaymmuri et al.(2023)]{2022ApJ...936L..15C}Chadayammuri, U., Bogdan, A., Oppenheimer, B., et al.\ 2023, \apjl, 936, 1. \url{https://doi.org/10.3847/2041-8213/ac8936}

\bibitem[Cohen, Cassinelli, \& MacFarlane(1997)]{1997ApJ...487..867C} Cohen, D.~H., Cassinelli, J.~P., \& MacFarlane, J.~J.\ 1997, \apj, 487, 867. \url{https://doi.org/10.1086/304636} 


\bibitem[Comparat et al.(2022)]{2022A&A...666A.156C} Comparat, J., Truong, N., Merloni, A., et al.\ 2022, \aap, 666, A156. \url{https://doi.org/10.1051/0004-6361/202243101}
\bibitem[Comparat et al.(2023)]{2023A&A...673A.122C} Comparat, J., Luo, W., Merloni, A., et al.\ 2023, \aap, 673, A122. \url{https://doi.org/10.1051/0004-6361/202245726}
\bibitem[Corral et al.(2011)]{2011A&A...530A..42C} Corral, A., Della Ceca, R., Caccianiga, A., et al.\ 2011, \aap, 530, A42. \url{https://doi.org/10.1051/0004-6361/201015227}

\bibitem[Crocker, Bicknell, Taylor, \& Carretti(2015)]{2015ApJ...808..107C} Crocker, R.~M., Bicknell, G.~V., Taylor, A.~M., \& Carretti, E.\ 2015, \apj, 808, 107. \url{https://doi.org/10.1088/0004-637X/808/2/107} 

\bibitem[Cumbee et al.(2014)]{2014ApJ...787L..31C} Cumbee, R.~S., Henley, D.~B., Stancil, P.~C., et al.\ 2014, \apjl, 787, L31. \url{https://doi.org/10.1088/2041-8205/787/2/L31} 

\bibitem[Di Mascolo et al.(2023)]{2023Natur.615..809D} Di Mascolo, L., Saro, A., Mroczkowski, T., et al.\ 2023, \nat, 615, 809. \url{https://doi.org/10.1038/s41586-023-05761-x}

\bibitem[Di Teodoro et al.(2018)]{2018ApJ...855...33D} Di Teodoro, E.~M., McClure-Griffiths, N.~M., Lockman, F.~J., et al.\ 2018, \apj, 855, 33. \url{https://doi.org/10.3847/1538-4357/aaad6a} 



\bibitem[Dolag et al.(2023)]{2023A&A...677A.169D} Dolag, K., Sorce, J.~G., Pilipenko, S., et al.\ 2023, \aap, 677, A169. \url{https://doi.org/10.1051/0004-6361/202346213}
\bibitem[Dolag, Komatsu, \& Sunyaev(2016)]{2016MNRAS.463.1797D} Dolag, K., Komatsu, E., \& Sunyaev, R.\ 2016, \mnras, 463, 1797. \url{https://doi.org/10.1093/mnras/stw2035}

\bibitem[Drake \& Stelzer(2023)]{2023hxga.book..132D} Drake, J.~J., \& Stelzer, B.\ 2023, Handbook of X-ray and Gamma-ray Astrophysics. Edited by Cosimo Bambi and Andrea Santangelo, 132. \url{https://doi.org/10.1007/978-981-16-4544-0\_78-1} 

\bibitem[Everett et al.(2008)]{2008ApJ...674..258E} Everett, J.~E., Zweibel, E.~G., Benjamin, R.~A., et al.\ 2008, \apj, 674, 258. \url{https://doi.org/10.1086/524766} 



\bibitem[Euclid Collaboration et al.(2019)]{2019A&A...631A..85E} Euclid Collaboration, Barnett, R., Warren, S.~J., et al.\ 2019, \aap, 631, A85. \url{https://doi.org/10.1051/0004-6361/201936427}

\bibitem[Faucher-Gigu{\`e}re \& Oh(2023)]{2023ARA&A..61..131F} Faucher-Gigu{\`e}re, C.-A., \& Oh, S.~P.\ 2023, \araa, 61, 131. \url{https://doi.org/10.1146/annurev-astro-052920-125203} 



\bibitem[Fleming \& Tagliaferri(1996)]{1996ApJ...472L.101F} Fleming, T.~A., \& Tagliaferri, G.\ 1996, \apjl, 472, L101. \url{https://doi.org/10.1086/310361} 

\bibitem[Fuller et al.(2023)]{2023ApJ...943...61F} Fuller, C.~A., Kaaret, P., Bluem, J., et al.\ 2023, \apj, 943, 61. \url{https://doi.org/10.3847/1538-4357/acaafc} 


\bibitem[Galeazzi et al.(2014)]{2014Natur.512..171G} Galeazzi, M., Chiao, M., Collier, M.~R., et al.\ 2014, \nat, 512, 171. \url{https://doi.org/10.1038/nature13525}

\bibitem[Gu, Mao, Costantini, \& Kaastra(2016)]{2016A&A...594A..78G} Gu, L., Mao, J., Costantini, E., \& Kaastra, J.\ 2016, \aap, 594, A78. \url{https://doi.org/10.1051/0004-6361/201628609} 



\bibitem[Gupta et al.(2012)]{2012ApJ...756L...8G} Gupta, A., Mathur, S., Krongold, Y., Nicastro, F., \& Galeazzi, M.\ 2012, \apjl, 756, L8. \url{https://doi.org/10.1088/2041-8205/756/1/L8} 

\bibitem[Gupta et al.(2023)]{2023NatAs...7..799G} Gupta, A., Mathur, S., Kingsbury, J., Das, S., \& Krongold, Y.\ 2023, Nature Astronomy, 7, 799. \url{https://doi.org/10.1038/s41550-023-01963-5} 





\bibitem[G{\"u}nther et al.(2012)]{2012ApJ...750...78G} G{\"u}nther, H.~M., Wolk, S.~J., Drake, J.~J., et al.\ 2012, \apj, 750, 78. \url{https://doi.org/10.1088/0004-637X/750/1/78} 



\bibitem[Hafen et al.(2022)]{2022MNRAS.514.5056H} Hafen, Z., Stern, J., Bullock, J., et al.\ 2022, \mnras, 514, 5056. \url{https://doi.org/10.1093/mnras/stac1603} 

\bibitem[Heitsch \& Putman(2009)]{2009ApJ...698.1485H} Heitsch, F., \& Putman, M.~E.\ 2009, \apj, 698, 1485. \url{https://doi.org/10.1088/0004-637X/698/2/1485} 


\bibitem[Henley \& Shelton(2013)]{2013ApJ...773...92H} Henley, D.~B., \& Shelton, R.~L.\ 2013, \apj, 773, 92. \url{https://doi.org/10.1088/0004-637X/773/2/92}

\bibitem[Henley, Shelton, \& Kwak(2014)]{2014ApJ...791...41H} Henley, D.~B., Shelton, R.~L., \& Kwak, K.\ 2014, \apj, 791, 41. \url{https://doi.org/10.1088/0004-637X/791/1/41} 



\bibitem[Heywood et al.(2019)]{2019Natur.573..235H} Heywood, I., Camilo, F., Cotton, W.~D., et al.\ 2019, \nat, 573, 235. \url{https://doi.org/10.1038/s41586-019-1532-5} 

\bibitem[Heywood et al.(2022)]{2022ApJ...925..165H} Heywood, I., Rammala, I., Camilo, F., et al.\ 2022, \apj, 925, 165. \url{https://doi.org/10.3847/1538-4357/ac449a} 

\bibitem[Hodges-Kluck, Miller, \& Bregman(2016)]{2016ApJ...822...21H} Hodges-Kluck, E.~J., Miller, M.~J., \& Bregman, J.~N.\ 2016, \apj, 822, 21. \url{https://doi.org/10.3847/0004-637X/822/1/21} 




\bibitem[Huchra et al.(2012)]{2012ApJS..199...26H} Huchra, J.~P., Macri, L.~M., Masters, K.~L., et al.\ 2012, \apjs, 199, 26. \url{https://doi.org/10.1088/0067-0049/199/2/26} 



\bibitem[Iwasawa et al.(2012)]{2012A&A...537A..86I} Iwasawa, K., Mainieri, V., Brusa, M., et al.\ 2012, \aap, 537, A86. \url{https://doi.org/10.1051/0004-6361/201118203}
\bibitem[Izmodenov \& Alexashov(2015)]{2015ApJS..220...32I} Izmodenov, V.~V. \& Alexashov, D.~B.\ 2015, \apjs, 220, 32. \url{https://doi.org/10.1088/0067-0049/220/2/32}

\bibitem[Kaaret et al.(2020)]{2020NatAs...4.1072K} Kaaret, P., Koutroumpa, D., Kuntz, K.~D., et al.\ 2020, Nature Astronomy, 4, 1072. \url{https://doi.org/10.1038/s41550-020-01215-w} 



\bibitem[Kahn et al.(2001)]{2001A&A...365L.312K} Kahn, S.~M., Leutenegger, M.~A., Cottam, J., et al.\ 2001, \aap, 365, L312. \url{https://doi.org/10.1051/0004-6361:20000093} 

\bibitem[Kataoka et al.(2013)]{2013ApJ...779...57K} Kataoka, J., Tahara, M., Totani, T., et al.\ 2013, \apj, 779, 57. \url{https://doi.org/10.1088/0004-637X/779/1/57} 

\bibitem[Katsuda et al.(2011)]{2011ApJ...730...24K} Katsuda, S., Tsunemi, H., Mori, K., et al.\ 2011, \apj, 730, 24. \url{https://doi.org/10.1088/0004-637X/730/1/24} 

\bibitem[Kerp et al.(1999)]{1999A&A...342..213K} Kerp, J., Burton, W.~B., Egger, R., et al.\ 1999, \aap, 342, 213. \url{https://doi.org/10.48550/arXiv.astro-ph/9810307} 




\bibitem[Khabibullin, Sazonov, \& Sunyaev(2014)]{2014MNRAS.437..327K} Khabibullin, I., Sazonov, S., \& Sunyaev, R.\ 2014, \mnras, 437, 327. \url{https://doi.org/10.1093/mnras/stt1889}
\bibitem[Kocevski et al.(2023)]{2023ApJ...946L..14K} Kocevski, D.~D., Barro, G., McGrath, E.~J., et al.\ 2023, \apjl, 946, L14. \url{https://doi.org/10.3847/2041-8213/acad00}
\bibitem[Koutroumpa et al.(2006)]{2006A&A...460..289K} Koutroumpa, D., Lallement, R., Kharchenko, V., et al.\ 2006, \aap, 460, 289. \url{https://doi.org/10.1051/0004-6361:20065250}
\bibitem[Koutroumpa(2012)]{2012AN....333..341K} Koutroumpa, D.\ 2012, Astronomische Nachrichten, 333, 341. \url{https://doi.org/10.1002/asna.201211666}
\bibitem[Koutroumpa(2023)]{Koutroumpa2023} Koutroumpa, D.\ 2023, Earth Planet. Phys., 8(1), 1–14. \url{https://doi.org/10.26464/epp2023056}
\bibitem[Koutroumpa, Lallement, Raymond, \& Kharchenko(2009)]{2009ApJ...696.1517K} Koutroumpa, D., Lallement, R., Raymond, J.~C., \& Kharchenko, V.\ 2009, \apj, 696, 1517. \url{https://doi.org/10.1088/0004-637X/696/2/1517}
\bibitem[Kraft et al.(2022)]{2022arXiv221109827K} Kraft, R., Markevitch, M., Kilbourne, C., et al.\ 2022, arXiv e-prints, arXiv:2211.09827. \url{https://doi.org/10.48550/arXiv.2211.09827}

\bibitem[Lacki(2014)]{2014MNRAS.444L..39L} Lacki, B.~C.\ 2014, \mnras, 444, L39. \url{https://doi.org/10.1093/mnrasl/slu107} 



\bibitem[Lallement et al.(2005)]{2005Sci...307.1447L} Lallement, R., Qu{\'e}merais, E., Bertaux, J.~L., et al.\ 2005, Science, 307, 1447. \url{https://doi.org/10.1126/science.1107953}
\bibitem[Lallement(2022)]{2022arXiv220301312L} Lallement, R.\ 2022, arXiv e-prints, arXiv:2203.01312. \url{https://doi.org/10.48550/arXiv.2203.01312}

\bibitem[LaRocca et al.(2020)]{2020ApJ...904...54L} LaRocca, D.~M., Kaaret, P., Kuntz, K.~D., et al.\ 2020, \apj, 904, 54. \url{https://doi.org/10.3847/1538-4357/abbdfd} 

\bibitem[Laureijs et al.(2011)]{2011arXiv1110.3193L} Laureijs, R., Amiaux, J., Arduini, S., et al.\ 2011, arXiv e-prints, arXiv:1110.3193. \url{https://doi.org/10.48550/arXiv.1110.3193}
\bibitem[Linsky \& Redfield(2021)]{2021ApJ...920...75L} Linsky, J.~L., \& Redfield, S.\ 2021, \apj, 920, 75. \url{https://doi.org/10.3847/1538-4357/ac1feb}






\bibitem[Liu et al.(2017)]{2017ApJ...834...33L} Liu, W., Chiao, M., Collier, M.~R., et al.\ 2017, \apj, 834, 33. \url{https://doi.org/10.3847/1538-4357/834/1/33} 

\bibitem[Lu \& Aschenbach(2000)]{2000A&A...362.1083L} Lu, F.~J., \& Aschenbach, B.\ 2000, \aap, 362, 1083. \url{https://doi.org/} 


\bibitem[Lyskova et al.(2023)]{2023MNRAS.525..898L} Lyskova, N., Churazov, E., Khabibullin, I.~I., et al.\ 2023, \mnras, 525, 898. \url{https://doi.org/10.1093/mnras/stad2305}
\bibitem[Marshall et al.(2020)]{2020MNRAS.499.3819M} Marshall, M.~A., Ni, Y., Di Matteo, T., et al.\ 2020, \mnras, 499, 3819. \url{https://doi.org/10.1093/mnras/staa2982}
\bibitem[Marshall et al.(2022)]{2022MNRAS.516.1047M} Marshall, M.~A., Watts, K., Wilkins, S., et al.\ 2022, \mnras, 516, 1047. \url{https://doi.org/10.1093/mnras/stac2111}

\bibitem[Mayer, Becker, Predehl, \& Sasaki(2023)]{2023A&A...676A..68M} Mayer, M.~G.~F., Becker, W., Predehl, P., \& Sasaki, M.\ 2023, \aap, 676, A68. \url{https://doi.org/10.1051/0004-6361/202346691}

\bibitem[McCammon et al.(2002)]{2002ApJ...576..188M} McCammon, D., Almy, R., Apodaca, E., et al.\ 2002, \apj, 576, 188. \url{https://doi.org/10.1086/341727}
\bibitem[McComas et al.(2008)]{2008GeoRL..3518103M} McComas, D.~J., Ebert, R.~W., Elliott, H.~A., et al.\ 2008, \grl, 35, L18103. \url{https://doi.org/10.1029/2008GL034896}

\bibitem[Merloni et al.(2012)]{2012arXiv1209.3114M} Merloni, A., Predehl, P., Becker, W., et al.\ 2012, arXiv e-prints, arXiv:1209.3114. \url{https://doi.org/10.48550/arXiv.1209.3114} 

 




\bibitem[Mernier et al.(2023)]{2023arXiv231004499M} Mernier, F., Su, Y., Markevitch, M., et al.\ 2023, arXiv e-prints, arXiv:2310.04499. \url{https://doi.org/10.48550/arXiv.2310.04499} 


\bibitem[Miller, Hodges-Kluck, \& Bregman(2016)]{2016ApJ...818..112M} Miller, M.~J., Hodges-Kluck, E.~J., \& Bregman, J.~N.\ 2016, \apj, 818, 112. \url{https://doi.org/10.3847/0004-637X/818/2/112} 


\bibitem[Miyata, Masai, \& Hughes(2008)]{2008PASJ...60..521M} Miyata, E., Masai, K., \& Hughes, J.~P.\ 2008, \pasj, 60, 521. \url{https://doi.org/10.1093/pasj/60.3.521} 



\bibitem[Negro et al.(2022)]{2022ApJ...927..225N} Negro, M., Fleischhack, H., Zoglauer, A., Digel, S., \& Ajello, M.\ 2022, \apj, 927, 225. \url{https://doi.org/10.3847/1538-4357/ac5326} 

\bibitem[Nelson et al.(2023)]{2023MNRAS.522.3665N} Nelson, D., Byrohl, C., Ogorzalek, A., et al.\ 2023, \mnras, 522, 3665. \url{https://doi.org/10.1093/mnras/stad1195} 



\bibitem[Oppenheimer(2018)]{2018MNRAS.480.2963O} Oppenheimer, B.~D.\ 2018, \mnras, 480, 2963. \url{https://doi.org/10.1093/mnras/sty1918} 


\bibitem[Owen \& Yang(2022)]{2022MNRAS.510.5834O} Owen, E.~R., \& Yang, H.-Y.~K.\ 2022, \mnras, 510, 5834. \url{https://doi.org/10.1093/mnras/stac119} 

\bibitem[Owen \& Yang(2022)]{2022MNRAS.516.1539O} Owen, E.~R., \& Yang, H.-Y.~K.\ 2022, \mnras, 516, 1539. \url{https://doi.org/10.1093/mnras/stac2289} 

\bibitem[Patnaude et al.(2023)]{2023patnaude} Patnaude, D., et al. 2023, ``Studies of Supernovae and Supernova Remnants with the Line Emission Mapper'', in preparation


\bibitem[Pease, Drake, \& Kashyap(2006)]{2006ApJ...636..426P} Pease, D.~O., Drake, J.~J., \& Kashyap, V.~L.\ 2006, \apj, 636, 426. \url{https://doi.org/10.1086/497888} 


\bibitem[Pillepich et al.(2021)]{2021MNRAS.508.4667P} Pillepich, A., Nelson, D., Truong, N., et al.\ 2021, \mnras, 508, 4667. \url{https://doi.org/10.1093/mnras/stab2779}

\bibitem[Planck Collaboration et al.(2016)]{2016A&A...594A...8P} Planck Collaboration, Adam, R., Ade, P.~A.~R., et al.\ 2016, \aap, 594, A8. \url{https://doi.org/10.1051/0004-6361/201525820}

\bibitem[Plucinsky et al.(1996)]{1996ApJ...463..224P} Plucinsky, P.~P., Snowden, S.~L., Aschenbach, B., et al.\ 1996, \apj, 463, 224. \url{https://doi.org/10.1086/177236} 


\bibitem[Ponti et al.(2019)]{2019Natur.567..347P} Ponti, G., Hofmann, F., Churazov, E., et al.\ 2019, \nat, 567, 347. \url{https://doi.org/10.1038/s41586-019-1009-6} 

\bibitem[Ponti et al.(2021)]{2021A&A...646A..66P} Ponti, G., Morris, M.~R., Churazov, E., Heywood, I., \& Fender, R.~P.\ 2021, \aap, 646, A66. \url{https://doi.org/10.1051/0004-6361/202039636} 


\bibitem[Ponti et al.(2023)]{2023A&A...670A..99P} Ponti, G., Sanders, J.~S., Locatelli, N., et al.\ 2023, \aap, 670, A99. \url{https://doi.org/10.1051/0004-6361/202244430} 

\bibitem[Ponti et al.(2023)]{2023A&A...674A.195P} Ponti, G., Zheng, X., Locatelli, N., et al.\ 2023, \aap, 674, A195. \url{https://doi.org/10.1051/0004-6361/202243992}

\bibitem[Predehl et al.(2020)]{2020Natur.588..227P} Predehl, P., Sunyaev, R.~A., Becker, W., et al.\ 2020, \nat, 588, 227. \url{https://doi.org/10.1038/s41586-020-2979-0}
\bibitem[Predehl et al.(2021)]{2021A&A...647A...1P} Predehl, P., Andritschke, R., Arefiev, V., et al.\ 2021, \aap, 647, A1. \url{https://doi.org/10.1051/0004-6361/202039313}

\bibitem[Putman, Peek, \& Joung(2012)]{2012ARA&A..50..491P} Putman, M.~E., Peek, J.~E.~G., \& Joung, M.~R.\ 2012, \araa, 50, 491. \url{https://doi.org/10.1146/annurev-astro-081811-125612}

\bibitem[Roberts \& Wang(2015)]{2015MNRAS.449.1340R} Roberts, S.~R., \& Wang, Q.~D.\ 2015, \mnras, 449, 1340. \url{https://doi.org/10.1093/mnras/stv319} 



\bibitem[Richardson \& Cane(2004)]{2004JGRA..109.9104R} Richardson, I.~G. \& Cane, H.~V.\ 2004, Journal of Geophysical Research (Space Physics), 109, A09104. \url{https://doi.org/10.1029/2004JA010598}
\bibitem[Roettgering et al.(1994)]{1994A&AS..108...79R} Roettgering, H.~J.~A., Lacy, M., Miley, G.~K., Chambers, K.~C., \& Saunders, R.\ 1994, \aaps, 108, 79. \url{https://doi.org/}


\bibitem[Savage \& Sembach(1996)]{1996ARA&A..34..279S} Savage, B.~D., \& Sembach, K.~R.\ 1996, \araa, 34, 279. \url{https://doi.org/10.1146/annurev.astro.34.1.279}

\bibitem[Schellenberger et al.(2023)]{2023arXiv230701259S} Schellenberger, G., Bogd{\'a}n, {\'A}., ZuHone, J.~A., et al.\ 2023, arXiv e-prints, arXiv:2307.01259. \url{https://doi.org/10.48550/arXiv.2307.01259} 

\bibitem[Schneider et al.(2006)]{2006A&A...458..855S} Schneider, N., Bontemps, S., Simon, R., et al.\ 2006, \aap, 458, 855. \url{https://doi.org/10.1051/0004-6361:20065088} 



\bibitem[Selig, Vacca, Oppermann, \& En{\ss}lin(2015)]{2015A&A...581A.126S} Selig, M., Vacca, V., Oppermann, N., \& En{\ss}lin, T.~A.\ 2015, \aap, 581, A126. \url{https://doi.org/10.1051/0004-6361/201425172}

\bibitem[Shelton, Kwak, \& Henley(2012)]{2012ApJ...751..120S} Shelton, R.~L., Kwak, K., \& Henley, D.~B.\ 2012, \apj, 751, 120. \url{https://doi.org/10.1088/0004-637X/751/2/120} 



\bibitem[Snowden et al.(1997)]{1997ApJ...485..125S} Snowden, S.~L., Egger, R., Freyberg, M.~J., et al.\ 1997, \apj, 485, 125. \url{https://doi.org/10.1086/304399} 


\bibitem[Stelzer et al.(2006)]{2006A&A...448..293S} Stelzer, B., Micela, G., Flaccomio, E., Neuh{\"a}user, R., \& Jayawardhana, R.\ 2006, \aap, 448, 293. \url{https://doi.org/10.1051/0004-6361:20053677} 


\bibitem[Stern et al.(2023)]{2023arXiv230600092S} Stern, J., Fielding, D., Hafen, Z., et al.\ 2023, arXiv e-prints, arXiv:2306.00092. \url{https://doi.org/10.48550/arXiv.2306.00092} 



\bibitem[Stone et al.(2005)]{2005Sci...309.2017S} Stone, E.~C., Cummings, A.~C., McDonald, F.~B., et al.\ 2005, Science, 309, 2017. \url{https://doi.org/10.1126/science.1117684}

\bibitem[Su, Slatyer, \& Finkbeiner(2010)]{2010ApJ...724.1044S} Su, M., Slatyer, T.~R., \& Finkbeiner, D.~P.\ 2010, \apj, 724, 1044. \url{https://doi.org/10.1088/0004-637X/724/2/1044} 


\bibitem[Sugiyama et al.(2023)]{2023arXiv230917246S} Sugiyama, H., Ueda, M., Fukushima, K., et al.\ 2023, arXiv e-prints, arXiv:2309.17246. \url{https://doi.org/10.48550/arXiv.2309.17246} 


\bibitem[Sunyaev et al.(2021)]{2021A&A...656A.132S} Sunyaev, R., Arefiev, V., Babyshkin, V., et al.\ 2021, \aap, 656, A132. \url{https://doi.org/10.1051/0004-6361/202141179}

\bibitem[Tozzi et al.(2022)]{2022A&A...667A.134T} Tozzi, P., Gilli, R., Liu, A., et al.\ 2022, \aap, 667, A134. \url{https://doi.org/10.1051/0004-6361/202244337}

\bibitem[Truemper(1982)]{1982AdSpR...2d.241T} Truemper, J.\ 1982, Advances in Space Research, 2, 241. \url{https://doi.org/10.1016/0273-1177(82)90070-9} 


\bibitem[Truong et al.(2023)]{2023MNRAS.525.1976T} Truong, N., Pillepich, A., Nelson, D., et al.\ 2023, \mnras, 525, 1976. \url{https://doi.org/10.1093/mnras/stad2216} 



\bibitem[Tumlinson, Peeples, \& Werk(2017)]{2017ARA&A..55..389T} Tumlinson, J., Peeples, M.~S., \& Werk, J.~K.\ 2017, \araa, 55, 389. \url{https://doi.org/10.1146/annurev-astro-091916-055240}

\bibitem[Uyan{\i}ker et al.(2001)]{2001A&A...371..675U} Uyan{\i}ker, B., F{\"u}rst, E., Reich, W., Aschenbach, B., \& Wielebinski, R.\ 2001, \aap, 371, 675. \url{https://doi.org/10.1051/0004-6361:20010387} 

\bibitem[Valentini et al.(2017)]{2017MNRAS.470.3167V} Valentini, M., Murante, G., Borgani, S., et al.\ 2017, \mnras, 470, 3167. \url{https://doi.org/10.1093/mnras/stx1352} 


\bibitem[Valentini et al.(2023)]{2023MNRAS.518.1128V} Valentini, M., Dolag, K., Borgani, S., et al.\ 2023, \mnras, 518, 1128. \url{https://doi.org/10.1093/mnras/stac2110}

\bibitem[Veilleux, Cecil, \& Bland-Hawthorn(2005)]{2005ARA&A..43..769V} Veilleux, S., Cecil, G., \& Bland-Hawthorn, J.\ 2005, \araa, 43, 769. \url{https://doi.org/10.1146/annurev.astro.43.072103.150610} 



\bibitem[Qu et al.(2020)]{2020ApJ...894..142Q} Qu, Z., Bregman, J.~N., Hodges-Kluck, E., Li, J.-T., \& Lindley, R.\ 2020, \apj, 894, 142. \url{https://doi.org/10.3847/1538-4357/ab774e} 


\bibitem[Welsh \& Lallement(2005)]{2005A&A...436..615W} Welsh, B.~Y., \& Lallement, R.\ 2005, \aap, 436, 615. \url{https://doi.org/10.1051/0004-6361:20042611}

\bibitem[Yang, Ruszkowski, \& Zweibel(2018)]{2018Galax...6...29Y} Yang, H.-Y., Ruszkowski, M., \& Zweibel, E.\ 2018, Galaxies, 6, 29. \url{https://doi.org/10.3390/galaxies6010029} 

\bibitem[Yang, Ruszkowski, \& Zweibel(2022)]{2022NatAs...6..584Y} Yang, H.-Y.~K., Ruszkowski, M., \& Zweibel, E.~G.\ 2022, Nature Astronomy, 6, 584. \url{https://doi.org/10.1038/s41550-022-01618-x} 


\bibitem[Zhang et al.(2023)]{2023arXiv231002225Z} Zhang, C., Zhuravleva, I., Markevitch, M., et al.\ 2023, arXiv e-prints, arXiv:2310.02225. \url{https://doi.org/10.48550/arXiv.2310.02225} 

\bibitem[Zucker et al.(2022)]{2022Natur.601..334Z} Zucker, C., Goodman, A.~A., Alves, J., et al.\ 2022, \nat, 601, 334. \url{https://doi.org/10.1038/s41586-021-04286-5} 



\bibitem[ZuHone et al.(2023)]{2023arXiv230701269Z} ZuHone, J.~A., Schellenberger, G., Ogorzalek, A., et al.\ 2023, arXiv e-prints, arXiv:2307.01269. \url{https://doi.org/10.48550/arXiv.2307.01269} 



%https://arxiv.org/abs/2310.04499
\end{thebibliography}
\end{document}